\documentclass[3p,12pt]{elsarticle}
\usepackage{threeparttable,booktabs,tabularx}
\usepackage[fleqn]{amsmath}
\usepackage{cases}
\usepackage{multirow}
\usepackage{xcolor}
\usepackage[normalem]{ulem}
\usepackage{tabularx}
\usepackage{siunitx}
\usepackage{comment}
\usepackage{algorithm}
\usepackage{algpseudocode}
\usepackage{algorithmicx}
\usepackage{grffile}
\usepackage{rotating}
\usepackage{array}
\usepackage{lineno}
\usepackage{hyperref}
\usepackage{makecell}
\usepackage{graphicx,enumerate}
\usepackage{amssymb}
\usepackage{bm,amsmath}
\allowdisplaybreaks  
\usepackage{subfigure}
\usepackage{caption,color}
\usepackage{threeparttable}
\usepackage{subeqnarray}
\usepackage{multirow}
\usepackage{lscape}
\usepackage{rotating}
\usepackage{graphics}
\floatname{algorithm}{Algorithm}
\usepackage{epstopdf}
\journal{}
\usepackage{minted}
\usepackage{listings}

\biboptions{numbers,sort&compress} 

\usepackage{natbib}
\AtBeginEnvironment{thebibliography}{%
  \footnotesize
  \interlinepenalty=10000
}

\usepackage{CJKutf8}

\begin{document}
\begin{CJK}{UTF8}{gbsn}

\begin{frontmatter}

\title{Parallel simulation of rarefied gas flows on unstructured meshes using the DIG-augmented DSMC method}

\author[]{Tao Huang }
\author[]{Liyan Luo }
\author[]{Hong Deng }
\author[]{Lei Wu\corref{Boss}}

\cortext[Boss]{Corresponding author: wul@sustech.edu.cn}

\address{Department of Mechanics and Aerospace Engineering, Southern University of Science and Technology, 518055 Shenzhen, China}

\begin{abstract}
While the direct simulation Monte Carlo (DSMC) is a mainstream stochastic particle method for simulating rarefied gas flows, it incurs excessively high computational costs in the near-continuum regime. As a hybrid acceleration approach coupling DSMC with macroscopic synthetic equations, the direct intermittent general synthetic iterative scheme (DIG) delivers fast convergence and asymptotic-preserving characteristics, which effectively alleviate the kinetic-scale limitations inherent to standard DSMC. In this study, we develop a parallel DIG-augmented DSMC solver for three-dimensional rarefied gas flow simulations on unstructured meshes. On top of the standard DSMC algorithms for particle transport and collisions, a reliable intermittent coupling framework is constructed to exchange macroscopic flow data between the stochastic DSMC module and deterministic macroscopic synthetic equations. For parallel execution on unstructured grids, we employ a hybrid MPI architecture equipped with virtual cells to enable local particle tracking and batch inter-rank particle migration. A graph-partitioning-based dynamic load balancing strategy is also integrated to mitigate uneven particle distribution over computational domains. Three-dimensional lid-driven cavity flow and hypersonic flows over a sphere, the Apollo reentry capsule, and the International Space Station are employed for numerical validation. Results demonstrate that the proposed solver achieves satisfactory agreement with the SPARTA DSMC. 
Leveraging DIG’s fast convergence and asymptotic-preserving properties, the required number of spatial cells and statistical sampling steps are drastically decreased, leading to substantial reductions in computational memory and runtime, e.g., the DIG solver achieves a roughly 360-fold reduction in total core hours and a 50-fold cut in memory consumption for cavity flow, with further improvements expected as the Knudsen number declines. This work presents an efficient high-performance numerical tool for high-fidelity simulations of rarefied flows over complex geometries. The C++ code is available in the developer’s repository at the following link: \url{https://github.com/ablackjoker/DIG-appolo}.

\end{abstract}


\end{frontmatter}

\section{Introduction}\label{sec:1}

Rarefied gas flows appear in many modern engineering applications, including spacecraft reentry~\cite{votta2013reentry,ivanov1998hypersonicRarefied}, vacuum systems in nuclear fusion~\cite{tantos2024vacuumDTT}, and extreme ultraviolet lithography~\cite{wang2024euvContamination}. In these problems, the molecular mean free path may become comparable to the characteristic flow length, and the local thermodynamic equilibrium assumption breaks down. The Navier--Stokes--Fourier (NSF) equations, which rely on the Newton law of stress and the Fourier law of heat conduction, cannot accurately describe the rarefied gas flow~\cite{boyd1995continuumBreakdown,lofthouse2008slipJump}. Kinetic description based on the Boltzmann equation provides a unified description over the whole range of gas rarefaction and become necessary when NSF equations lose validity.

Numerical solution of the kinetic equation is required to characterize rarefied gas flows in engineering contexts. Boltzmann solvers fall into two major categories: deterministic discrete velocity schemes and stochastic particle approaches. Discrete velocity methods solve the distribution function in both physical and molecular velocity spaces~\cite{aristov2001directBoltzmann,wang2018DVMComparison}. For three-dimensional (3D) hypersonic flows, especially when internal energy modes are considered, the required molecular velocity space range and internal energy variables may lead to large memory consumption and computational cost~\cite{schwartzentruber2007modularHybrid}. 
In contrast, the direct simulation Monte Carlo (DSMC) method replaces gas molecules using a limited set of simulation particles, eliminating the need to construct a velocity and internal energy grid. This particle formulation renders DSMC well-suited for high-speed flows featuring pronounced nonequilibrium phenomena and intricate physico-chemical collision processes. Despite its modest memory footprint, DSMC demands extensive statistical averaging, trading computational time for reduced memory overhead~\cite{bird1994dsmc,white2018dsmcFoam,borgnakke1975polyatomic}.

The main limitation of conventional DSMC appears in near-continuum flow regime~\cite{boyd1995continuumBreakdown,kannenberg2000particleResolution}, where the molecular mean free path is much smaller than the characteristic flow length. Since molecular transport and intermolecular collisions are treated separately, the spatial cell size and time step are usually required to be smaller than the local molecular mean free path and mean collision time, respectively. This kinetic scale restriction leads to substantial increases in the numbers of spatial cells and decreases in time step~\cite{kannenberg2000particleResolution,sun2011properCellDimension}. In addition, macroscopic quantities are obtained from statistical particle samples, and long time averaging is often needed to reduce stochastic noise~\cite{chen1996statisticalErrorDSMC,hadjiconstantinou2003statisticalError,sun2004macroscopicProperties}. 

Over the past two decades, numerous multiscale kinetic algorithms have been proposed to mitigate the computational overhead of rarefied gas flow simulations~\cite{wu2022rarefiedGasDynamics}. Hybrid particle-continuum frameworks cut particle-based computations by deploying continuum solvers in near-continuum zones while retaining particle schemes for rarefied domains~\cite{schwartzentruber2007modularHybrid}. Asymptotic-preserving particle approaches ease stringent kinetic-scale constraints, either via modified particle evolution rules~\cite{pareschi2001trmc,ren2014apmc,fei2023trmcNS} or macroscopic flow-guided corrections~\cite{degond2010mgmc}. Additionally, a stochastic particle solver has been formulated for the Fokker–Planck equation, substituting binary collisions with Langevin dynamics to permit enlarged time steps within continuum flow regimes~\cite{gorji2011fokkerplanck}.
The unified gas-kinetic scheme~\cite{xu2010ugks,zhu2016implicitUGKS,guo2013dugks} couples molecular transport and collision in the numerical flux based on the analytical solution of the BGK kinetic model, so that the mesh size can be much larger than the kinetic scale. The unified gas-kinetic wave-particle method~\cite{liu2020ugkwp} performs wave-particle decomposition according to local collision scales, using deterministic hydrodynamic waves to describe macroscopic evolution mainly in continuum regions while retaining stochastic particle transport in rarefied regions, thus providing another efficient route for multiscale simulation~\cite{chen2020ugkwp3d,long2024ugkwp}. The general synthetic iterative scheme (GSIS) accelerates steady-state simulations by concurrently solving mesoscopic kinetic equations and macroscopic synthetic relations. Rigorously derived and closed using kinetic numerical solutions, these synthetic relations enable global exchange of macroscopic flow quantities, delivering fast convergence and asymptotic NSF preserving properties~\cite{su2020steadyGSIS,su2020syntheticIteration}.

Given DSMC’s inherent capability to resolve complex physico-chemical processes, retaining its native particle framework while incorporating macroscopic flow data as an acceleration strategy remains an appealing route to boost simulation efficiency. The direct intermittent GSIS-DSMC coupling method (DIG) was developed from this idea~\cite{luo2024directIntermittentDIG}. In this method, DSMC is first advanced for hundreds particle time steps, after which the sampled macroscopic moments and nonequilibrium closure information are supplied to the macroscopic synthetic equations. The corrected macroscopic state obtained from the synthetic equations is then fed back into the particle distribution, so that the DSMC solution is guided toward steady-state convergence while the basic particle procedure is retained. 
Owing to the tailored constitutive relations embedded within its macroscopic synthetic equations, the DIG framework distinguishes itself from moment-guided DSMC~\cite{degond2010mgmc}. Beyond enhanced numerical stability~\cite{SIAM2025}, DIG exhibits rapid convergence and asymptotic preservation of the NSF limit. In the near-continuum flow regime, this characteristic lowers sampling costs for both transient and steady-state computations and permits mesh cells substantially larger than the local molecular mean free path.

Prior DIG work has verified its acceleration efficacy for monatomic~\cite{luo2024directIntermittentDIG}, polyatomic~\cite{luo2026polyatomicDIG}, mixed~\cite{luo2026mixtureDIG} and reactive gas flows~\cite{Deng2026Chemical}, centering on acceleration and asymptotic-preserving mechanisms rather than high-performance parallel frameworks. 
This work presents a fully parallel DIG solver for efficient rarefied flow simulations on 3D unstructured-mesh. It preserves standard DSMC molecular collision, transport, surface interaction and energy relaxation routines~\cite{bird1994dsmc,borgnakke1975polyatomic} with periodic macroscopic synthetic corrections~\cite{luo2024directIntermittentDIG,luo2026polyatomicDIG}. Key advances include tightly integrated DSMC–macroscopic coupling, stability enhancement, and parallel pipelines for cross-process particle transfer and dynamic load balancing.

The remainder of this paper is organized as follows. Section~\ref{sec:DIG_method} introduces the basic DSMC formulation and the macroscopic synthetic equations. Section~\ref{sec:Coupling strategy} describes the intermittent coupling strategy, including extraction of high-order closure terms beyond the NSF approximation, particle distribution adjustment, and statistical treatment. Section~\ref{sec:Solver} presents the 3D parallelization strategy, including local mesh construction, particle migration, MPI communication, and dynamic load balancing. Section~\ref{sec:Numerical results} presents numerical examples and discusses the accuracy and efficiency of the proposed parallel DIG solver. Section~\ref{sec:conclusion} summarizes the main contributions and outlines future extensions.

\section{The DIG method}\label{sec:DIG_method}

This section outlines the formulation of the DIG method for polyatomic gase flows. We first introduce the DSMC particle solver, followed by the derivation of the associated macroscopic synthetic equations. These equations govern the evolution of low-order flow moments and incorporate the high-order terms (HoTs) to capture non-equilibrium phenomena.

\subsection{The DSMC method}

In the absence of external forces, the kinetic behavior of a rarefied gas can be described by the Boltzmann-type equation \cite{WangCS}
\begin{equation}
    \frac{\partial f}{\partial t}
    +
    \boldsymbol{\xi}\cdot\nabla_{\mathbf{x}} f
    =
    \mathcal{Q}(f,f),
    \label{eq:boltzmann_dsmc}
\end{equation}
where $f(\mathbf{x},\boldsymbol{\xi}, I, t)$ is the molecular distribution function, $t$ is the time, $\mathbf{x}$ is the physical space, $\boldsymbol{\xi}$ is the molecular velocity, $I$ is the molecular internal energy (e.g., rotational, vibrational, and electronic energies), and $\mathcal{Q}(f,f)$ denotes the binary collision operator. The high-dimensional nonlinear collision integral poses significant challenges to numerical simulation of rarefied gas flows.

The DSMC method serves as a stochastic particle approximation for Eq.~\eqref{eq:boltzmann_dsmc}. The DSMC method mimics the molecular motion and collision processes of a rarefied gas by a finite number of simulation particles. Within each time step, simulation particles are first moved in physical space according to their molecular velocities. Then, particle pairs are randomly selected within the same computational cell, and collisions are sampled according to the prescribed molecular collision model.  Without loss of generality, we consider polyatomic gases with $d_r$ rotational degrees of freedom. In DSMC, each simulation particle contains not only its position and translational velocity, but also its rotational internal energy. The state of the $p$-th simulation particle is expressed as
\begin{equation}
    \mathcal{P}_p
    =
    \left(
    \mathbf{x}_p,\mathbf{v}_p,I_{r,p};N_{\mathrm{eff}}
    \right),
    \label{eq:dsmc_particle_state}
\end{equation}
where $\mathbf{x}_p$ and $\mathbf{v}_p$ are the position and velocity of the $p$-th particle, $I_{r,p}$ is its rotational internal energy, and $N_{\mathrm{eff}}$ is the number of real molecules represented by one simulation particle. The translational collision process is treated by Bird's no-time-counter (NTC) method, together with the variable soft sphere (VSS) model for determining the collision probability and post-collisional velocity. The exchange between translational and rotational energies is described by the Borgnakke--Larsen model \cite{borgnakke1975polyatomic}. When the VSS model is combined with Bird's NTC collision-pair selection procedure, the probability of translational--rotational energy exchange is written as
\begin{equation}
    P_{\mathrm{inelastic}}
    =
    \frac{
        \alpha(5-2\omega)(7-2\omega)
    }{
        5(\alpha+1)(\alpha+2)Z_r
    },
    \label{eq:inelastic_probability}
\end{equation}
where $\alpha$ is the angular scattering parameter in the VSS model \cite{bird1994dsmc}, $\omega$ is the viscosity index, and $Z_r$ is the rotational collision number. When an inelastic energy exchange event is selected, the available energy of the particle pair is redistributed between the translational and rotational modes according to the Borgnakke--Larsen sampling procedure. Otherwise, only the post-collisional translational velocities are updated, while the rotational internal energy remains unchanged. As a result, in a spatially homogeneous problem, the energy exchange between the translational and rotational modes follows from the Jeans--Landau relaxation relation,
\begin{equation}
    \frac{\mathrm{d}T_r}{\mathrm{d}t}
    =
    \frac{T-T_r}{Z_r\tau},
    \label{eq:jeans_landau}
\end{equation}
where $\tau={\mu}/{p_t}$ is the mean collision time ($\mu$ is the shear viscosity and $p_t$  is the translational pressure). 

Macroscopic quantities in DSMC are obtained from particle averages inside each computational cell.
For a computational cell with volume $V_{\mathrm{c}}$ and $N_{p,c}$ simulation particles, the density, velocity, stress tensor, translational temperature, rotational temperature, translational heat flux, and rotational heat flux are sampled as
\begin{equation}    \label{eq:dsmc_temperatures}
\begin{aligned}
    \rho
    =&
    \frac{N_{\mathrm{eff}}}{V_{\mathrm{c}}}N_{p,c},
    \quad
    u_i
    =
    \frac{1}N_{p,c}
    \sum_{p=1}^{N_{p,c}} v_{i,p},\\
    T_t
    =&
    \frac{1}{3N_{p,c}}
    \sum_{p=1}^{N_{p,c}}
    \left|\mathbf{v}_p-\mathbf{u}\right|^2,
    \quad
    T_r
    =
    \frac{1}{d_rN_{p,c}}
    \sum_{p=1}^{N_{p,c}}
    I_{r,p},\\
    \sigma^{\mathrm{DSMC}}_{ij}
    =&
    \frac{N_{\mathrm{eff}}}{V_{\mathrm{c}}}
    \sum_{p=1}^{N_{p,c}}
    \left[
    \left(v_{i,p}-u_i\right)
    \left(v_{j,p}-u_j\right)
    -
    \frac{\delta_{ij}}{3}
    \left|\mathbf{v}_p-\mathbf{u}\right|^2
    \right],\\
    q^{\mathrm{DSMC}}_{t,i}
    =&
    \frac{N_{\mathrm{eff}}}{2V_{\mathrm{cell}}}
    \sum_{p=1}^{N_{p,c}}
    \left(v_{i,p}-u_i\right)
    \left|\mathbf{v}_p-\mathbf{u}\right|^2,
    \quad
    q^{\mathrm{DSMC}}_{r,i}
    =
    \frac{N_{\mathrm{eff}}}{V_{\mathrm{c}}}
    \sum_{p=1}^{N_{p,c}}
    \left(v_{i,p}-u_i\right)I_{r,p}.
\end{aligned}
\end{equation}
Here, $\delta_{ij}$ is the Kronecker delta. In the above formulas, $\rho$, $\mathbf{u}$, $T_t$, and $T_r$ describe the cell averaged macroscopic state, while $\boldsymbol{\sigma}^{\mathrm{DSMC}}$, $\mathbf{q}^{\mathrm{DSMC}}_t$, and $\mathbf{q}^{\mathrm{DSMC}}_r$ characterize the nonequilibrium moments determined by the peculiar velocities and rotational internal energies of simulation particles. 
The equilibrium temperature associated with the translational and rotational modes is defined as
\begin{equation}
  T=\frac{3T_t+d_rT_r}{3+d_r}. 
\end{equation}

Note that dimensionless variables are adopted, with the reference length, density, temperature, and velocity taken as $L_0$, $\rho_0$, $T_0$, and $\sqrt{RT_0}$, respectively. The rotational internal energy $I_{r,p}$ is nondimensionalized by $k_BT_0$. Therefore, the molecular mass and the Boltzmann constant do not appear explicitly in the following sampling formulas.

\subsection{Macroscopic synthetic equations}

Taking the velocity moments of Eq.~\eqref{eq:boltzmann_dsmc}, the macroscopic conservation equations for the polyatomic gas are obtained:
\begin{equation}\label{eq:macro}
\begin{aligned}
     \frac{\partial \rho}{\partial t}
    +
    \nabla\cdot(\rho\mathbf{u})
    =
    0,\\
    \frac{\partial(\rho\mathbf{u})}{\partial t}
    +
    \nabla\cdot(\rho\mathbf{u}\mathbf{u})
    +
    \nabla\cdot
    \left(
    \rho T_t\mathbf{I}
    +
    \boldsymbol{\sigma}
    \right)
    =
    0, \\
     \frac{\partial(\rho e)}{\partial t}
    +
    \nabla\cdot(\rho e\mathbf{u})
    +
    \nabla\cdot
    \left(
    \rho T_t\mathbf{u}
    +
    \boldsymbol{\sigma}\cdot\mathbf{u}
    +
    \mathbf{q}_t
    +
    \mathbf{q}_r
    \right)
    =
    0,\\
     \frac{\partial(\rho e_r)}{\partial t}
    +
    \nabla\cdot(\rho e_r\mathbf{u})
    +
    \nabla\cdot\mathbf{q}_r
    =
    \frac{d_r\rho}{2}
    \frac{T-T_r}{Z_r\tau},
\end{aligned}
\end{equation}
where $e$ is the specific total energy and $e_r$ is the specific rotational energy, given by
\begin{equation}
    e
    =
    \frac{3T_t}{2}
    +
    \frac{|\mathbf{u}|^2}{2} + e_r,
    \quad
    e_r
    =
    \frac{d_rT_r}{2}.
    \label{eq:energy_definitions}
\end{equation}

The stress tensor $\boldsymbol{\sigma}$, translational heat flux $\mathbf{q}_t$, and rotational heat flux $\mathbf{q}_r$ are required to close the macroscopic equations. In the continuum limit, they reduce to the NSF constitutive relations as per the Chapman-Enskog expansion \cite{chapman1970mathematical,mason1962heat}: 
\begin{equation}\label{eq:nsf_heat_flux} 
\begin{aligned}
     \boldsymbol{\sigma}^{\mathrm{NSF}}
    &=
    -\mu
    \left[
    \nabla\mathbf{u}
    +
    (\nabla\mathbf{u})^T
    -
    \frac{2}{3}
    (\nabla\cdot\mathbf{u})\mathbf{I}
    \right],\\
     \mathbf{q}^{\mathrm{NSF}}_t
    &=
    -\kappa_t\nabla T_t,
    \quad
    \mathbf{q}^{\mathrm{NSF}}_r
    =
    -\kappa_r\nabla T_r.   
\end{aligned}
\end{equation}
The translational and rotational thermal conductivities are determined by the heat-flux relaxation matrix:
\begin{equation}
    \begin{bmatrix}
        \kappa_t\\
        \kappa_r
    \end{bmatrix}
    =
    \frac{\mu}{2}
    \begin{bmatrix}
        A_{tt} & A_{tr}\\
        A_{rt} & A_{rr}
    \end{bmatrix}^{-1}
    \begin{bmatrix}
        5\\
        d_r
    \end{bmatrix},
    \label{eq:thermal_conductivity_matrix}
\end{equation}
where $A_{tt}$, $A_{tr}$, $A_{rt}$, and $A_{rr}$ are the relaxation-rate coefficients associated with the translational and rotational heat fluxes \cite{Li2021Uncertainty}. 

For rarefied gas flows, the stress and heat flux not only contains the NSF constitutive relations, but also HoTs that account for the rarefaction effects. In general, they are written as \cite{luo2024directIntermittentDIG,luo2026mixtureDIG}
\begin{equation}
    \boldsymbol{\sigma}
    =
    \boldsymbol{\sigma}^{\mathrm{NSF}}
    +
    \mathrm{HoT}_{\sigma},
    \quad
    \mathbf{q}_t
    =
    \mathbf{q}^{\mathrm{NSF}}_t
    +
    \mathrm{HoT}_{q_t},
    \quad
    \mathbf{q}_r
    =
    \mathbf{q}^{\mathrm{NSF}}_r
    +
    \mathrm{HoT}_{q_r},
    \label{eq:synthetic_closure}
\end{equation}
where $\mathrm{HoT}_{\sigma}$, $\mathrm{HoT}_{q_t}$, and $\mathrm{HoT}_{q_r}$ represent the higher-order terms (HoT) corresponding to the stress tensor, translational heat flux, and rotational heat flux, respectively. These terms vary with flow conditions, and their expressions can only be numerically retrieved from DSMC simulations; further details are provided in the subsequent sections.
Substituting Eq.~\eqref{eq:synthetic_closure} into Eq.~\eqref{eq:macro} yields the macroscopic synthetic governing equations for polyatomic gases.

The core procedure of the DIG method relies on intermittent two-way coupling between the stochastic DSMC scheme and deterministic macroscopic synthetic equations. Specifically, within each outer coupling cycle, conventional DSMC is advanced for $N_s$ steps. The HoTs extracted from DSMC simulations then supply closure terms for the macroscopic synthetic equations. These equations are subsequently solved over the inner coupling cycle, and the resulting updated low-order macroscopic states are fed back to resample DSMC particles. This coupled framework proceeds via sequential iteration of three stages: DSMC statistical sampling, macroscopic equation solution, and particle distribution recalibration. The last two stages are detailed in Refs.~\cite{luo2026mixtureDIG,luo2026polyatomicDIG} and will not be restated herein. The DSMC sampling will be elaborated below. 

\section{Statistical treatment}\label{sec:Coupling strategy}

DSMC itself exhibits strong robustness, yet it can produce inaccurate results when spatial and temporal resolutions are inadequate.  The DIG eliminates these deficiencies with the help of macroscopic synthetic equation~\cite{luo2024directIntermittentDIG}. Accordingly, the stability of the DIG framework hinges on its macroscopic solver, particularly the quality of HoTs sampled from DSMC data. This section details the procedure for extracting HoTs from DSMC simulations.

\subsection{Extraction of HoT corrections}\label{Sec:extraction}

Let the $k$-th outer coupling cycle contain $N_s$ DSMC time steps, whose indices are $m=(k-1)N_s+1,\ldots,kN_s$. At each DSMC time step, the particle moments in the cell $c$ are accumulated into an instantaneous summation variable $\Phi_c^m$. For polyatomic gas, $\Phi_c^m$ contains the moments required to evaluate the density, momentum, translational and rotational temperatures, stress tensor, and translational/rotational heat flux.

The outer coupling interval $N_s$ governs both the statistical sampling quality and the update frequency of the macroscopic governing equations~\cite{luo2024directIntermittentDIG}. Increasing $N_s$ widens the sampling window for each outer coupling cycle but reduces the correction rate of the macroscopic synthetic equations, which can exacerbate the buildup of numerical dissipation in DSMC. Conversely, a small $N_s$ suppresses such numerical dissipation at the cost of introducing severe statistical noise into the sampled quantities passed to the macroscopic solver. In practice, we typically adopt $N_s=100$, and employ an exponentially weighted moving time average to reduce statistical noise:~\cite{jenny-2009}
\begin{equation}
	\overline{\Phi}_c^m
	=
	\frac{n_a-1}{n_a}
	\overline{\Phi}_c^{m-1}
	+
	\frac{1}{n_a}
	\Phi_c^m ,
	\label{eq:exponential_average_moments_ch3}
\end{equation}
where $\overline{\Phi}_c^m$ is the smoothed summation variable, and $n_a$ denotes the number of effective samples in the moving average. A larger $n_a$ gives more weight to previous samples and suppresses statistical noise more strongly, whereas a smaller $n_a$ makes the averaged quantity respond more rapidly to the current DSMC samples. Usually we choose $n_a=1000$.

According to the synthetic closure in Eq.~\eqref{eq:synthetic_closure}, the HoT corrections are determined by the difference between the stress and heat flux sampled from DSMC and the corresponding NSF constitutive quantities. At the end of the $k$-th outer coupling cycle, the exponential averaging provides the low-order macroscopic variables in cell $c$ as
\begin{equation}
	M_c^{k+\frac{1}{2}}
	=
	\left(
	\rho_c^{k+\frac{1}{2}},
	\mathbf{u}_c^{k+\frac{1}{2}},
	T_{t,c}^{k+\frac{1}{2}},
	T_{r,c}^{k+\frac{1}{2}}
	\right),
	\label{eq:sampled_macro_state_ch3}
\end{equation}
where the superscript $k+\frac{1}{2}$ denotes the intermediate state after DSMC sampling and before the macroscopic synthetic equations are updated. The NSF stress tensor $\boldsymbol{\sigma}_c^{\mathrm{NSF},k+\frac{1}{2}}$ and heat fluxes ($\mathbf{q}_{t,c}^{\mathrm{NSF},k+\frac{1}{2}}, \mathbf{q}_{r,c}^{\mathrm{NSF},k+\frac{1}{2}}$) are calculated from the same sampled macroscopic variables and their spatial gradients.
The corresponding quantities sampled from DSMC are $
\boldsymbol{\sigma}_c^{\mathrm{DSMC},k+\frac{1}{2}}$,
$\mathbf{q}_{t,c}^{\mathrm{DSMC},k+\frac{1}{2}}$,
$\mathbf{q}_{r,c}^{\mathrm{DSMC},k+\frac{1}{2}}$.
Therefore, the HoT corrections are calculated as
\begin{equation}
	\begin{aligned}
		\mathrm{HoT}_{\sigma,c}^{k+\frac{1}{2}}
		&=
		\boldsymbol{\sigma}^{\mathrm{DSMC},k+\frac{1}{2}}_c
		-
		\boldsymbol{\sigma}^{\mathrm{NSF},k+\frac{1}{2}}_c,\\
		\mathrm{HoT}_{q_t,c}^{k+\frac{1}{2}}
		&=
		\mathbf{q}^{\mathrm{DSMC},k+\frac{1}{2}}_{t,c}
		-
		\mathbf{q}^{\mathrm{NSF},k+\frac{1}{2}}_{t,c},\\
		\mathrm{HoT}_{q_r,c}^{k+\frac{1}{2}}
		&=
		\mathbf{q}^{\mathrm{DSMC},k+\frac{1}{2}}_{r,c}
		-
		\mathbf{q}^{\mathrm{NSF},k+\frac{1}{2}}_{r,c}.
	\end{aligned}
	\label{eq:hot_extraction_ch3}
\end{equation}

The stress tensor and heat fluxes used in the macroscopic synthetic equations are expressed as
\begin{equation}
	\boldsymbol{\sigma}_c
	=
	\boldsymbol{\sigma}_c^{\mathrm{NSF}}
	+
	\mathrm{HoT}_{\sigma,c}^{k+\frac{1}{2}},
	\qquad
	\mathbf{q}_{t,c}
	=
	\mathbf{q}_{t,c}^{\mathrm{NSF}}
	+
	\mathrm{HoT}_{q_t,c}^{k+\frac{1}{2}},
	\qquad
	\mathbf{q}_{r,c}
	=
	\mathbf{q}_{r,c}^{\mathrm{NSF}}
	+
	\mathrm{HoT}_{q_r,c}^{k+\frac{1}{2}} .
	\label{eq:hot_used_in_macro_solver_ch3}
\end{equation}
Note that when solving the macroscopic synthetic equations, these intermediate terms are fixed, while the density, velocity, translational/rotational temperature (denoted by $M^{k+1}$) are updated when solving the synthetic equation towards the steady state \cite{luo2026mixtureDIG}. 

Once the flow reaches steady state, cell-averaged moments are computed only after discarding the initial transient evolution, that is, all macroscopic quantities sampled in the subsequent DSMC iterations are retained for averaging. Let $N_{\mathrm{ss}}$ be the number of steps discarded
before statistical averaging. When the DSMC evolution step $n$ satisfies $n>N_{\mathrm{ss}}$, the statistical average of the
cell moment is calculated as 
 \begin{equation}
 	\frac{1}{n-N_{\mathrm{ss}}}
 	\sum_{m=N_{\mathrm{ss}}+1}^{n}
	\Phi_c^m.
 	\label{eq:steady_average_moments_ch3}
 \end{equation}

Compared with conventional DSMC, the DIG method produces lower statistical noise, thereby requiring fewer statistical samples \cite{wu2026accelerating}.

\subsection{Stability enhancement} 
\label{subsec:stability_enhancement}

On highly irregular unstructured meshes, tiny computational cells may hold only a sparse population of particles over a single coupling interval, yielding substantial statistical noise. Feeding these noisy sampled moments directly into the macroscopic synthetic equations risks triggering numerical instabilities. Accordingly, a per-cell reliability criterion is enforced prior to utilizing the extracted macroscopic states. 

Given that this issue originates from the particle count within each computational cell, the local Knudsen number serves as a reliable metric for stability assessment. We first define the global Knudsen number as
\begin{equation}
	Kn
	=
	\frac{\lambda_0}{L_0}
	=
	\frac{\mu_0}{p_0L_0}
	\sqrt{\frac{\pi RT_0}{2}},
	\label{eq:kn_definition_stability}
\end{equation}
where \(\lambda_0\) is the reference mean free path, \(\mu_0\) is the dynamic viscosity at the reference temperature \(T_0\), and \(p_0=\rho_0T_0\) is the reference pressure. With the viscosity temperature exponent \(\omega\) and the non-dimensional translational pressure \(p_{t,c}=\rho_cT_{t,c}\), the local molecular mean free path in cell \(c\) is evaluated as
$\lambda_c
	=Kn
	T_{t,c}^{\omega-\frac{1}{2}}/{\rho_c}$.
For an unstructured cell with volume \(V_c\), an equivalent cell diameter $d_c = 2 \left( {3V_c}/{4\pi} \right)^{1/3}$ is introduced by approximating the cell as a sphere with the same volume. 
We then define the local cell Knudsen number as
\begin{equation}
	Kn_{\Delta,c}
	=
	\frac{\lambda_c}{d_c}.
	\label{eq:cell_kn_stability}
\end{equation}

To prevent direct coupling of DSMC samples from very small cells, the sampled low-order macroscopic state in cell \(c\), stated in the previous sub-section, is accepted by the macroscopic solver only when 
\begin{equation}\label{criterion}
    Kn_{\Delta,c} \leq Kn_{\Delta,\max},
\end{equation} 
where the threshold \(Kn_{\Delta,\max}\) is a numerical parameter selected according to the range of the local cell Knudsen number in the target flow field.

\begin{figure}[t]
	\centering
	\includegraphics[width=0.8\textwidth]{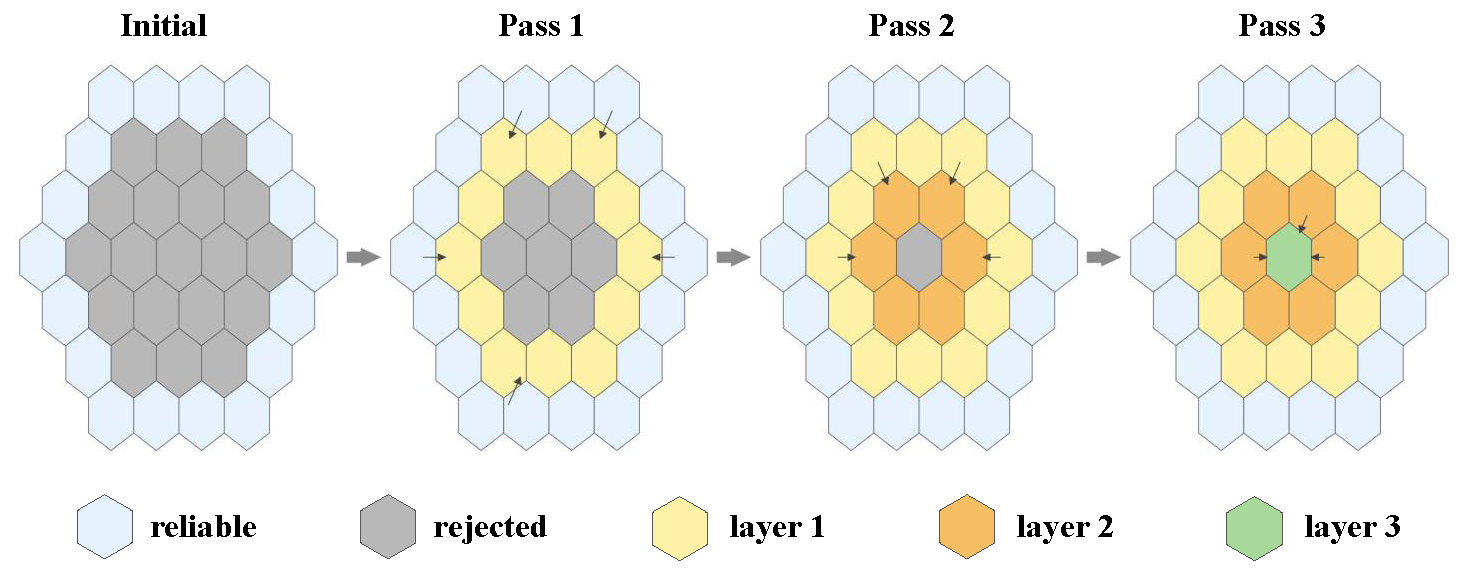}
    \caption{Illustration of the local reconstruction procedure for cells rejected by criterion \eqref{criterion}. Blue cells denote reliable cells whose macroscopic states are directly available, and gray cells denote rejected cells. In the first local pass, each layer 1 cell (yellow) is reconstructed from the available neighboring cell states that share a face with it, with the same weight assigned to each contributing neighbor. Once reconstructed, these layer 1 cells are treated as available in the next pass. The same procedure is then applied successively to the newly adjacent rejected cells, forming layers 2 and 3 and propagating the macroscopic information progressively toward the interior rejected region.}
	\label{fig:data-diffusion-local-passes}
\end{figure}

For cells satisfying the criterion \eqref{criterion}, macroscopic quantities are evaluated from the window statistics obtained through the intermittent sampling procedure and are supplied directly to the macroscopic synthetic equations in the current outer coupling cycle, see description in Section \ref{Sec:extraction}. For cells that do not satisfy the criterion \eqref{criterion}, the sampled quantities in cell \(c\) are not supplied directly to the macroscopic solver. Instead, macroscopic state are reconstructed by equal-weight averaging over reliable neighboring cells and are then used as the input to the macroscopic synthetic equations. This reconstruction can be applied for several local passes, so that available macroscopic information is propagated from reliable cells to adjacent rejected cells, as illustrated in Fig.~\ref{fig:data-diffusion-local-passes}. Meanwhile, the cell \(c\) enters a local accumulation state, where macroscopic quantities are accumulated over subsequent DSMC sampling intervals. The underlying reason is that DSMC exhibits little numerical dissipation on fine grid cells, so that all DSMC computed data are available for use, rather than merely retaining information from the prior $N_s$ steps.

\subsection{Overall DIG iteration procedure}

\begin{algorithm}[t!]
	\caption{Outer coupling procedure of the DIG method}
	\label{alg:overall-dig-iteration}
	\begin{algorithmic}[1]
		\Require Initial macroscopic field $M^0$ or existing DSMC particle state $\mathcal{P}^0$; outer coupling interval $N_s$; maximum number of macroscopic iterations $N_m$; residual tolerance $\epsilon_m$.
		\Ensure Converged macroscopic field and DSMC statistical quantities.
		
		\State Initialize the macroscopic state and DSMC particle state in each computational cell.
		\State Set $k=0$.
		
		\While{k $\leq$ MaxSteps} 
		\State Starting from $\mathcal{P}^{k}$, perform $N_s$ standard DSMC time steps.
		
		\State Obtain $M^{k+\frac{1}{2}}$ and $\mathrm{HoT}^{k+\frac{1}{2}}$ from the current sampling window.
		
		\State Solve the macroscopic synthetic equations with $M^{k+\frac{1}{2}}$ and $\mathrm{HoT}^{k+\frac{1}{2}}$ as input, and obtain $M^{k+1}$.
		
		\State Modify the DSMC particle number, velocity, and rotational internal energy according to $M^{k+1}$, and obtain $\mathcal{P}^{k+1}$.
		
		\State Set $k\leftarrow k+1$.
		\EndWhile
	\end{algorithmic}
\end{algorithm}

The outer coupling procedure of the DIG method is summarized in Algorithm~\ref{alg:overall-dig-iteration}.
In each outer coupling cycle, the HoT corrections are determined from the current DSMC sampling results and are treated as known closure terms during the inner iterations of the macroscopic synthetic equations. The macroscopic solver starts from $M^{k+\frac{1}{2}}$ and stops when the residual is lower than $\epsilon_m$ or the number of inner iterations reaches $N_m$, yielding the updated macroscopic state $M^{k+1}$. The DSMC particle number, velocity, and rotational internal energy are then adjusted according to $M^{k+1}$, so that the low-order moments of the particle sample are consistent with the macroscopic solution~\cite{luo2026mixtureDIG,luo2026polyatomicDIG}. The nonequilibrium quantities of higher-order are not directly imposed during the feedback step; instead, they are regenerated and sampled through the DSMC particle evolution in the next outer coupling cycle.

\section{Efficient parallelization strategy for DIG method}\label{sec:Solver}

Our DIG solver over unstructured grids is implemented in C++ with an MPI-based parallel framework, where both the macroscopic synthetic solver and the mesoscopic particle solver are parallelized. The parallel implementation of the macroscopic solver in the finite-volume framework, including the domain decomposition strategy and the solution procedure of the macroscopic equations, has been presented in our previous work~\cite{zhang2024parallelGSIS}. However, unlike the deterministic GSIS solver, in which the distribution function is directly stored and evolved on the computational mesh, the DIG method describes the distribution function by simulation particles. Consequently, efficient management of particle evolution and migration becomes the key issue in the parallel implementation of DIG method. Since particles continuously migrate across subdomain boundaries during the simulation, frequent MPI communication is inevitable, and its efficiency is closely coupled with the mesh partitioning strategy, largely determining the overall performance of the parallel DIG solver. Accordingly, this section presents the parallelization strategy for the particle solver, including the MPI process organization, MPI communication for particle migration, and dynamic load balancing process during the simulation.

\subsection{MPI process and local data organization}
\label{sec:mpi-local-mesh}

To efficiently organize particle data and support large-scale parallel simulations, the MPI processes are divided into two functional groups: one storage process, denoted by $r_{\mathrm{s}}$, and a set of all compute processes, 
\begin{equation}
    \mathcal{R}_{\mathrm{c}}=\{r_1,r_2,\ldots,r_{N_{\mathrm{c}}}\},
\end{equation}
where $N_{\mathrm{c}}$ is the total number of compute processes. The storage process maintains the global mesh together with the workload information gathered from all compute processes, which is used to support dynamic domain partitioning, load balancing and data output. Each compute process manages the local mesh and particle data within its assigned subdomain, performs DSMC particle evolution and macroscopic synthetic equation solving, and exchanges particle and mesh data with neighboring processes across partition interfaces.

\begin{center}
\begin{minipage}{0.98\linewidth}
\captionof{listing}{C++ implementation of the local mesh package data structure based on an array-based layout.}
\label{lst:dsmc-local-mesh-package}
\vspace{-0.2cm}
\begin{lstlisting}[frame=single, breaklines=true, tabsize=4]
struct DSMCLocalMeshPackage
{
    int my_owned_ncell;                           // number of owned cells
    int my_ghost_ncell;                           // number of ghost cells
    // mutual mappings between global ids on r_s and local indices on r_i
    std::unordered_map<int, int> gid2local;       // global cell id to local cell index
    std::unordered_map<int, int> face_gid2local;  // global face id to local face index
    std::vector<DsmcCell> cells;                  // cell information
    std::vector<DsmcEdge> faces;                  // face information
};
\end{lstlisting}
\end{minipage}
\end{center}

\begin{figure}[t]
	\centering
	\begin{tabular}{@{}cc@{}}
		\includegraphics[scale=0.65]{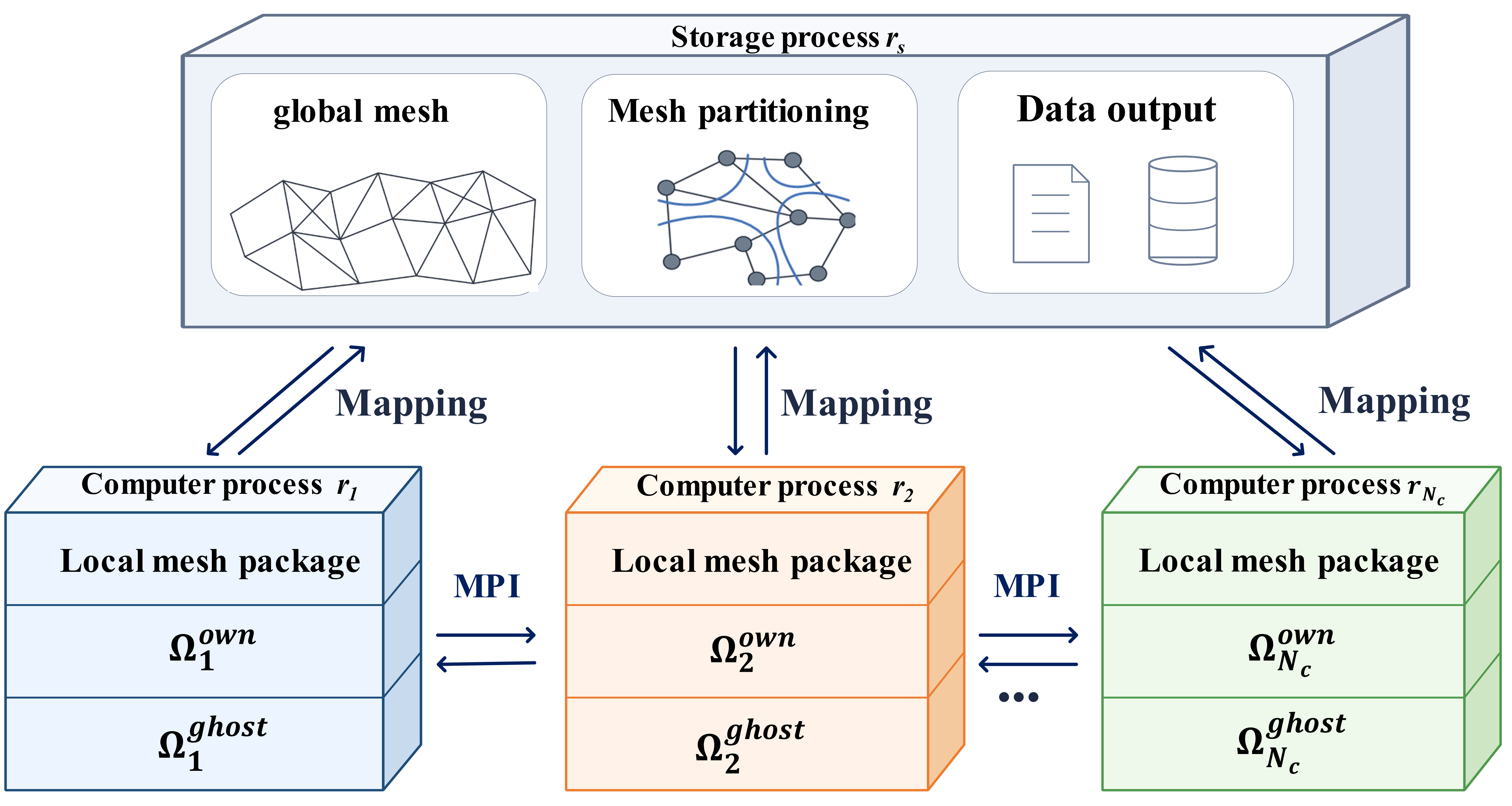}

	\end{tabular}
	\caption{MPI process organization in the DIG solver. The global mesh is distributed over $N_c$ compute processes through mesh partitioning, while neighboring processes communicate through MPI.}
	\label{fig:parallel-organization-overall}
\end{figure}

After domain decomposition performed by the storage process $r_\text{s}$, a local mesh package is generated for each compute process $r_i \in \mathcal{R}_\text{c}$ and distributed accordingly. The core data members of this package are summarized in the Listing~\ref{lst:dsmc-local-mesh-package}. The package stores all local mesh information required for computation and communication. As illustrated in Fig.~\ref{fig:parallel-organization-overall}, the local mesh package consists of two different cell sets, namely the owned-cell set $\Omega_i^{\text{own}}$ and the ghost-cell set $\Omega_i^{\text{ghost}}$. The owned-cell set $\Omega_i^{\text{own}}$ contains all computational cells assigned to the compute process $r_i$, where particle advection, collision, statistical sampling and modification of particle information are performed. The ghost-cell set $\Omega_i^{\text{ghost}}$ contains the neighboring cells information owned by adjacent compute processes, which identifies the destination process when particles migrate out of the local process. In addition, the local mesh package maintains mappings between global and local indices for cells, faces, and nodes, allowing local mesh entities to be efficiently mapped to their corresponding global entities maintained by $r_\text{s}$ or to neighboring entities owned by adjacent compute processes.

Once the local mesh has been constructed in $r_i$, the DSMC particles are organized and stored in cell buckets, with one bucket assigned to each owned cell in $\Omega_i^{\text{own}}$. As illustrated in the Listing~\ref{lst:particle-bucket-soa}, within the particle bucket, particle data are organized using a structure-of-arrays layout, where each particle attribute is stored in a dedicated contiguous array containing the information of all particles in the bucket. Specifically, the positions, velocities, rotational internal energies, cell indices, process indices, and remaining tracing times of all particles in the bucket are stored in separate contiguous arrays.

It should be noted that, in DIG, the macroscopic solver and the DSMC particle solver employ different domain decompositions on the same global mesh because their computational load characteristics are fundamentally different. The former adopts a conventional mesh partition generated by METIS without additional weighting, whereas the latter relies on particle information to characterize the computational load. These two partitions represent different data ownership arrangements on the same global mesh rather than different physical regions. Their correspondence is established through the global cell index, which enables data exchange between the macroscopic solver and the corresponding particle buckets during the DIG coupling procedure.

\begin{center}
\begin{minipage}{0.98\linewidth}
\captionof{listing}{C++ implementation of the particle bucket data structure based on a structure-of-arrays layout.}
\label{lst:particle-bucket-soa}
\vspace{-0.2cm}
\begin{lstlisting}[frame=single, breaklines=true, tabsize=4]
struct ParticleBucketSoA
{
    std::vector<int> p_serial;       // particle index
    std::vector<int> p_rank_serial;  // owner MPI process
    std::vector<int> p_mesh_serial;  // local cell index
    std::vector<double> vx, vy, vz;  // particle velocity
    std::vector<double> px, py, pz;  // particle position
    std::vector<double> p_Ir;        // rotational internal energy
    std::vector<double> dt_left;     // remaining tracing time
    void push_back(particle& part);  // append other particle to the bucket
    void pop_back();                 // remove one particle from the bucket
    // append all particles from another bucket (used during load balancing)
    void append_bucket(ParticleBucketSoA& other);
};
\end{lstlisting}
\end{minipage}
\end{center}

\subsection{Particle migration and MPI communication strategy}

\begin{figure}[ht]
	\centering
	\includegraphics[width=0.95\textwidth]{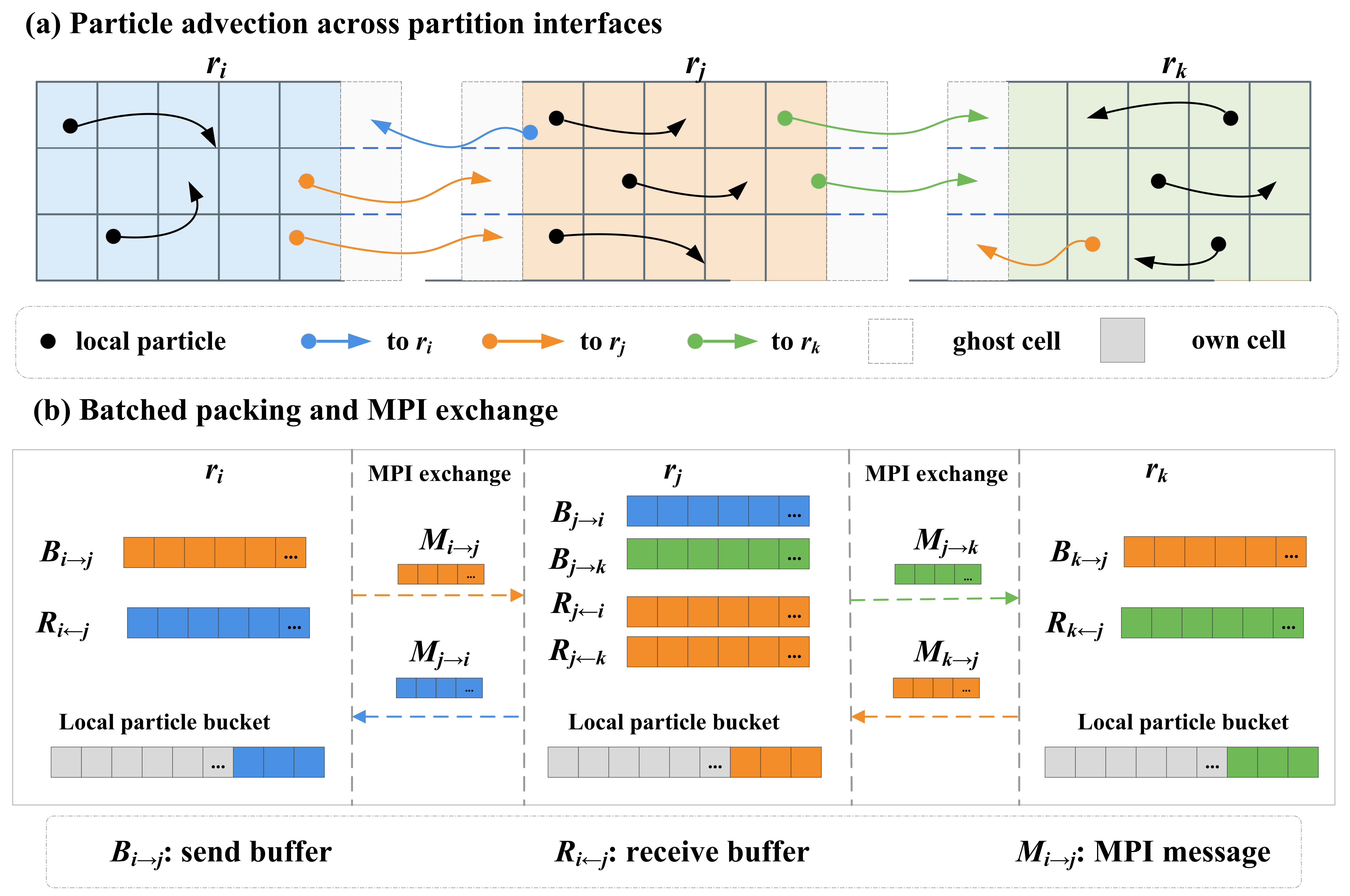}
	\caption{Particle migration across partition interfaces and batched MPI communication. (a) Particle advection near partition interfaces. (b) Batched inter-process MPI communication. After the destination process of each outgoing particle is determined, particles are packed into destination-specific send buffers, exchanged through MPI, and inserted into the corresponding receive buffers.}
	\label{fig:Particle_advection_figure}
\end{figure}

In our DIG solver, within one DSMC time step, every particle is first traced on the local unstructured mesh using a face-based tracking algorithm~\cite{chorda2002particleLocating,macpherson2009particleTracking} until its final target cell is identified.
During particle transport, particles may move across partition interfaces and enter subdomains owned by neighboring MPI processes. Therefore, an efficient particle migration strategy is required to maintain the consistency of particle ownership while minimizing communication overhead.
As illustrated in Fig.~\ref{fig:Particle_advection_figure}(a), when a particle crosses a partition interface during tracing on compute process $r_i$, the geometric information stored in the ghost-cell set $\Omega_i^\text{ghost}$ allows the particle trajectory to be completed locally without interrupting the tracking procedure. Once the final target cell has been identified, its owner process is obtained from the cell ownership mapping. If the target cell is still owned by process $r_i$, the particle is inserted directly into the corresponding local particle bucket. Otherwise, the particle is packed into the send buffer associated with the destination process for the subsequent MPI communication.

\begin{algorithm}[t!]
	\caption{Pseudocode for particle migration and batch MPI communication}
	\label{alg:particle-migration-communication}
	\begin{algorithmic}[1]
		\Require local mesh package on process $r_i$, current particle buckets of the owned cell set $\Omega_i^{\text{own}}$, and time step $\Delta t$
		\Ensure particle buckets of $\Omega_i^{\text{own}}$ after migration, together with migration statistics
		
		\State Clear the destination particle buckets, the send buffers $B_{i\rightarrow j}$, and the receive buffers $R_{i\leftarrow j}$.
		\For{each cell $c\in\Omega_i^{\text{own}}$}
			\For{each particle $\mathcal{P}$ in the current bucket}
				\State Particle advection within $\Delta t$ using the local geometry of $\Omega_i^{\text{own}}\cup\Omega_i^{\text{ghost}}$.
				\State Determine the final target cell.
				\If{the target cell is still owned by $r_i$}
					\State Insert the particle into the particle bucket of its target cell in $\Omega_i^{\text{own}}$.
				\Else
					\State Pack the particle state and the target global cell index into $B_{i\rightarrow j}$.
				\EndIf
			\EndFor
		\EndFor
		\State Allocate the required receive buffers $R_{i\leftarrow j}$ according to the migration counts.
		\State Transfer the packed particle data as MPI messages $M_{i\rightarrow j}$.
		\State Insert the received particles into the particle buckets of their target cells in $\Omega_i^{\text{own}}$.
		\State Record the number of migrated particles and the communication volume.
	\end{algorithmic}
\end{algorithm}

After all particles on compute process $r_i$ have completed the local tracing procedure, those whose final target cells are not owned by process $r_i$ are packed into the corresponding send buffers for subsequent MPI communication. As illustrated in Fig.~\ref{fig:Particle_advection_figure}(b), particles destined for process $r_j$ are collected in the send buffer $B_{i\rightarrow j}$, where each communication packet contains the complete particle state together with the target global cell index. The packed particle data are then transferred as the MPI message $M_{i\rightarrow j}$ to process $r_j$ and temporarily stored in the corresponding receive buffer $R_{i\leftarrow j}$. After all particle packets have been received, the target global cell index is converted into the corresponding local cell index, and the particles are inserted into the particle buckets of the associated owned cells.
It should be noted that, only after all received particles have been reconstructed into the local particle buckets does the solver proceed to the collision and sampling stages.
The overall implementation of the particle migration and MPI communication procedure is summarized in Algorithm~\ref{alg:particle-migration-communication}.

\subsection{Dynamic load balancing strategy}
\label{sec:dynamic-load-balancing}

\begin{figure}[t]
	\centering
	\includegraphics[width=0.96\textwidth]{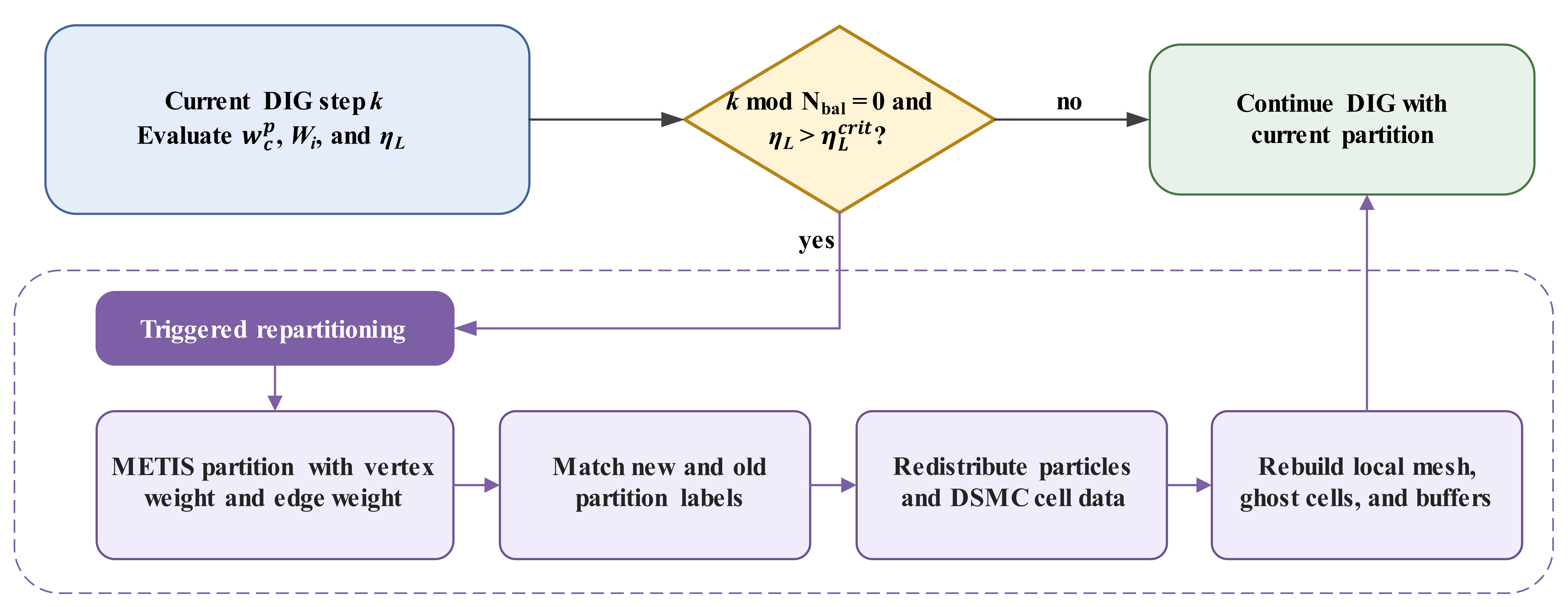}
	\caption{The flowchart of the proposed dynamic mesh re-partitioning procedure. At every $N_\text{bal}$ DIG steps, the load imbalance ratio is evaluated to determine whether mesh re-partitioning is required. Once the re-partitioning criterion is satisfied, METIS graph partitioning, partition-label matching, particle and cell data redistribution, and local mesh reconstruction are performed before resuming the DIG simulation. }
	\label{fig:dynamic-load-balancing-flow}
\end{figure}

In the DIG method, the macroscopic solver and the DSMC particle solver employ different data ownership arrangements on the same global mesh. 
The mesh partition for the macroscopic solver is initialized once and remains fixed throughout the simulation, whereas the DSMC particle partition is dynamically updated as the particle distribution evolves.
Consequently, the computational workload of the DSMC particle solver varies continuously during the simulation and may become severely imbalanced among compute processes~\cite{wu2005dynamicDSMC}.
In 3D hypersonic simulations, strong density variations often lead to severe load imbalance among compute processes.
Since the overall execution time is determined by the slowest process, dynamic load balancing is employed to improve parallel efficiency~\cite{zhang2022loadDecouplingDSMC,shamseddine2019adaptive3DDSMC,li2014dynamicLoadDSMC}.
This requires first evaluating the workload imbalance to determine whether re-partitioning is necessary, followed by a weighted graph re-partitioning procedure to generate a new particle partition with balanced workload and reduced communication overhead. The general flowchart of dynamic load balancing process is illustrated in Fig.~\ref{fig:dynamic-load-balancing-flow} and details will be given in the following subsections.

\subsubsection{Workload evaluation and re-partitioning criteria}

To evaluate the load imbalance among compute processes, the computational workload of each cell is first estimated from its particle population. In the $c$-th computational cell in compute process $r_i$, the corresponding particle workload $w_c^{p}$ is defined as,
\begin{equation}
	w_c^{p}
	=
	\max\left(N_{p,c},1\right),
	\label{eq:particle-number-weight}
\end{equation}
where the lower bound ensures that cells containing no particles are also assigned a positive integer weight.
The total workload of compute process $r_i$, denoted by $W_i$, is defined as the sum of the particle workloads of all owned mesh cells,
\begin{equation}
W_i=\sum_{c\in\Omega_i^{\mathrm{own}}}w_c^p.
\end{equation}
It should be noted that, the ghost cells are excluded from the workload evaluation since they contain no particles after particle migration has been completed. Therefore, based on the workloads of all compute processes, the global load imbalance ratio $\eta_L$ is defined as,
\begin{equation}
	\eta_L
	=
	\frac{
		\max_{r_i\in\mathcal{R}_{\mathrm{c}}}W_i
	}{
		\overline{W}
	},\,\, ~~\text{with}\,\,~~ \overline{W}
	= \frac{1}{N_{\mathrm{c}}}
	\sum_{r\in\mathcal{R}_{\mathrm{c}}}
	W_i.
	\label{eq:load-imbalance-ratio}
\end{equation}
Note that the value of $\eta_L$ close to unity indicates a balanced workload among compute processes. However, in practice, an exactly balanced partition is rarely achieved since the workload balance is not the only objective considered during the partitioning. Therefore, a load imbalance ratio threshold, denoted by $\eta_L^\text{crit}$, is introduced to determine whether mesh re-partitioning is necessary. Moreover, evaluating the workload imbalance and performing graph re-partitioning at every DIG step would introduce unnecessary computational cost. Therefore, the imbalance ratio is evaluated  every $N_\text{bal}$ DSMC steps. As illustrated in Fig.~\ref{fig:dynamic-load-balancing-flow}, the mesh re-partitioning process is activated at the $k$-the DIG time step only when 
\begin{equation}
    \text{mod}(k,N_\text{bal}) = 0 \quad \text{and}\quad 
    \eta_L > \eta_L^\text{crit}. 
\end{equation}
In practical computations, $\eta_L^{\text{crit}}$ and $N_\text{bal}$ are selected according to the load fluctuation level of the case and the computational overhead introduced by re-partitioning. Unless otherwise stated, $\eta_L^\text{crit}=1.05$ and $N_\text{bal}=10$ are adopted throughout the present work, allowing up to a 5\% workload imbalance before mesh re-partitioning is activated.

\subsubsection{Weighted graph re-partitioning}

\begin{figure}[t!]
	\centering
	\subfigure[Graph representation and METIS partitioning]{{\includegraphics[width=0.96\textwidth]{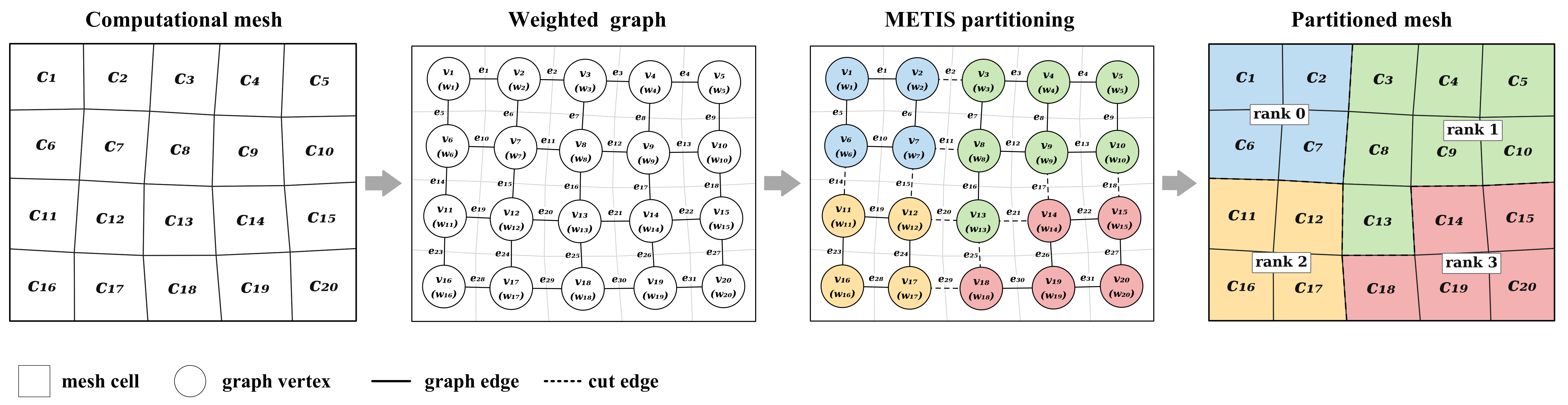}}}\\
	\subfigure[Before the re-partition]{\includegraphics[width=0.4\textwidth]{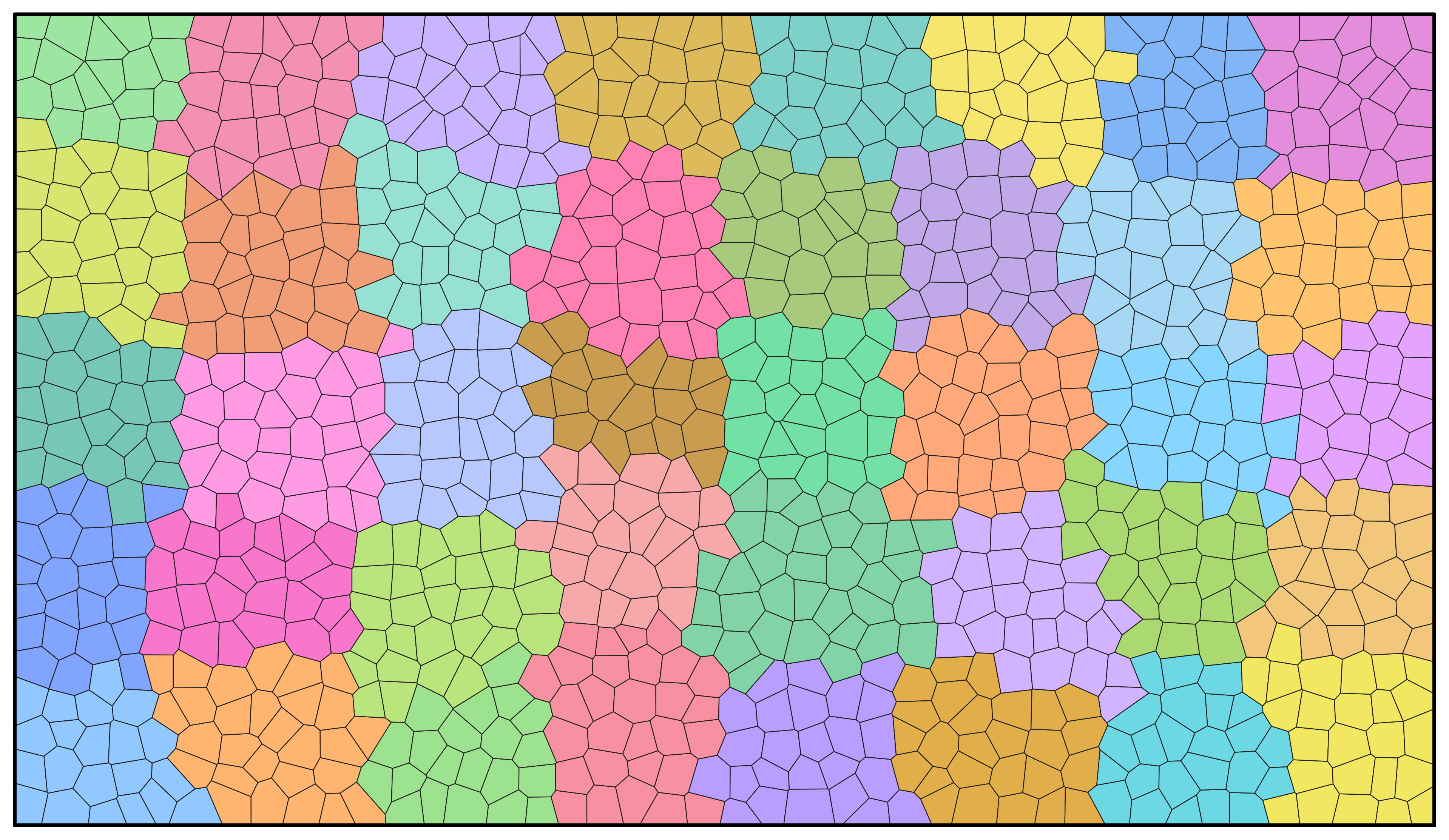}}
	\subfigure[After the re-partition]{\includegraphics[width=0.4\textwidth]{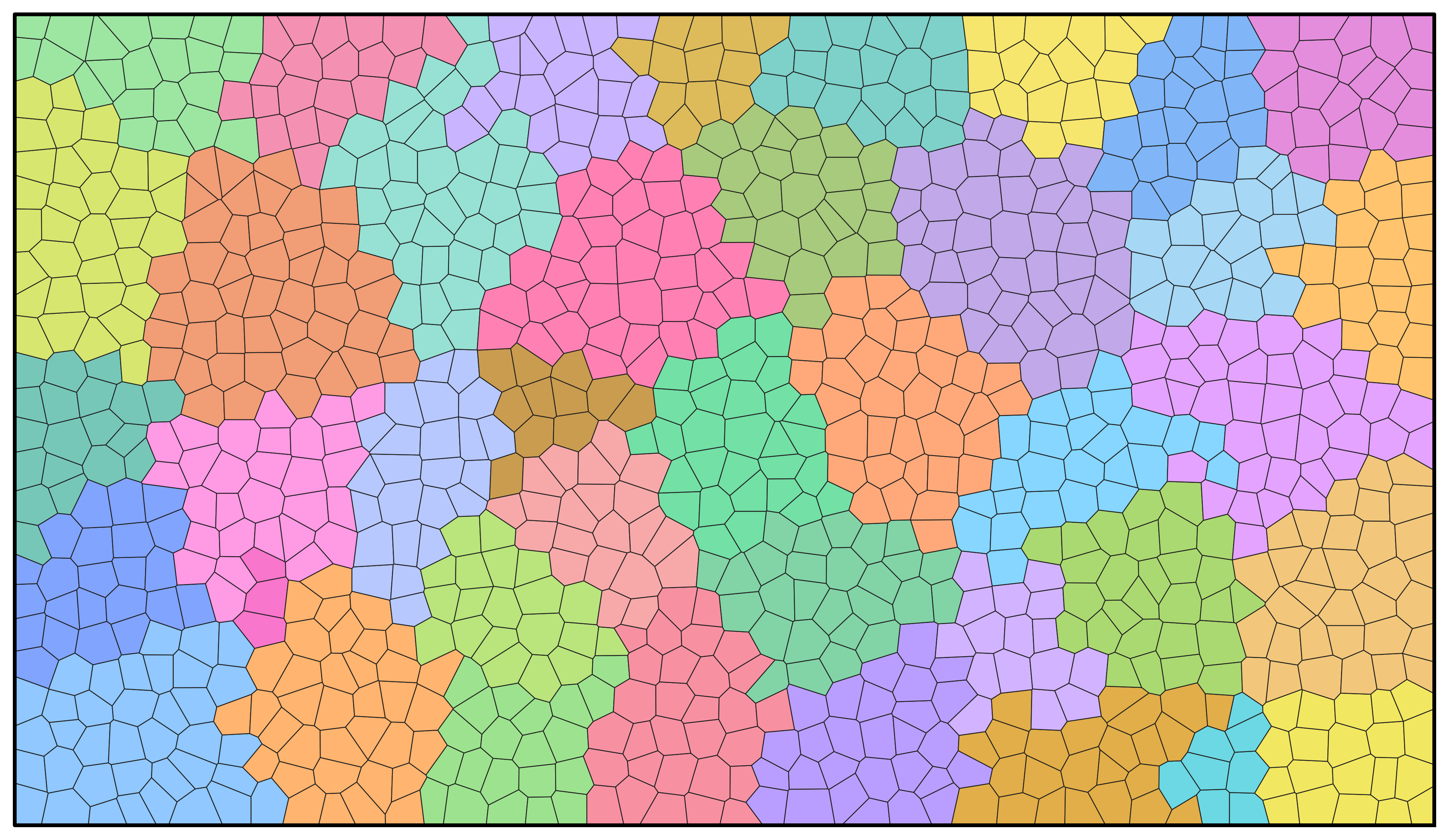}}
	\caption{Graph-based mesh re-partitioning using METIS in the DIG method. (a) Schematic illustration of the graph representation and METIS partitioning procedure, where $c_i$, $v_i$, $w_i$ and $e_i$ denote the mesh cell, graph vertex, vertex weight, and edge weight, respectively. (b) Mesh partition before re-partitioning. (c) Re-partitioned mesh by employing the Kuhn-Munkres algorithm \cite{kuhn-1955,munkres-1957}.}
	\label{fig:ch5_sphere_partition_comparison}
\end{figure}

Once the mesh re-partitioning criterion is satisfied, the current computational mesh is converted into a weighted graph for partitioning by METIS, as illustrated in Fig.~\ref{fig:ch5_sphere_partition_comparison}(a).
Specifically, each mesh cell is represented by a graph vertex, and two graph vertices are connected by an edge if the corresponding mesh cells share a common face.
By default, the particle workload $w_c^p$ defined in Eq.~(\ref{eq:particle-number-weight}) is directly assigned as the weight of the corresponding graph vertex, meaning particle count serves as the baseline metric for quantifying cell computational load during graph partitioning.

Despite being a simple and practical load estimator, particle population alone cannot accurately capture the true runtime overhead of individual cells.
Cells containing similar numbers of particles may still require different computational costs because the runtime also depends on particle transport, face crossings, collision processing, and particle state updates.
For this reason, a timing-aware load metric is adopted whenever accurate wall-clock time measurements for each cell can be retrieved.
For the $c$-th cell, the actual wall-clock time $t_c^{(m)}$ is accumulated over $m$-th mesh re-partitioning process consisting of $N_\text{bal}$ DIG steps.
To reduce unnecessary variations in the generated mesh partitions caused by stochastic fluctuations in the measured runtime, the accumulated wall-clock time is smoothed using an exponential moving average. The resulting smoothed runtime estimate, denoted by $\tilde{t}_c^{(m)}$, is employed to construct the timing-based vertex weight for the $m$-th mesh re-partitioning and is given by,
\begin{equation}
\tilde{t}_c^{(m)}=
\begin{cases}
t_c^{(1)}, & m=1,\\[4pt]
\alpha t_c^{(m)}
+
(1-\alpha)\tilde{t}_c^{(m-1)}, & m>1,
\end{cases}
\label{eq:ema-timing}
\end{equation}
where $\tilde{t}_c^{(m-1)}$ is the smoothed runtime estimate from
the previous mesh re-partitioning process, and $\alpha=0.5$ is the smoothing factor. Based on the smoothed runtime estimate, the timing-based vertex weight $w_c^t$ is constructed as
\begin{equation}
 w_c^{t}=\max\!\left(1,\mathrm{round}\!\left(100\,\tilde t_c/\bar t\right)\right),
\end{equation} 
where $\bar{t}$ represents global mean smoothed runtime over all mesh cells.
It should be noted that, since METIS prefers integer vertex weights, the normalized runtime is multiplied by a constant scaling factor of 100 before rounding to preserve approximately two decimal places of precision.
In the present work, the timing-based vertex weight $w_c^t$ is adopted as the vertex weight for METIS graph partitioning.

Besides the vertex weights used to balance the computational workload, edge weights are also introduced in METIS graph partitioning to reduce inter-process communication caused by particle migration.
In the present work, the edge weight is defined as the total number of particles crossing the shared face between two neighboring mesh cells during the load-balancing interval.
Consequently, interfaces with more frequent particle crossings are assigned larger edge weights and are therefore less likely to be cut during graph partitioning.
However, the partition labels assigned by METIS are not guaranteed to remain consistent between consecutive mesh re-partitioning operations.
To address this issue, the Kuhn-Munkres algorithm~\cite{kuhn-1955,munkres-1957} is employed to match the newly generated partitions with the original compute processes by maximizing the number of mesh cells that remain assigned to the same process. As shown in Fig.~\ref{fig:ch5_sphere_partition_comparison}(b) and (c), the partition labels are largely preserved after mesh re-partitioning, with only locally modified regions being assigned new process labels. Consequently, unnecessary inter-process particle migration and cell data redistribution are significantly reduced over repeated mesh re-partitioning operations throughout the simulation.

\section{Numerical results}\label{sec:Numerical results}

The performance of the present parallel DIG is tested in four representative flows: the low-speed lid-driven cavity flow, the hypersonic flow over a sphere,  a reentry capsule and the International Space Station. 
Nitrogen is used as the working gas, and intermolecular collisions are described by variable hard sphere model. The molecular mass is $m=4.65\times10^{-26}\,\mathrm{kg}$, the reference molecular diameter is $d=4.17\times10^{-10}\,\mathrm{m}$, the viscosity index is $\omega=0.74$. The gas viscosity is evaluated from the power-law relation $\mu=\mu_0(T_t/T_0)^\omega$. The rotational degree of freedom of nitrogen is taken as $d_r=2$, and the translational--rotational energy exchange is modeled by the Borgnakke--Larsen procedure, with a fixed  rotational collision number of $Z_r=2.59$.  Fully diffuse reflection is prescribed for gas--surface interactions at solid walls. The Knudsen number is defined in Eq.~\eqref{eq:kn_definition_stability}, where the characteristic length will be specified for each configuration. Simulations are performed on an Intel Xeon Gold 6226R workstation (2.90 GHz).

\subsection{3D lid-driven cavity flow}
\label{subsec:3d_lid_driven_cavity}

A 3D lid-driven cavity flow is considered as a low-speed benchmark to assess the convergence acceleration and computational efficiency of the DIG. The computational domain is a unit cube, $-0.5 \leq x \leq 0.5$, $-0.5 \leq y \leq 0.5$, and $-0.5 \leq z \leq 0.5$, with the cavity length used as the characteristic length $L_0$. The upper wall at $y=0.5$ moves in the positive $x$ direction with a constant velocity $\mathbf{u}_w=(U_w,0,0)$, where $U_w=\sqrt{2}$ is the lid speed, while the remaining five walls are stationary. All solid walls are isothermal with $T_w=1$.

\begin{figure}[t!]
	\centering
	\setlength{\tabcolsep}{2pt}
	\begin{tabular}{@{}ccc@{}}
		\includegraphics[width=0.315\textwidth,trim={30 10 10 30}]{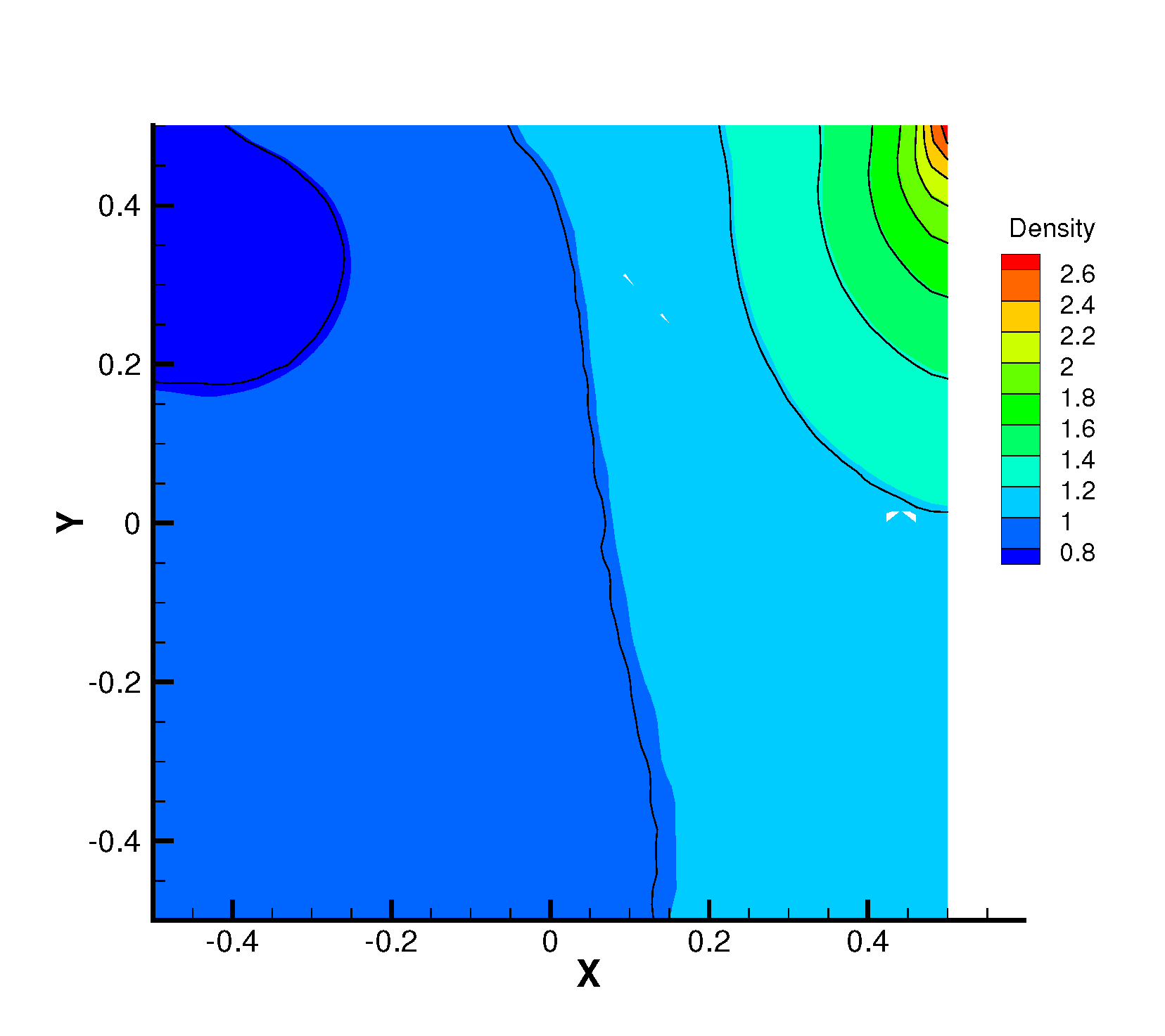}
		&
		\includegraphics[width=0.315\textwidth,trim={30 10 10 30}]{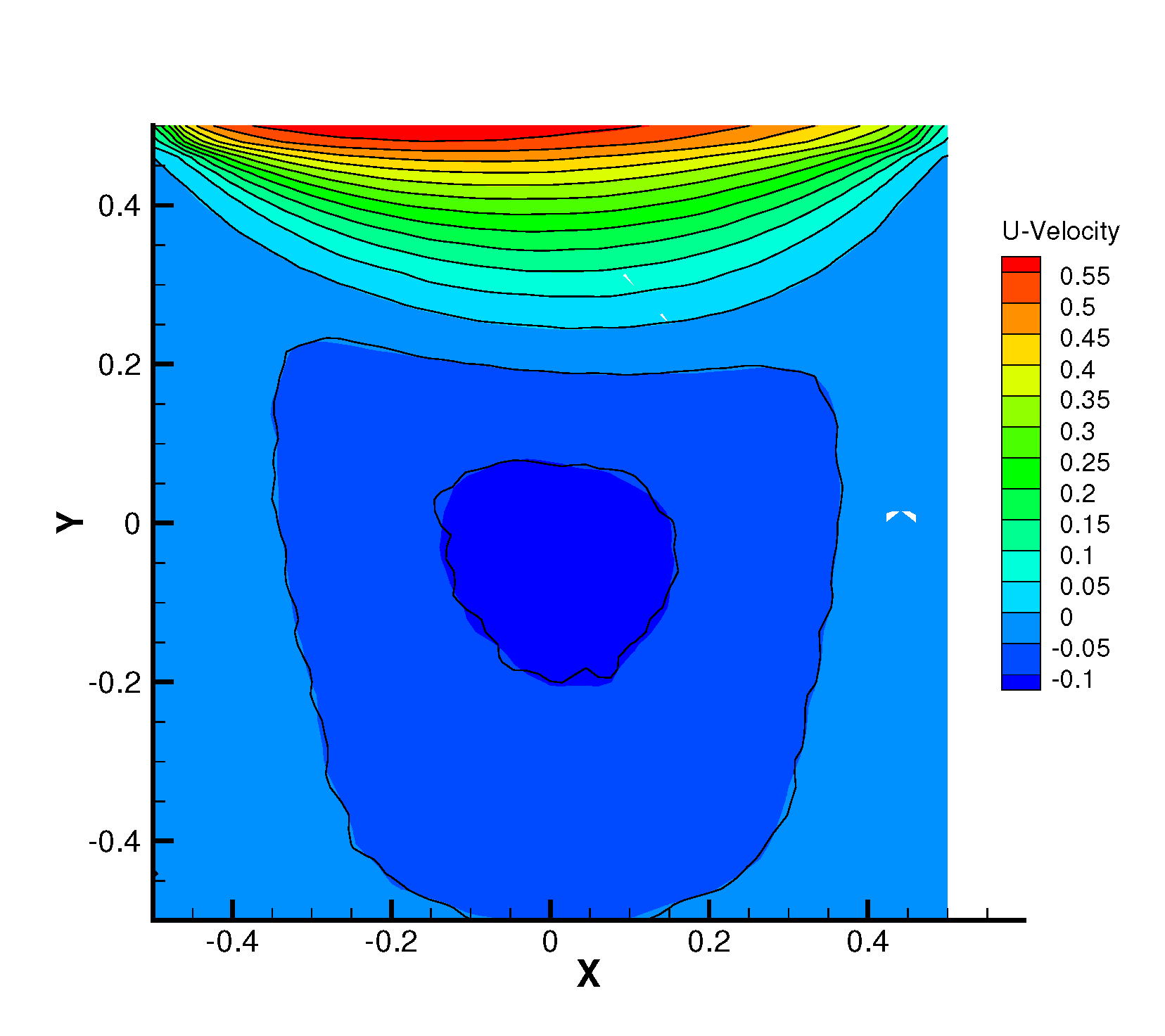}
		&
		\includegraphics[width=0.315\textwidth,trim={30 10 10 30}]{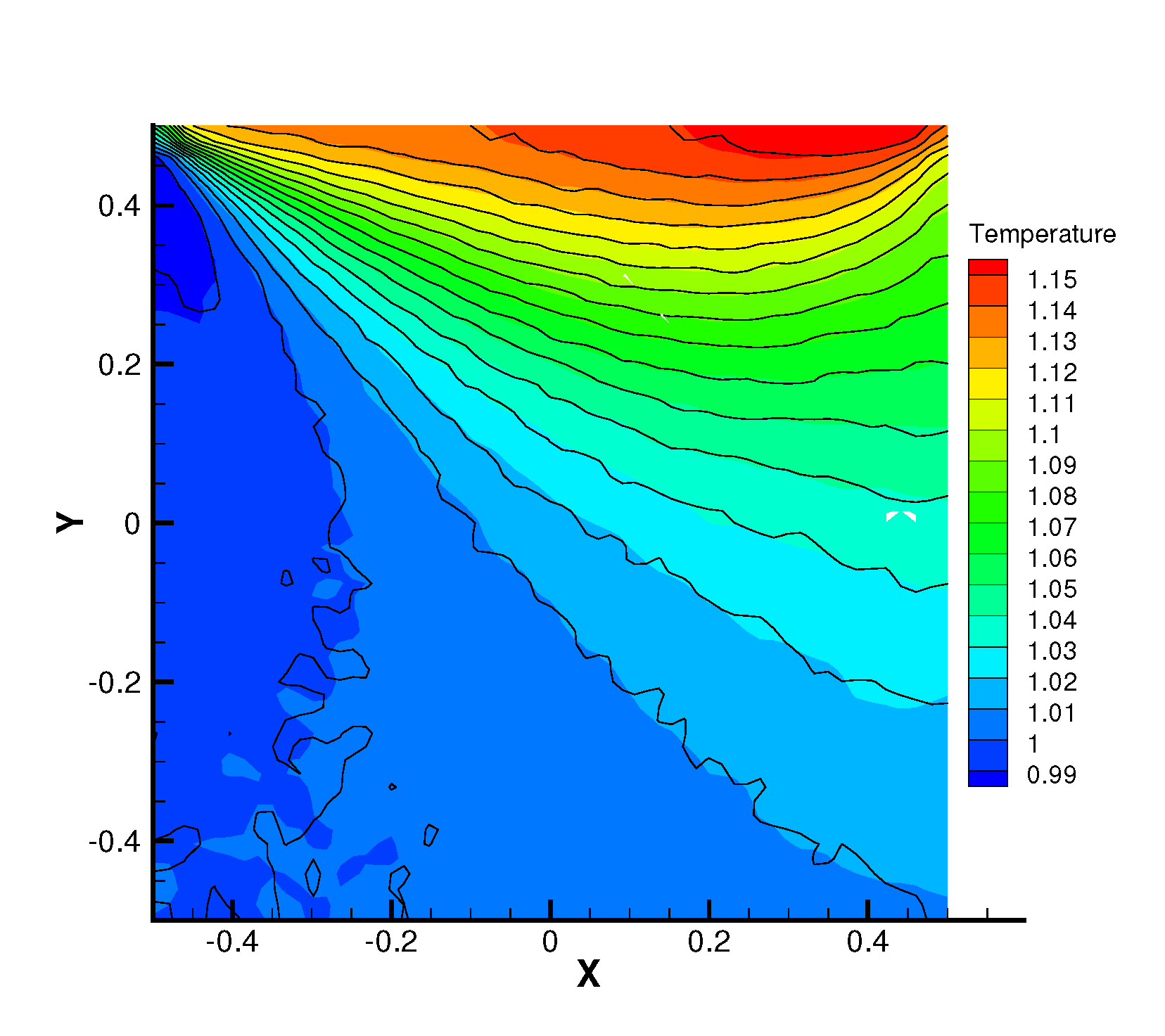}
		\\[-0.15cm]
		\multicolumn{3}{c}{$Kn=1$}\\[0.15cm]
		\includegraphics[width=0.315\textwidth,trim={30 10 10 30}]{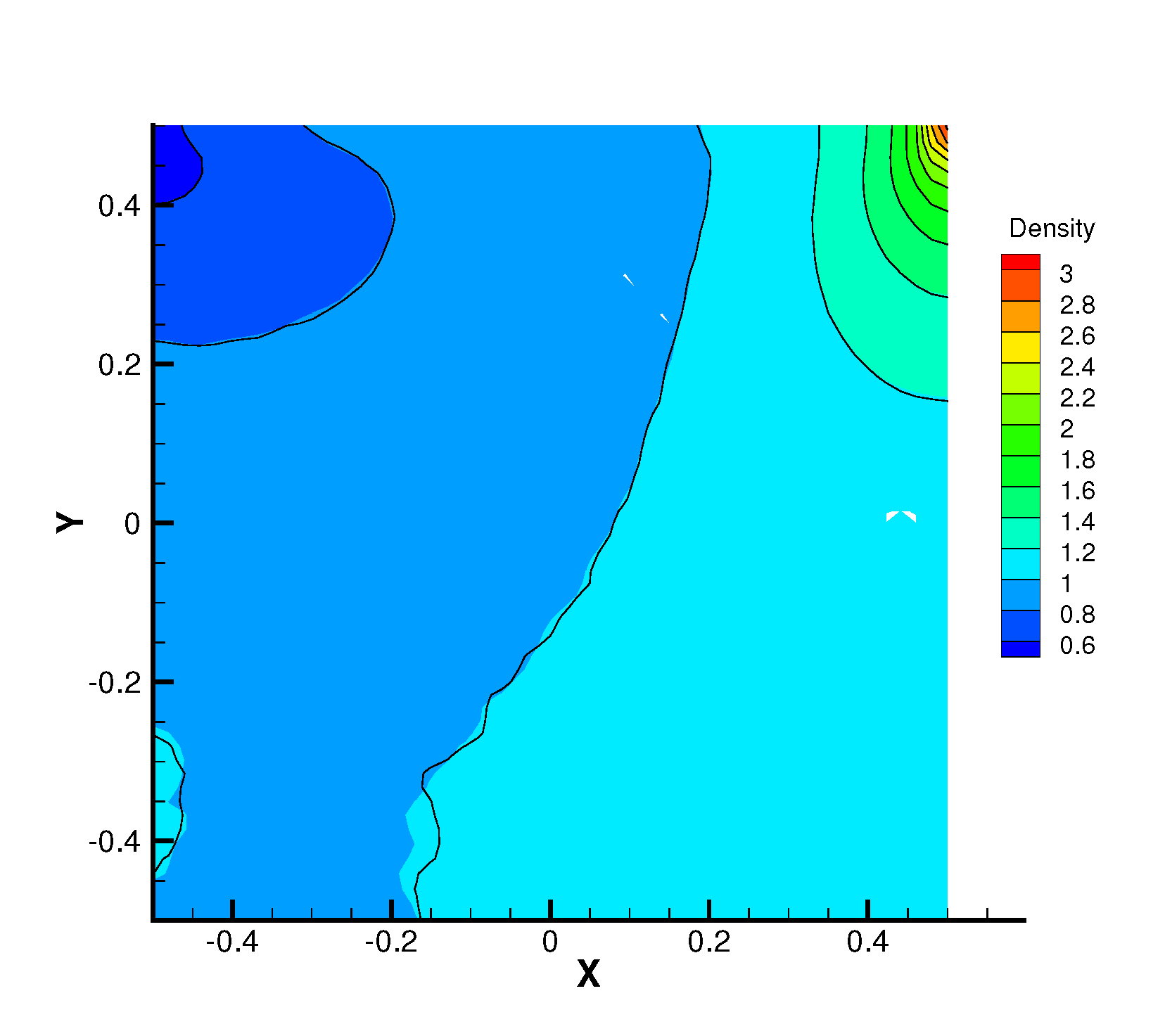}
		&
		\includegraphics[width=0.315\textwidth,trim={30 10 10 30}]{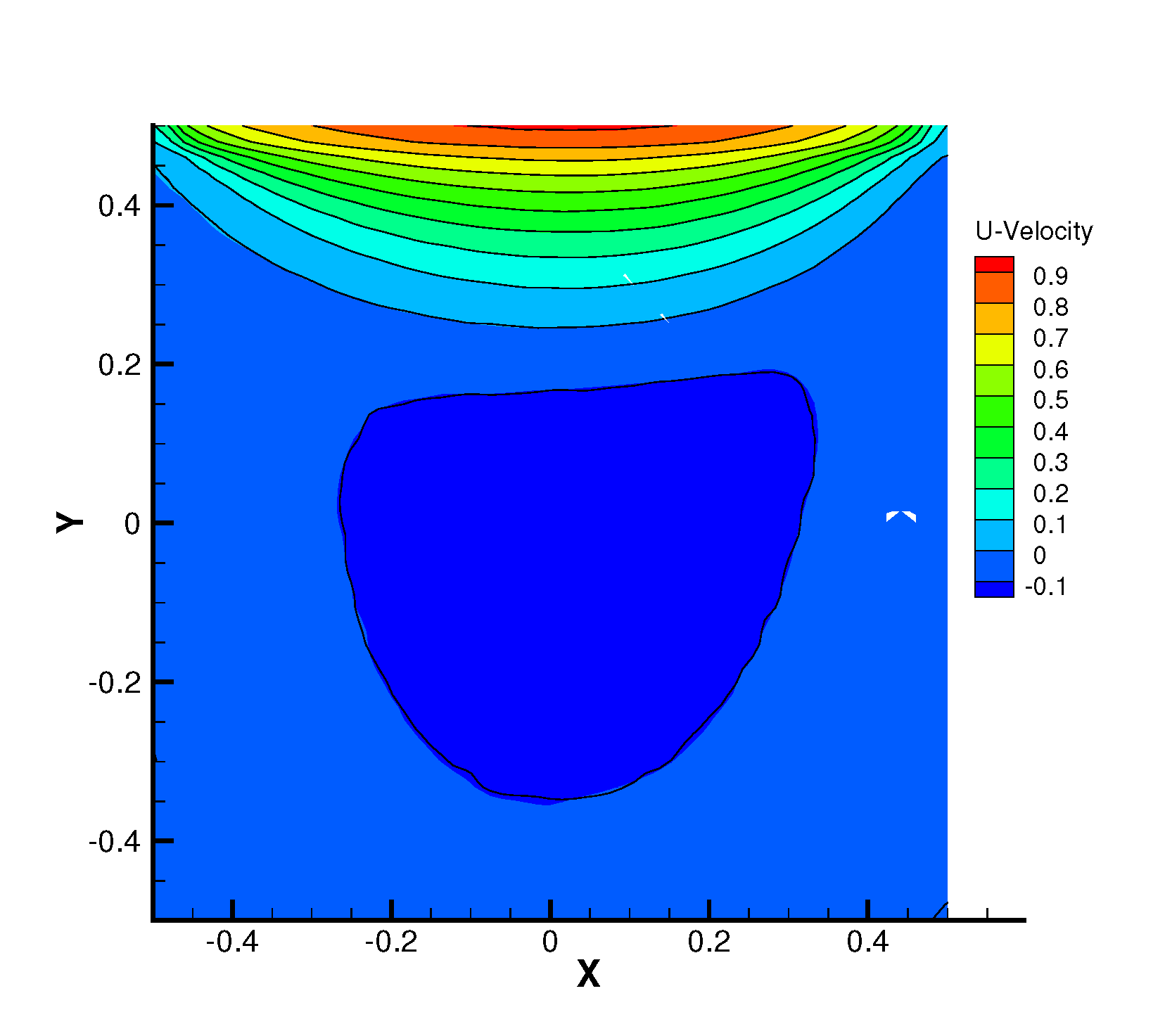}
		&
		\includegraphics[width=0.315\textwidth,trim={30 10 10 30}]{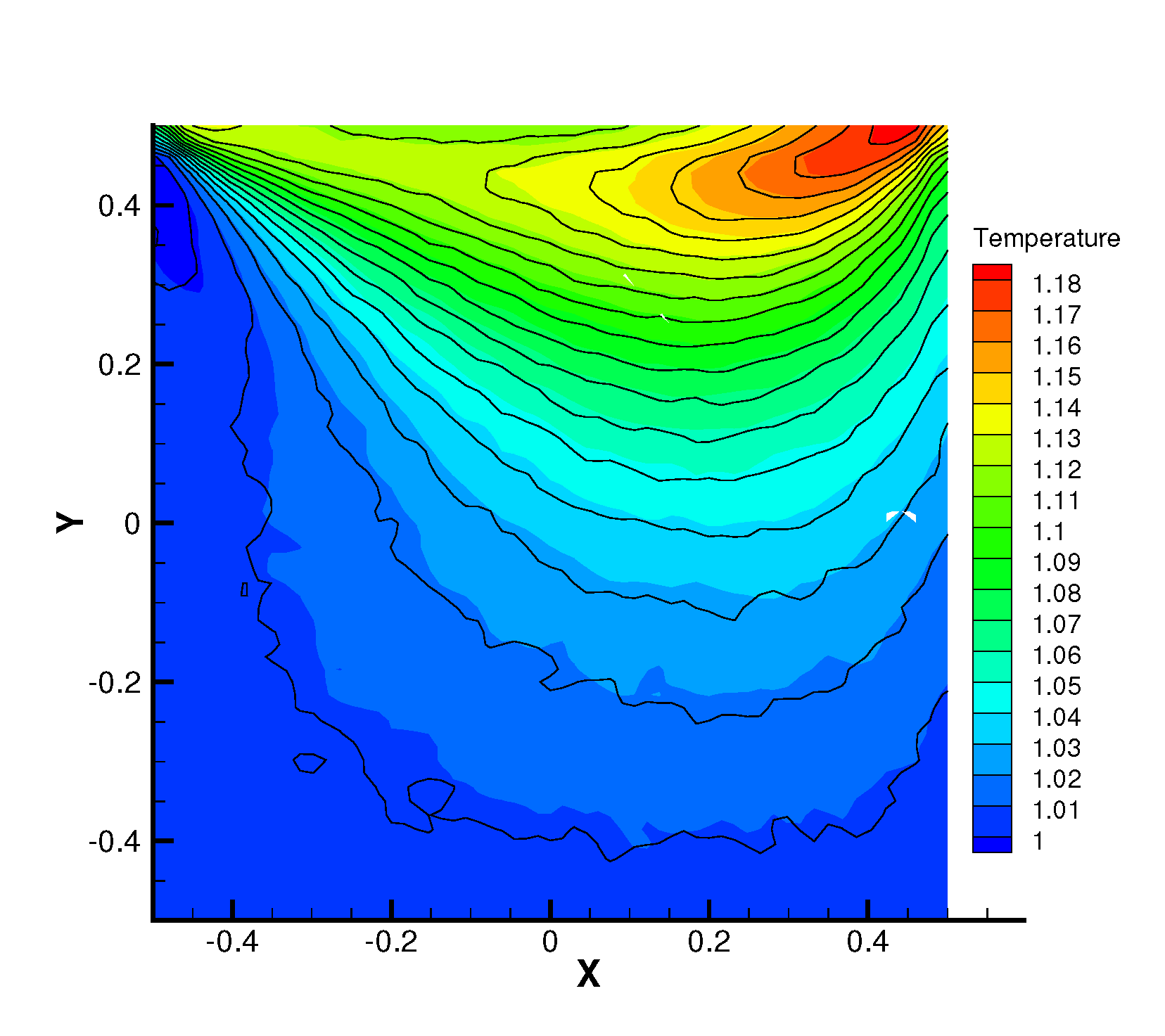}
		\\[-0.15cm]
		\multicolumn{3}{c}{$Kn=0.1$}\\[0.15cm]
		\includegraphics[width=0.315\textwidth,trim={30 10 10 30}]{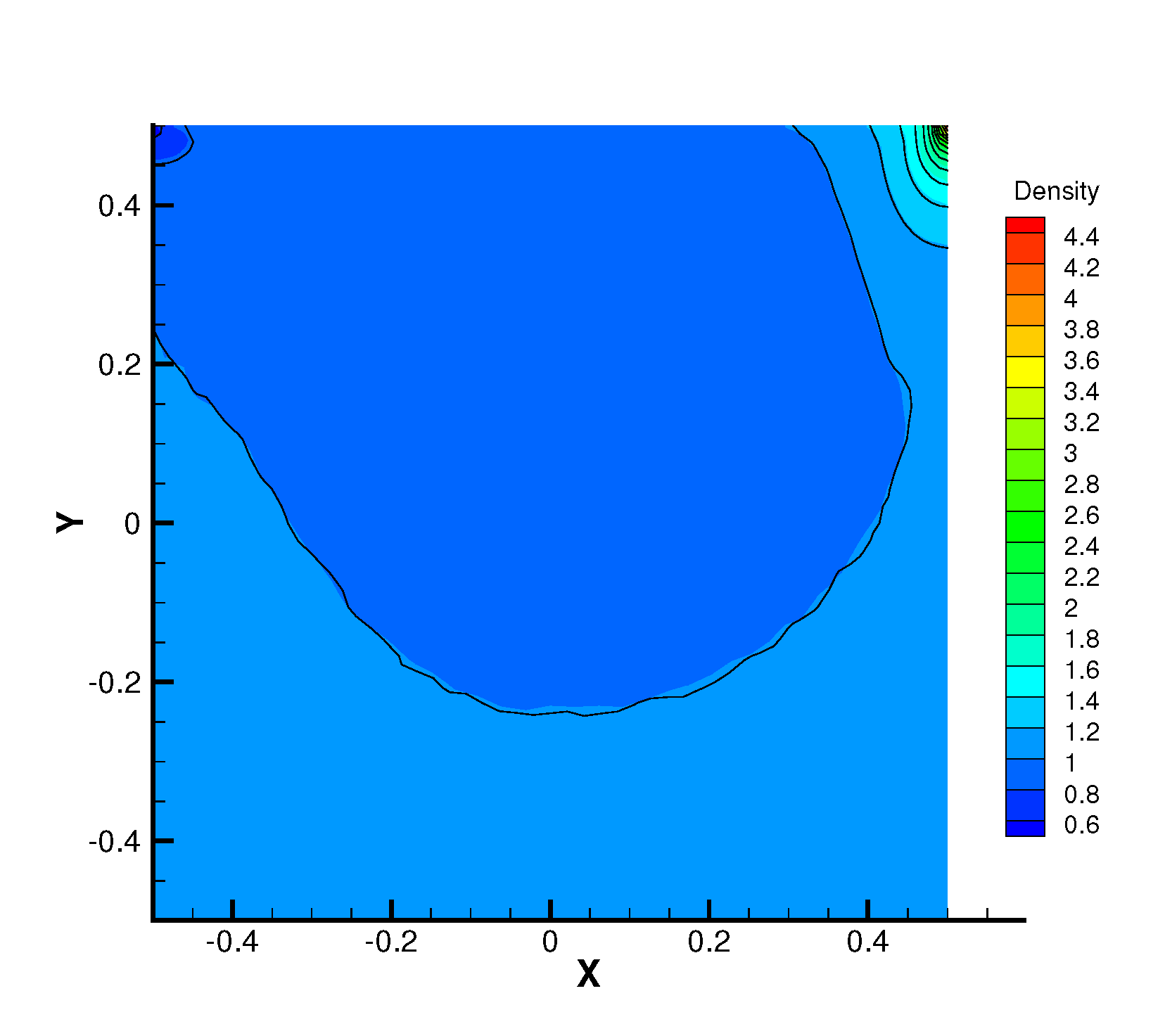}
		&
		\includegraphics[width=0.315\textwidth,trim={30 10 10 30}]{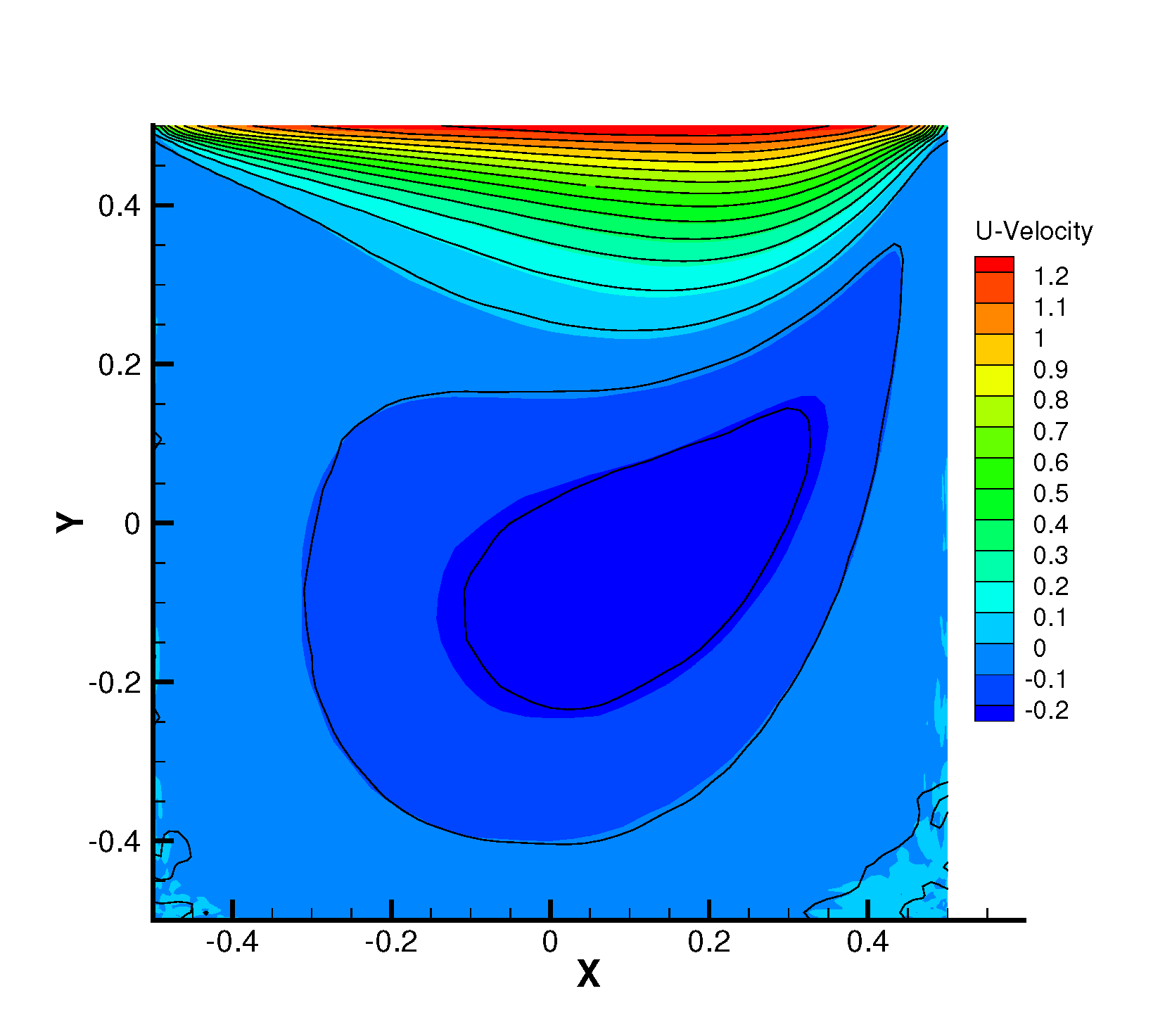}
		&
		\includegraphics[width=0.315\textwidth,trim={30 10 10 30}]{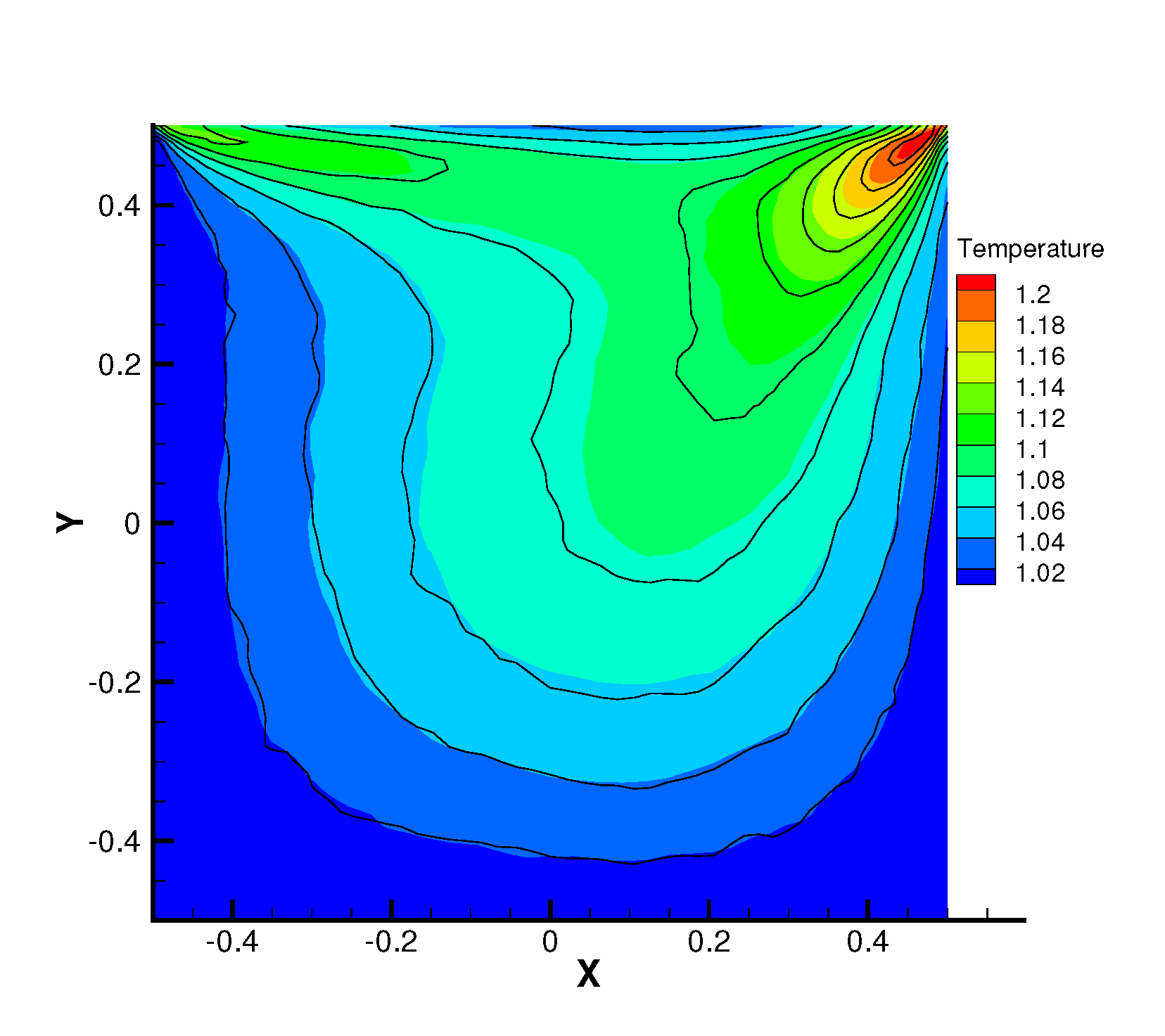}
		\\[-0.15cm]
		\multicolumn{3}{c}{$Kn=0.01$}
	\end{tabular}
	\caption{Comparison of density (first column), velocity magnitude (second column), and total temperature (third column) on the middle plane $z=0$ in the 3D lid-driven cavity flow. Colored contours and black contour lines  denote DSMC and DIG results, respectively. 
    }
	\label{fig:ch5_cavity_midplane}
\end{figure}

The initial flow field is in global equilibrium state, with the non-dimensional density $\rho=1$, velocity $\mathbf{u}=\mathbf{0}$, translational temperature $T_t=1$, and rotational temperature $T_r=1$. Three Knudsen numbers, $Kn=1$, $Kn=0.1$, and $Kn=0.01$, are considered.
In DIG, the domain is discretized by $60 \times 60 \times 60$ cells. The mesh is refined near the solid walls, with the minimum wall normal spacing of $0.01$, to capture the Knudsen layer. The CFL number of the particle solver is set to 0.2, based on the minimum cell size. In SPARTA, the cell resolution is selected according to the DSMC spatial resolution requirement at each Knudsen number, and the same particle time step as in DIG is adopted. In both DIG and SPARTA, each computational cell is initially populated with 100 simulation particles on average.

\begin{figure}[t]
	\centering
	\begin{minipage}[t]{0.32\textwidth}
		\centering
		\includegraphics[width=\linewidth,trim={30 10 50 50}]{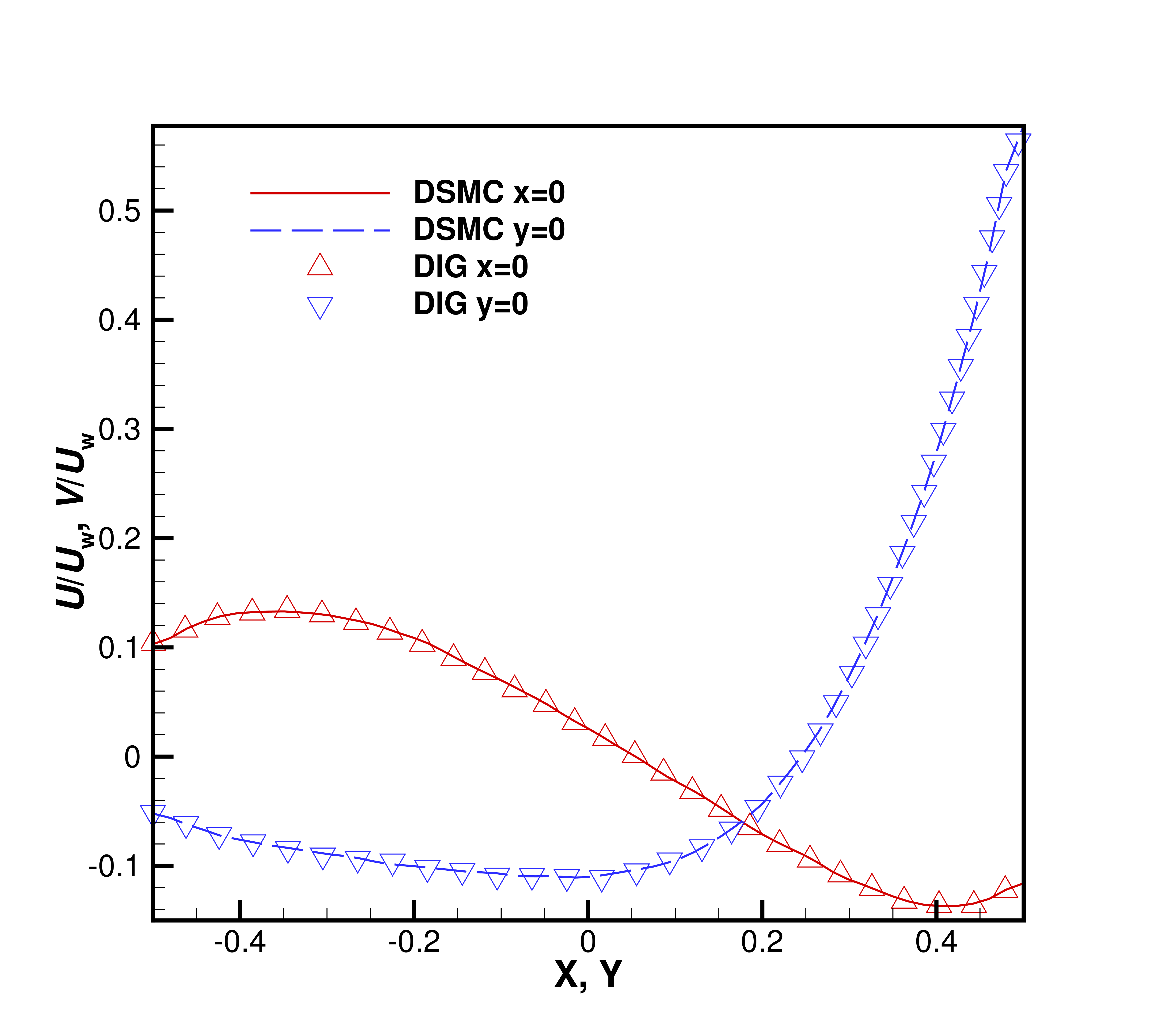}\\[-0.1cm]
		(a) $Kn=1$
	\end{minipage}
	\hfill
	\begin{minipage}[t]{0.32\textwidth}
		\centering
		\includegraphics[width=\linewidth,trim={30 10 50 50}]{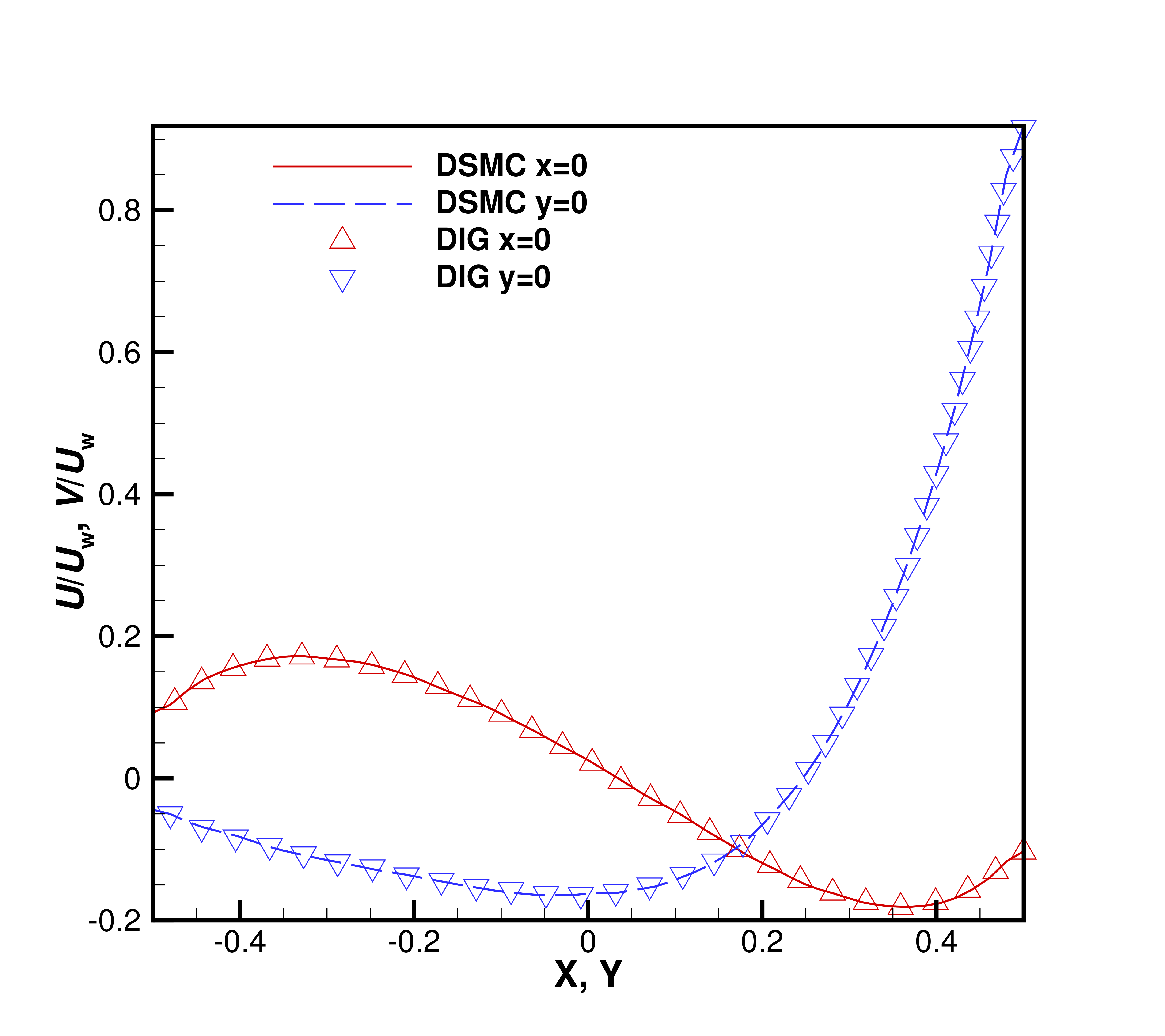}\\[-0.1cm]
		(b) $Kn=0.1$
	\end{minipage}
	\hfill
	\begin{minipage}[t]{0.32\textwidth}
		\centering
		\includegraphics[width=\linewidth,trim={30 10 50 50}]{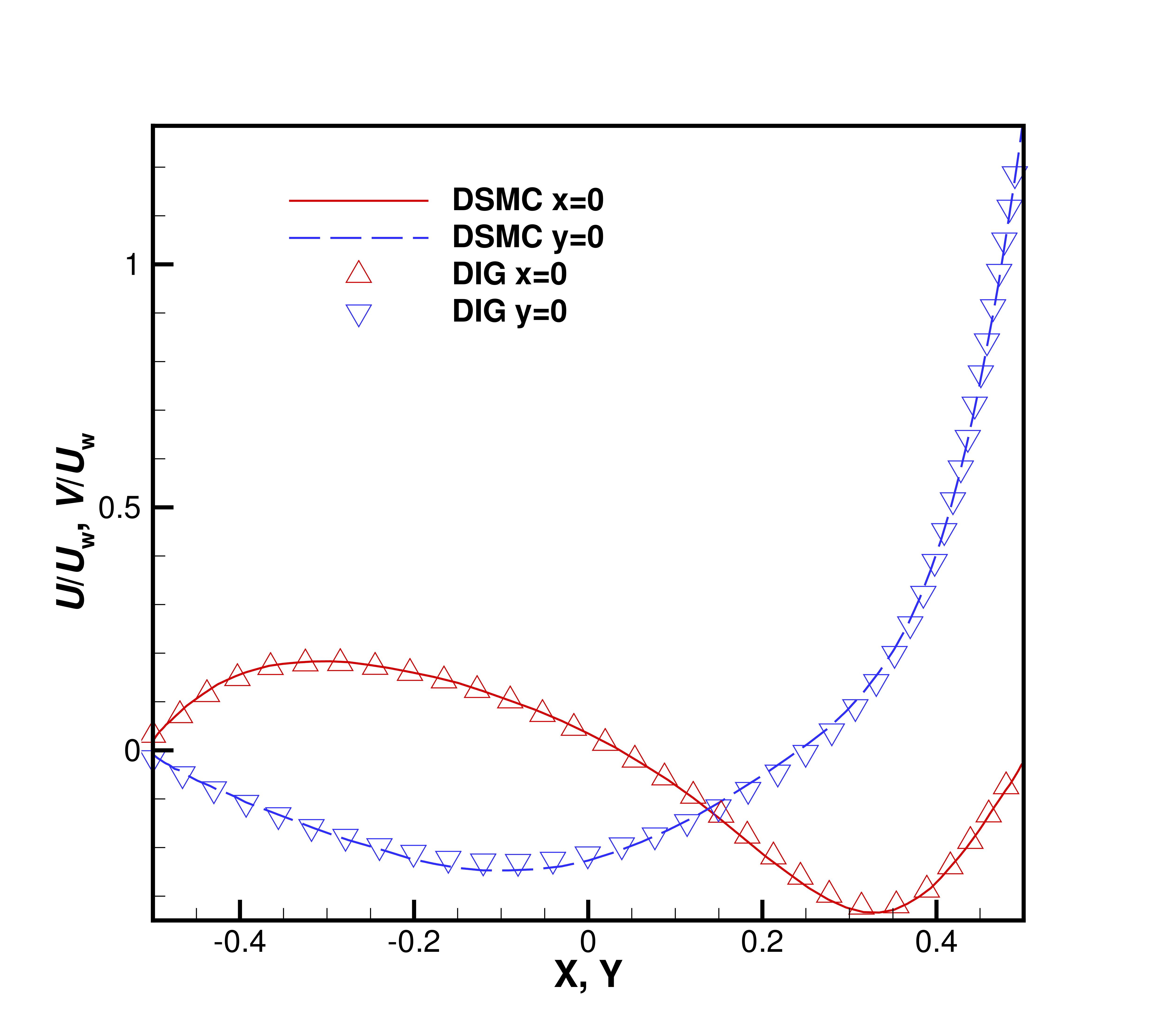}\\[-0.1cm]
		(c) $Kn=0.01$
	\end{minipage}
	\caption{Normalized horizontal and vertical velocity profiles on the centerlines of the middle plane $z=0$ in the 3D lid-driven cavity flow. The horizontal velocity $u/U_w$ is extracted along the vertical centerline with $x=z=0$, and the vertical velocity $v/U_w$ is extracted along the horizontal centerline with $y=z=0$. }
	\label{fig:ch5_cavity_centerline}
\end{figure}

Figure~\ref{fig:ch5_cavity_midplane} compares the statistically averaged density, velocity magnitude, and total temperature on the middle plane $z=0$, while Fig.~\ref{fig:ch5_cavity_centerline} compares the normalized velocity profiles on two centerlines of the same middle plane. Good agreement between the DIG and DSMC macroscopic fields is seen for all three Knudsen numbers.

\begin{table}[t]
	\centering
	\caption{
		Computational cost comparison between the DIG and SPARTA in the 3D lid-driven cavity flow. $N_{\mathrm{cell}}$, $N_{\mathrm{core}}$, $N_{\mathrm{step}}$, $t_w$, and $N_p$ denote the number of computational cells, CPU cores, evolution steps, wall-clock time, and total number of simulation particles in the sampled state, respectively. For SPARTA, the reported total number of computational cells is counted from the final adaptively refined mesh. For entries written as $a+b$ in $N_{\mathrm{step}}$, $a$ denotes the initial transient steps before final statistical averaging, and $b$ denotes the final sampling steps.
	}
	\label{tab:ch5_cavity_efficiency}
	\small
	\begin{tabular}{ccccccc}
		\hline
		$Kn$ & Method & $N_{\mathrm{cell}}$ & $N_{\mathrm{core}}$ & $N_{\mathrm{step}}$ & $t_w$ (h) & $N_p$ \\
		\hline
		\multirow{2}{*}{$1$}
		& SPARTA
		& $125{,}000$
		& $160$
		& $2000+28000$
		& $0.19$
		& $12{,}500{,}000$ \\
		{}
		& DIG
		& $60 \times 60 \times 60$
		& $80$
		& $2000+10000$
		& $0.64$
		& $22{,}032{,}000$ \\
		\hline
		\multirow{2}{*}{$0.1$}
		& SPARTA
		& $125{,}000$
		& $160$
		& $3000+27000$
		& $0.17$
		& $12{,}500{,}000$ \\
		{}
		& DIG
		& $60 \times 60 \times 60$
		& $80$
		& $2000+10000$
		& $0.77$
		& $21{,}600{,}000$ \\
		\hline
		\multirow{2}{*}{$0.01$}
		& SPARTA
		& $27{,}000{,}000$
		& $1000$
		& $50000+60000$
		& $41.1$
		& $2{,}700{,}000{,}000$ \\
		{}
		& DIG
		& $60 \times 60 \times 60$
		& $80$
		& $3000+10000$
		& $1.42$
		& $21{,}600{,}000$ \\
		\hline
	\end{tabular}
	
	\vspace{0.05cm}

\end{table}

Table~\ref{tab:ch5_cavity_efficiency} summarizes the computational cost of the DIG and SPARTA. The reported wall-clock time excludes mesh generation and post processing. For DIG, this time includes the DSMC particle evolution and the macroscopic correction based on the synthetic equations. 
Two regimes of computational behavior can be observed from the table. For $Kn=1$ and $Kn=0.1$, both SPARTA and DIG reach statistically steady macroscopic fields after approximately $2000$ time steps. The computational costs of SPARTA and DIG are comparable.
For $Kn=0.01$, the computational behavior changes significantly. SPARTA requires on the order of $5.0\times10^4$ particle time steps to reach a comparable statistically steady state, and the total number of evolution steps increases to $110{,}000$. Meanwhile, the number of cells used by SPARTA increases to $27{,}000{,}000$ because the smaller mean free path imposes a much stricter DSMC spatial resolution requirement. In contrast, DIG obtains a consistent macroscopic result using only a $60 \times 60 \times 60$ mesh, reaches a statistically steady macroscopic state after approximately $3000$ particle time steps, and uses $13{,}000$ evolution steps in total. This behavior is consistent with the asymptotic preserving property of the DIG formulation and the fast convergence introduced by the intermittent macroscopic correction \cite{hu2025fastconverging}.
In terms of computational time for the 
$Kn=0.01$ case, SPARTA consumes approximately 41,100 core hours, while the DIG method requires merely 113.6 core hours, corresponding to a speedup factor of 362. Meanwhile, SPARTA exhibits a peak memory usage of  914.1 GB, whereas DIG consumes only 
18.6 GB, achieving a memory reduction factor of 50.

\subsection{Hypersonic flow over a sphere}
\label{subsec:ch5_sphere}

The hypersonic flow over a sphere is computed in a spherical external domain. The sphere diameter is taken as the characteristic length, namely $D=L_0$, and the sphere center is located at the origin. The freestream flow is imposed along the positive $x$ direction. The outer boundary is placed at $R_{\rm out}=10L_0$, and the sphere surface forms the inner boundary with radius $R_s=0.5L_0$. Three Knudsen numbers, $Kn=1$, $Kn=0.1$, and $Kn=0.01$, are considered. The freestream Mach number is $Ma_{\infty}=5$. The freestream density and temperature are used as the reference density and reference temperature, respectively, and both the freestream and sphere wall temperatures are set to $273.15\,\mathrm{K}$.

The DIG uses a 3D unstructured mesh with $230,400$ computational cells. The mesh is refined near the sphere surface and in the bow shock region, with the minimum wall normal spacing set to $0.001$. Each cell is initialized with 100 simulation particles on average, and the particle CFL number is set to $0.2$. The SPARTA uses the same freestream condition, wall temperature, initial particle number, and particle CFL number, while the Cartesian grid is adaptively refined near the sphere surface and in the bow shock region.

\begin{figure}[t!]
	\centering
\subfigure[density]{\includegraphics[width=0.4\linewidth]{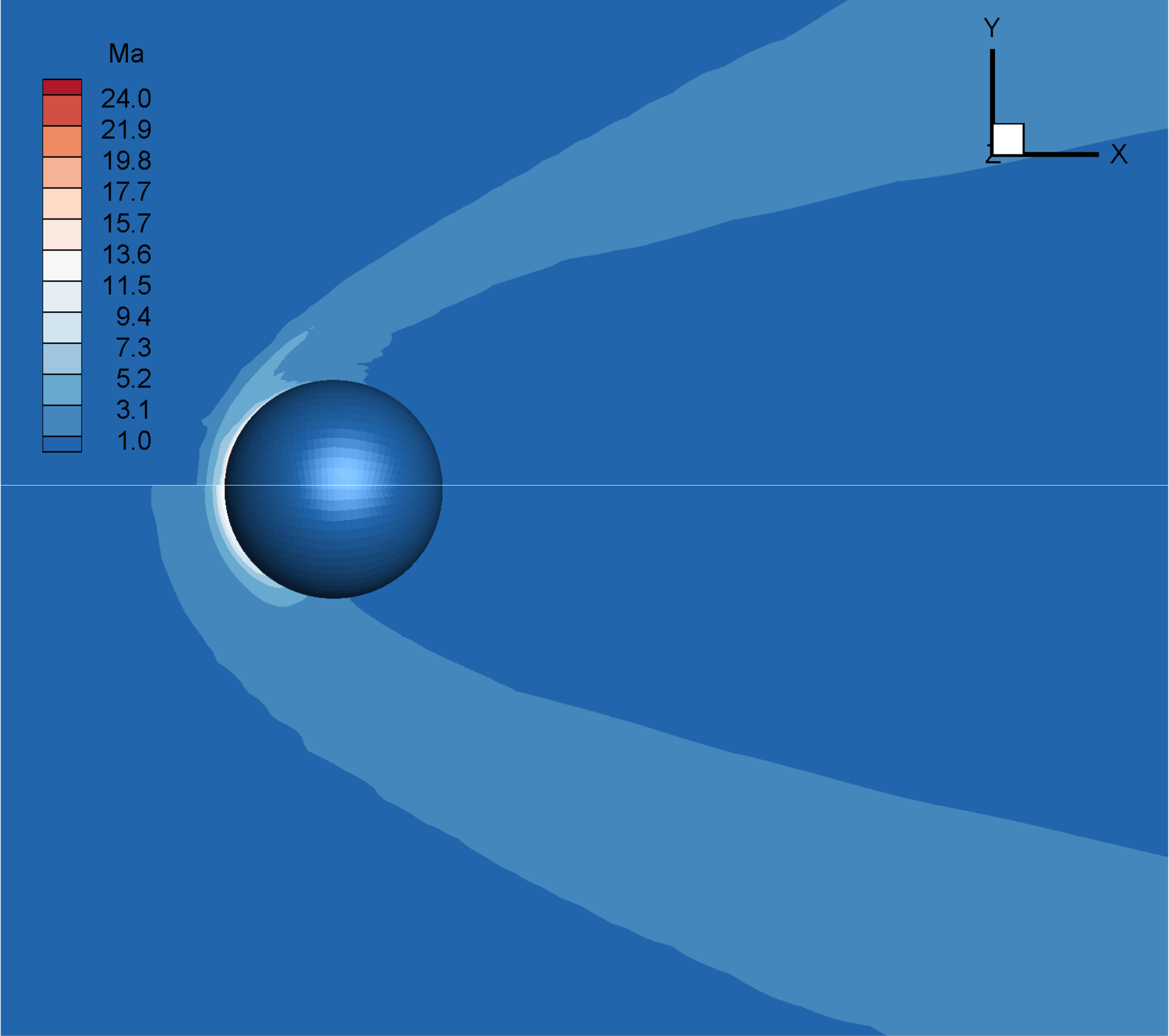}}
\subfigure[local Mach number]{\includegraphics[width=0.4\linewidth]{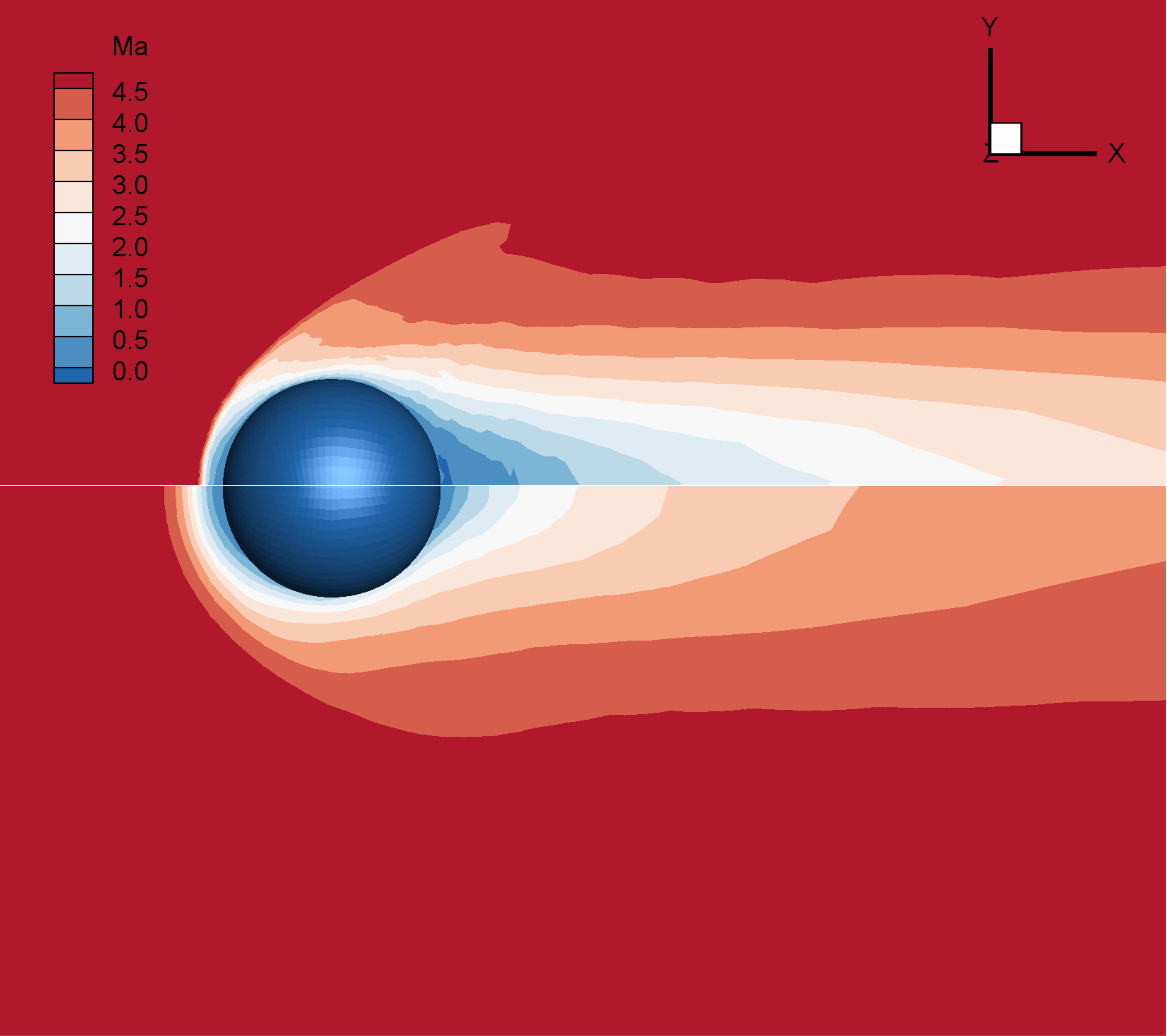}}
\subfigure[translational temperature]{\includegraphics[width=0.4\linewidth]{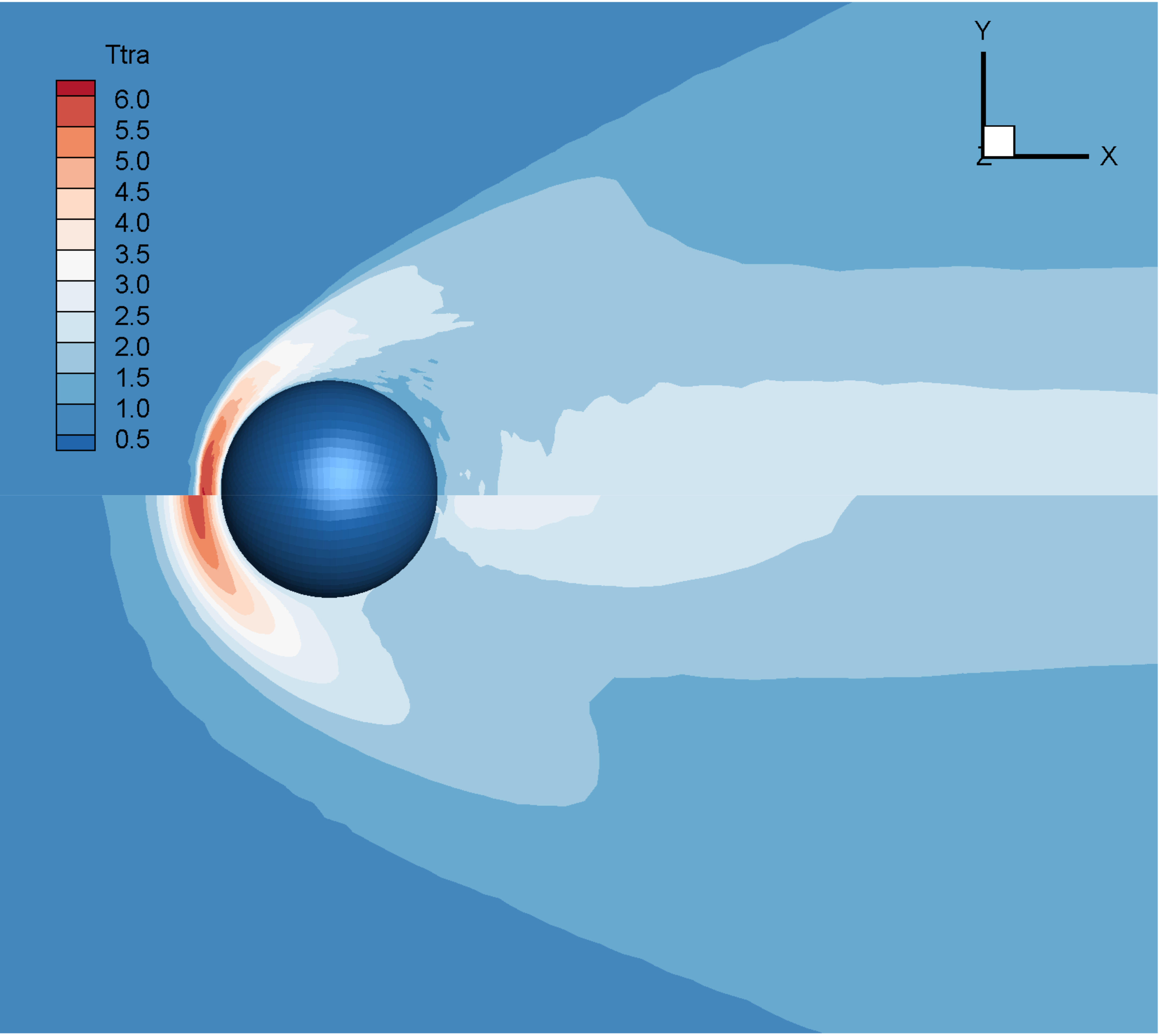}}
\subfigure[rotational temperature]{\includegraphics[width=0.4\linewidth]{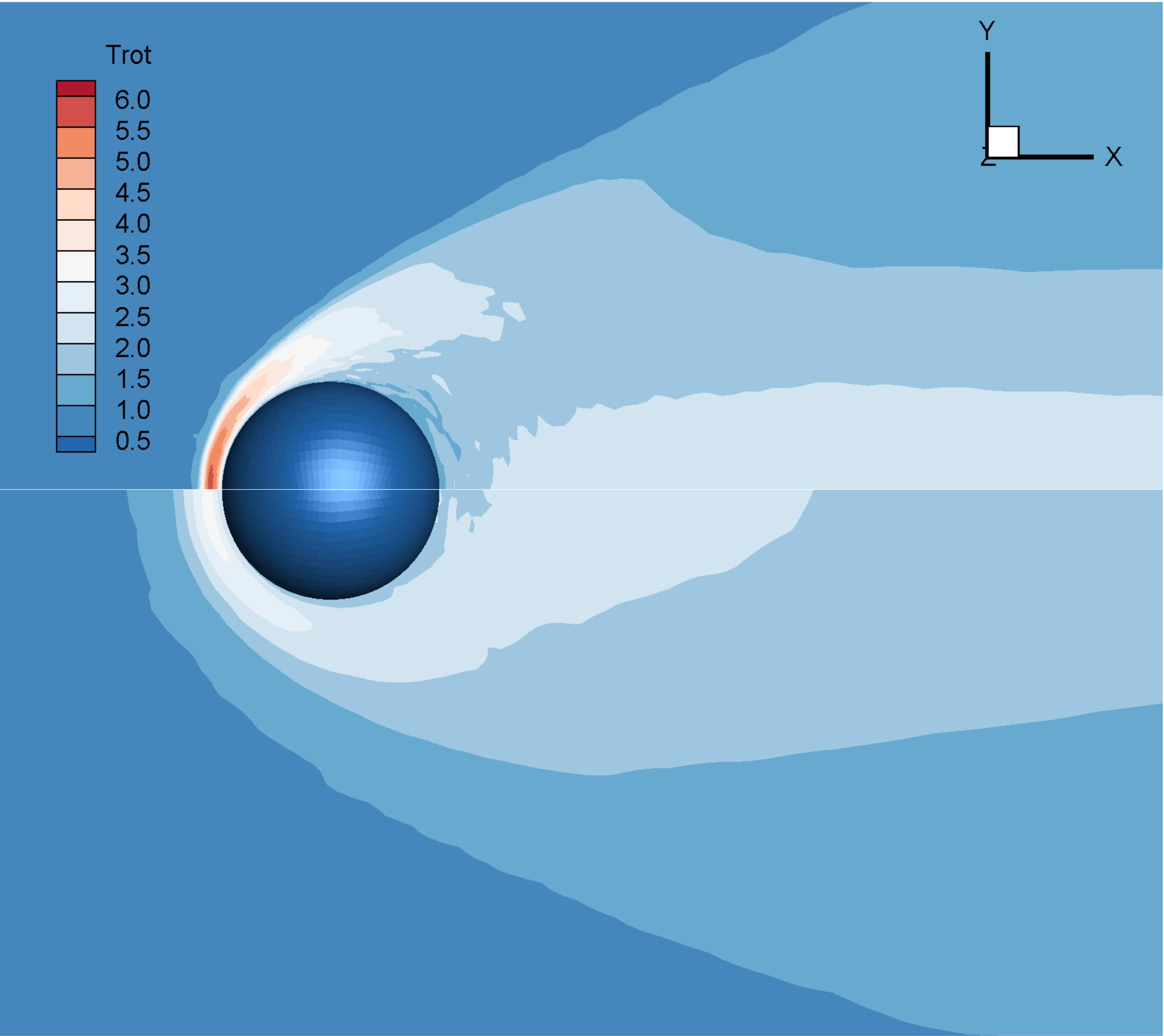}}
	\caption{Macroscopic quantities in the hypersonic flow over a sphere on the $z=0$ plane. In each figure, the top and bottom half region shows the results of Kn=0.01 and 0.1, respectively. }
	\label{fig:ch5_sphere_flowfield_kn01}
\end{figure}

Figure~\ref{fig:ch5_sphere_flowfield_kn01} shows the density, local Mach number, translational temperature, and rotational temperature for $Kn=0.1$ and $Kn=0.01$, respectively. For $Kn=0.1$, a distinct compressed region has formed in front of the sphere, and the variations of density, Mach number, and temperature begin to concentrate near the windward side. For $Kn=0.01$, the compressed region becomes thinner, and the gradients of density, Mach number, and temperature near the windward side become sharper. The difference between translational and rotational temperatures is mainly observed inside the shock layer, and at larger Knudsen number, indicating thermal nonequilibrium caused by strong compression and finite rotational relaxation.

\begin{figure}[t]
	\centering
\subfigure[$Kn=1$]{\includegraphics[width=0.32\linewidth,trim={30 10 50 50}]{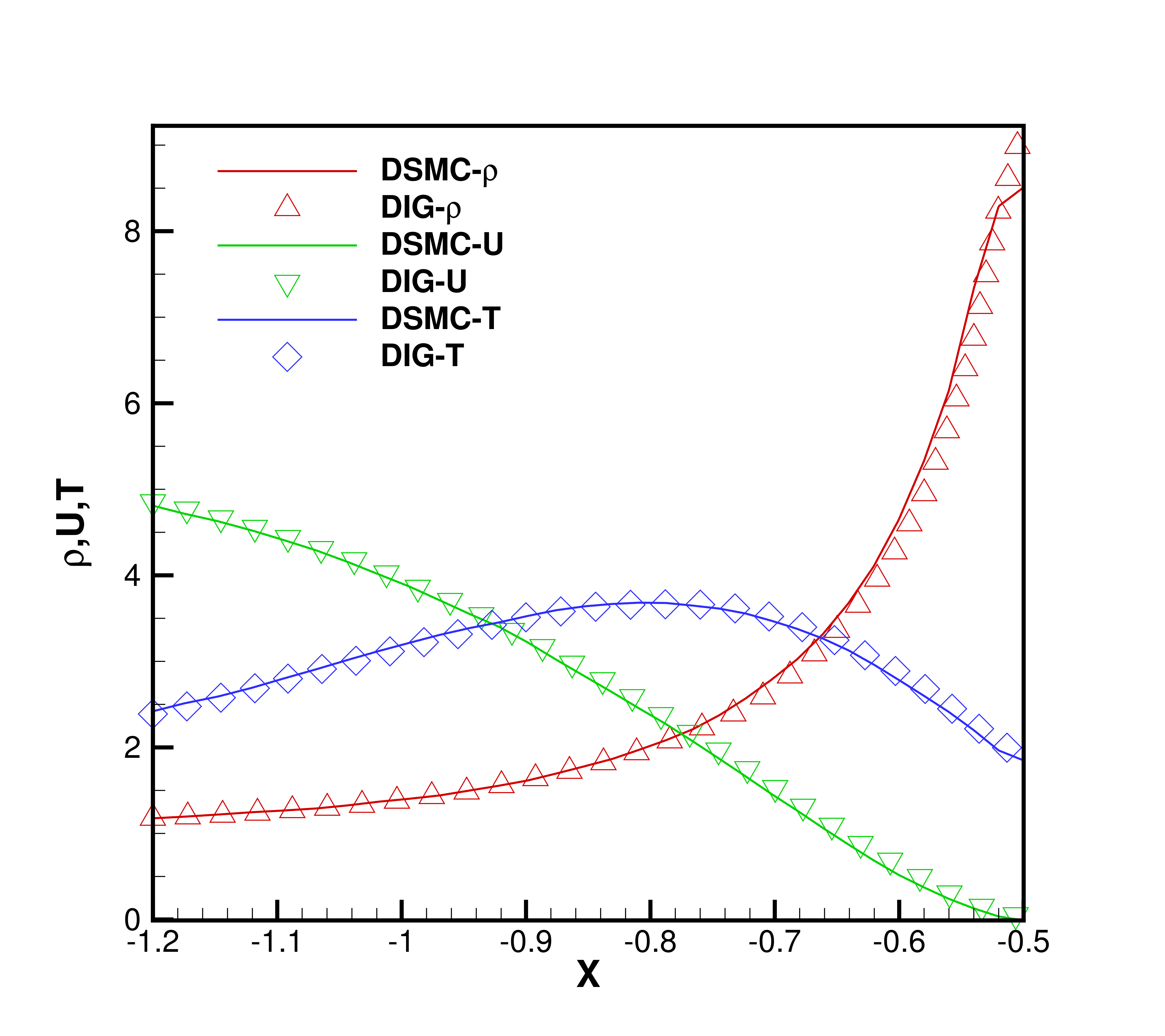}}
	\subfigure[$Kn=0.1$]{\includegraphics[width=0.32\linewidth,trim={30 10 50 50}]{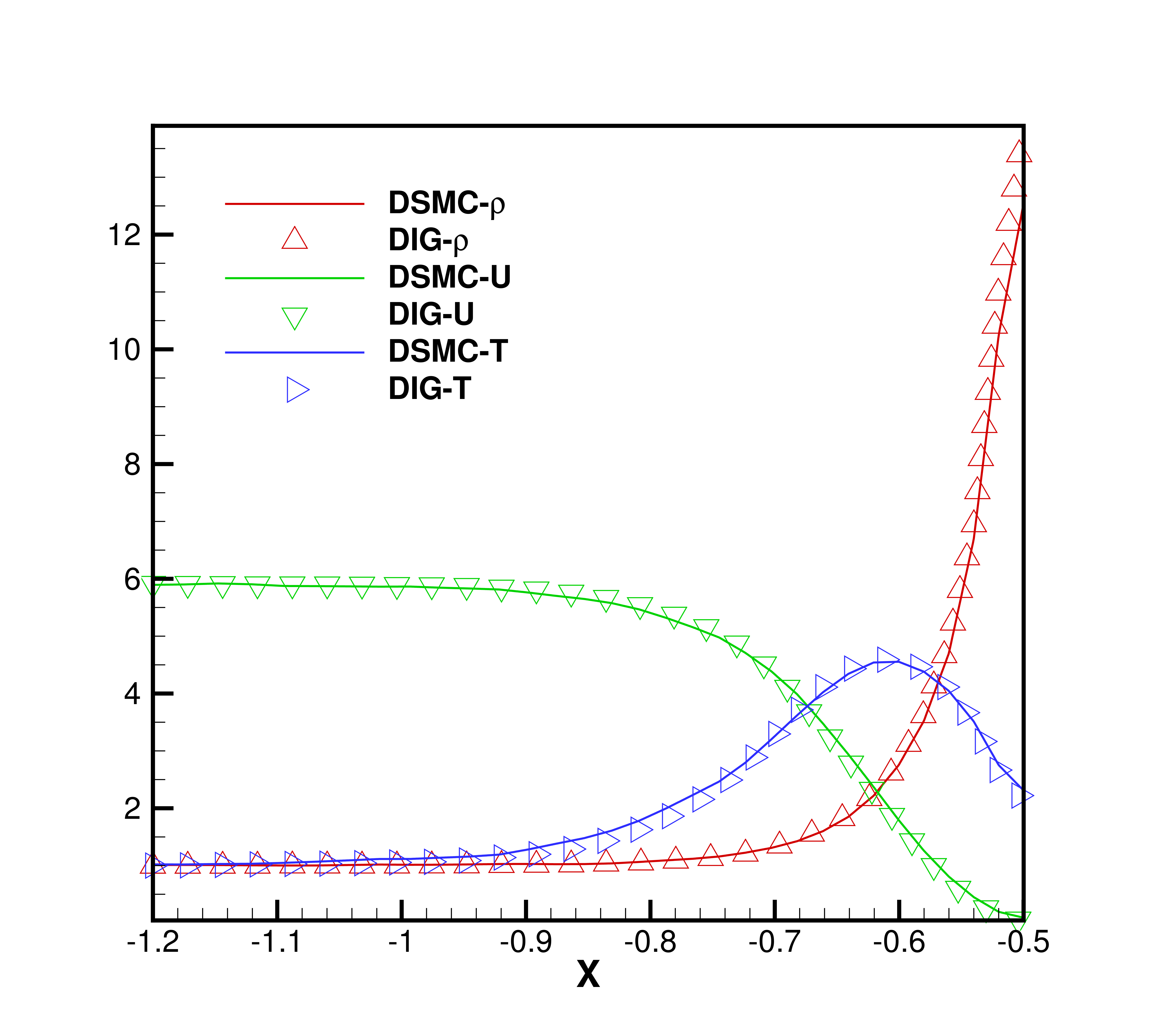}}
	\subfigure[$Kn=0.01$]{\includegraphics[width=0.32\linewidth,trim={30 10 50 50}]{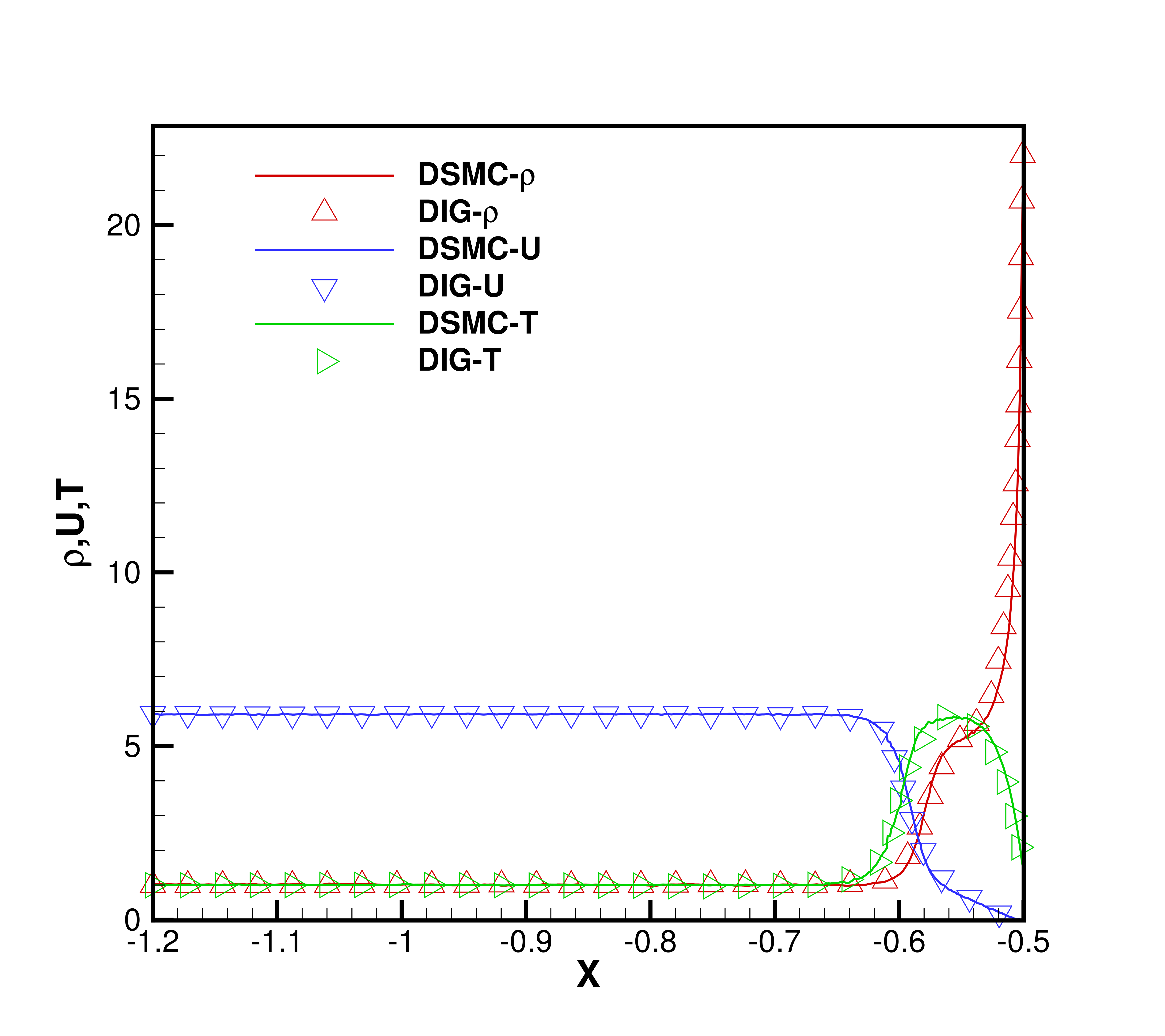}}\\
    \subfigure[$Kn=0.1$]{\includegraphics[width=0.4\linewidth,trim={10 10 30 30}]{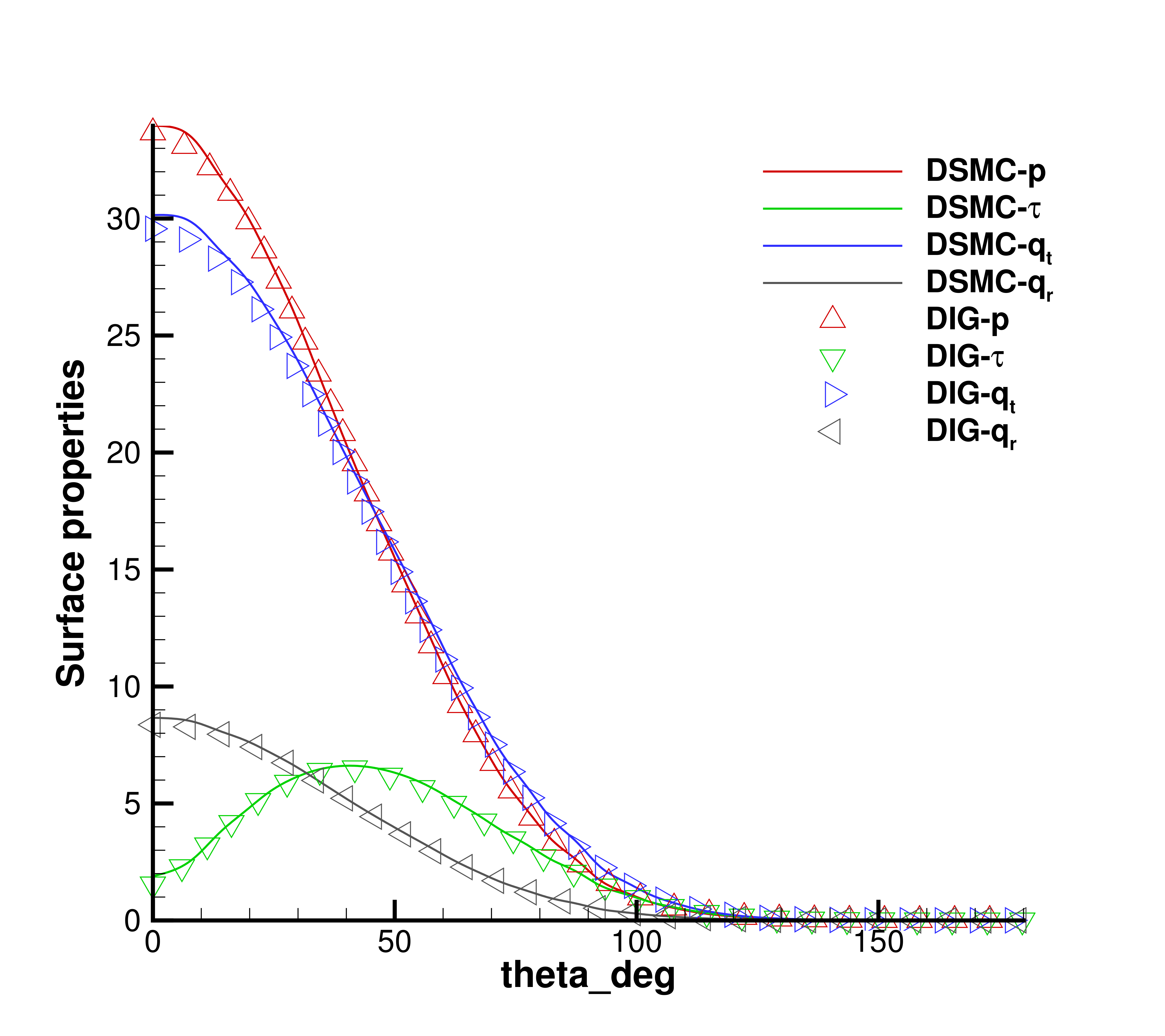}}
		\subfigure[$Kn=0.01$]{\includegraphics[width=0.4\linewidth,trim={10 10 30 30}]{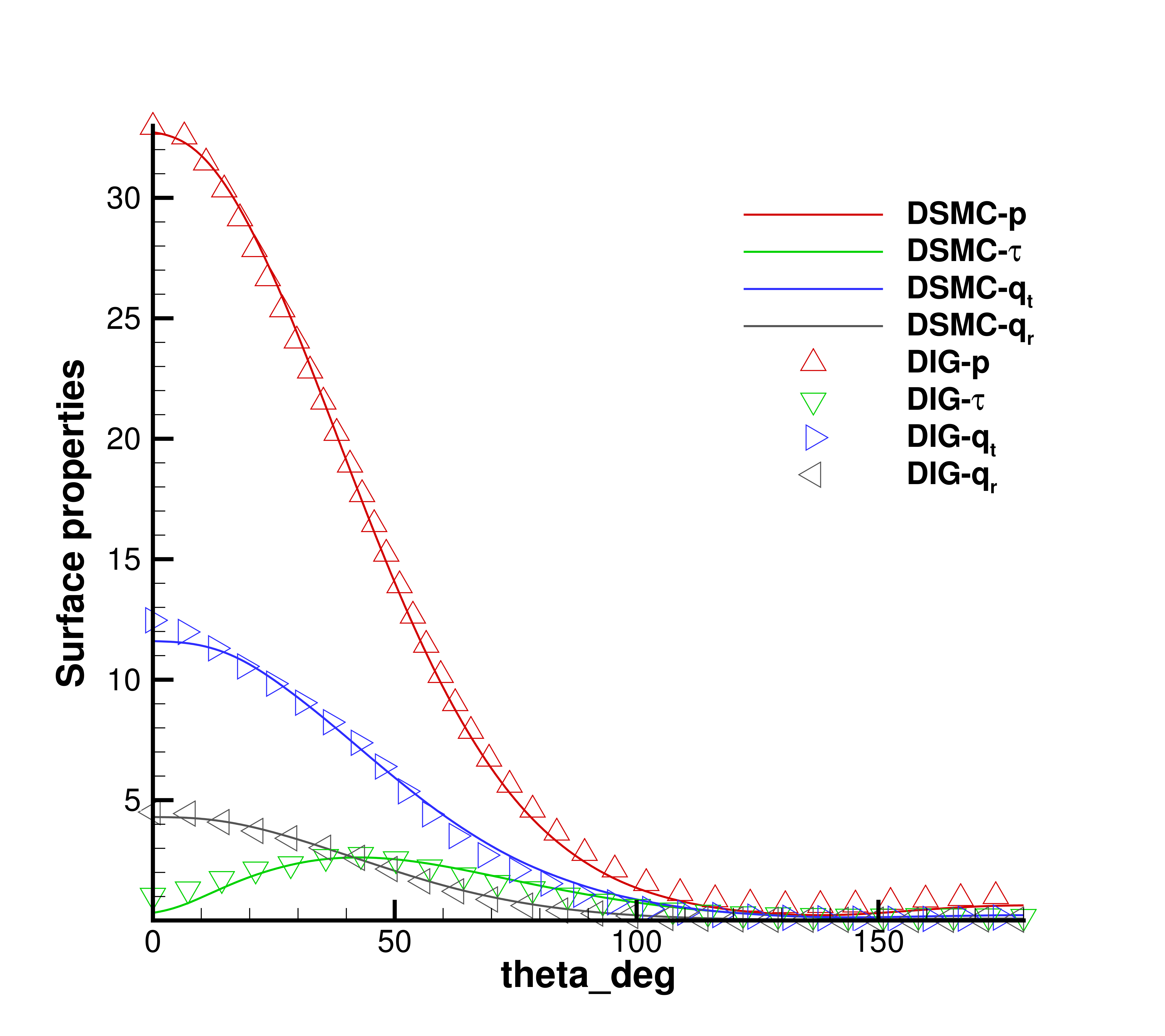}}
	\caption{(a-c) The density, streamwise velocity, and temperature along the windward stagnation line in the hypersonic flow over a sphere, sampled from $(-1.2,0,0)$ to $(-0.5,0,0)$. 
    (d,e) The pressure, wall shear stress, translational heat flux, and rotational heat flux distributions on the sphere surface. }
	\label{fig:ch5_sphere_stagnation_line}
\end{figure}

The density, streamwise velocity, and translational/rotational temperature along the windward stagnation line are compared between the SPARTA and DIG in Fig.~\ref{fig:ch5_sphere_stagnation_line}. The line passes through the sphere center and the windward stagnation point.  For $Kn=1$, the profiles vary smoothly because the rarefaction effect is strong and the compressed layer in front of the sphere is relatively diffuse. As the Knudsen number decreases, the variations of density and temperature become more concentrated near the windward side, and the streamwise velocity drops within a thinner shock layer. The DIG results show overall agreement with the DSMC profiles for the three Knudsen numbers, especially in the shock layer location, the density increase, and the translational and rotational temperature variations in the compressed region.

Figure~\ref{fig:ch5_sphere_stagnation_line} also compares the surface quantities on the sphere for $Kn=0.1$ and $Kn=0.01$. The polar angle $\theta$ is measured from the upstream stagnation point, with $\theta=0^{\circ}$ corresponding to the windward stagnation point and $\theta=180^{\circ}$ corresponding to the rear side. The compared quantities include surface pressure, wall shear stress, translational heat flux, and rotational heat flux. The surface pressure reaches its maximum near the stagnation point and decreases along the sphere surface. The wall shear stress remains small near the stagnation point because of the local symmetry of the flow, increases along the windward surface, and then decreases toward the downstream region. The translational heat flux is mainly concentrated near the windward side, where the compression and temperature variation are strongest. In the present cases, the rotational heat flux is smaller than the translational heat flux over most of the surface and varies more mildly along the polar angle. For both Knudsen numbers, the DIG results agree well with the DSMC data over most of the surface, indicating that the surface aerodynamic and heat transfer quantities are reasonably captured by the 3D DIG calculation.

\begin{table}[t]
	\centering
	\caption{
		Computational cost comparison between the DIG and SPARTA in the hypersonic flow over a sphere. 
		Note that when $Kn=0.01$, the SPARTA adopts the axisymmetric geometry to save the computational cost.
	}
	\label{tab:ch5_sphere_efficiency}
	\small
	\begin{tabular}{ccccccc}
		\hline
		$Kn$ & Method & $N_{\mathrm{cell}}$ & $N_{\mathrm{core}}$ & $N_{\mathrm{step}}$ & $t_w$ (h) & $N_p$ \\
		\hline
		\multirow{2}{*}{$1$}
		& SPARTA
		& $788{,}493$
		& $160$
		& $3000+22000$
		& $0.64$
		& $40{,}213{,}143$ \\
		{}
		& DIG
		& $230{,}400$
		& $160$
		& $2000+2000$
		& $0.12$
		& $23{,}270{,}400$ \\
		\hline
		\multirow{2}{*}{$0.1$}
		& SPARTA
		& $2{,}951{,}843$
		& $160$
		& $5000+25000$
		& $1.18$
		& $44{,}277{,}645$ \\
		{}
		& DIG
		& $230{,}400$
		& $160$
		& $3000+2000$
		& $0.19$
		& $23{,}270{,}400$ \\
		\hline
		\multirow{2}{*}{$0.01$}
		& SPARTA (axisymmetric)
		& $625{,}076$ 
		& $320$
		& $50000+50000$
		& $0.34$
		& $11{,}251{,}368$ \\
		{}
		& DIG
		& $230{,}400$
		& $160$
		& $5000+5000$
		& $0.42$
		& $23{,}270{,}400$ \\
		\hline
	\end{tabular}
\end{table}

Table~\ref{tab:ch5_sphere_efficiency} summarizes the computational cost of the DIG and SPARTA. The DIG uses $230,400$ computational cells for the three Knudsen numbers, while the SPARTA Cartesian grid is refined according to the local DSMC resolution requirement. For $Kn=0.01$, a direct 3D SPARTA calculation would require prohibitive computational resources under the DSMC spatial resolution requirement. Therefore, only the axisymmetric SPARTA calculation is used.
For $Kn=1$, DIG requires fewer cells and fewer evolution steps than SPARTA, reducing the wall-clock time from $0.64$~h to $0.12$~h under the same number of CPU cores. For $Kn=0.1$, SPARTA uses $2{,}951{,}843$ cells and $30000$ evolution steps, whereas DIG uses $230,400$ cells and $5000$ evolution steps. The wall-clock time is reduced from $1.18$~h to $0.19$~h. In the same case, the memory usage of DIG is $37$~GB, which is slightly lower than the $43$~GB in SPARTA. These results show that the intermittent macroscopic correction reduces the number of evolution steps required by the particle calculation, while the unstructured mesh used in the DIG calculation keeps the spatial resolution concentrated around the sphere and the bow shock region. For $Kn=0.01$, the DIG calculation is completed with $230,400$ cells and $10,000$ total evolution steps. Compared with the larger Knudsen number cases, the wall-clock runtime slightly increases owing to elevated collision computational overhead and amplified stiffness introduced by macroscopic corrections. In contrast, SPARTA demands an extremely large mesh count. Even when its 2D axisymmetric solver is deployed, the simulation requires 625,076 cells and approximately 109 core hours, which is larger than the 67 core hours consumed by the 3D DIG.

\subsubsection{Parallel efficiency and dynamic load balancing results}

The parallel scalability of the present DIG is examined, for the $Kn=1$ case after the flow reaches a statistically steady state. Starting from this steady particle state, $10000$ DSMC time steps are advanced, and the corresponding wall-clock time is recorded. This test is used only to assess the parallel performance of the particle evolution stage on unstructured mesh, and it is not the total DIG cost reported in Table~\ref{tab:ch5_sphere_efficiency}. To reduce the cost of the scaling test, the particle load is set to $10\%$ of that used in the production calculation.

The results are listed in Table~\ref{tab:ch5_sphere_parallel_efficiency}. The actual speedup ratio $S_p$ and parallel efficiency $\eta_p$ are defined using the 10 core result as the reference,
\begin{equation}
	S_p
	=
	\frac{t_{10}}{t_p},
	\qquad
	\eta_p
	=
	\frac{S_p}{p/10}\times 100\%,
	\label{eq:ch5_sphere_parallel_efficiency}
\end{equation}
where $t_{10}$ and $t_p$ are the wall-clock times measured with 10 and $p$ CPU cores, respectively. 
As shown in Table~\ref{tab:ch5_sphere_parallel_efficiency}, the wall-clock time decreases from $10344$ s on 10 cores to $706$ s on 200 cores, corresponding to an actual speedup of $14.7$ and a parallel efficiency of $73.3\%$. The efficiency remains above $70\%$ up to 200 cores, indicating that the particle bucket organization, local mesh construction, and particle migration strategy provide effective parallel performance for the present 3D unstructured mesh. When the number of CPU cores is further increased, the efficiency decreases because the number of cells and particles assigned to each process becomes smaller, and the relative cost of MPI communication, particle migration, and synchronization becomes more important.

\begin{table}[t]
\centering
\caption{Parallel efficiency of DIG for hypersonic flow over a sphere at $Kn=1$.
		The wall-clock time is measured for $10000$ DSMC time steps after the flow reaches a statistically steady state. 
		The particle load is set to $10\%$ of that used in the production calculation.
        }
        \label{tab:ch5_sphere_parallel_efficiency}
\begin{tabular}{lccccccccc}
\hline
Cores & $10$ & $40$ & $80$ & $120$ & $160$ & $200$ & $240$ & $280$ & $320$ \\
\hline
$t_w$ (s) & 10,344 & 2,410 & 1,481 & 1,105 & 869 & 706 & 658 & 550 & 597 \\
$S_p$ & 1 & 4.3 & 7.0 & 9.4 & 11.9 & 14.7 & 15.7 & 18.8 & 17.3 \\
$\eta_p$ (\%) & 100 & 107.3 & 87.3 & 78.0 & 74.4 & 73.3 & 65.5 & 67.2 & 54.2 \\
\hline
\end{tabular}
\label{tab:parallel_transpose}
\end{table}

When the CPU cores are raised from 10 to 40, the parallel efficiency slightly exceeds 100\%. This phenomenon originates from the stochastic nature of the particle method and nonuniform particle distribution in hypersonic sphere flow. Even if the macroscopic flow attains a statistically steady state, the particle count within each subdomain and the quantity of migrating particles still fluctuate across sampled DSMC steps. Such particle fluctuations lead to small deviations in measured wall-clock time during strong scaling tests, which becomes more prominent when the particle load is reduced to 10\%, leaving each process with relatively little computational workload. For the same reason, the wall-clock time rises slightly instead when the core number further increases from 280 to 320.

\begin{figure}[t!]
	\centering
    \subfigure[Initial state mesh partition]{\includegraphics[width=0.48\textwidth]{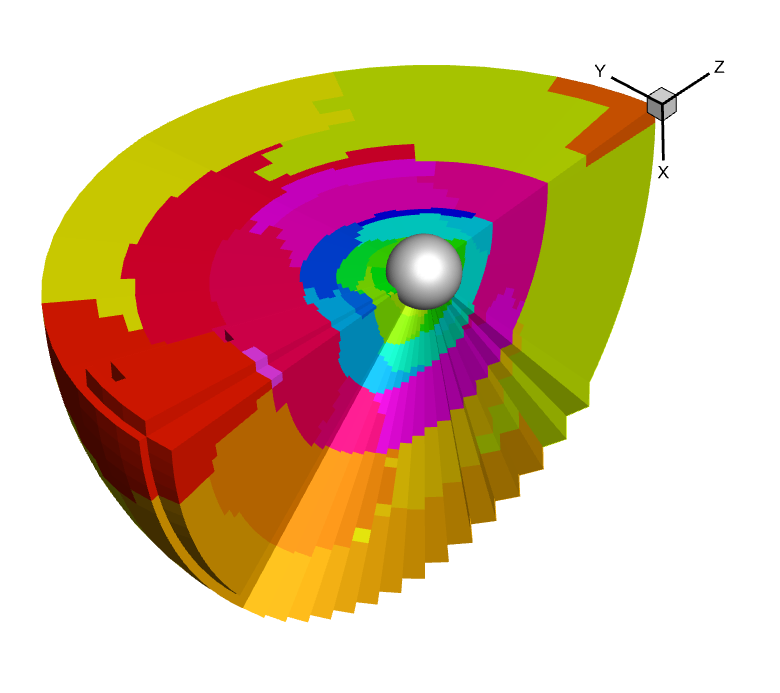}}
	\subfigure[Final state mesh partition]{\includegraphics[width=0.48\textwidth]{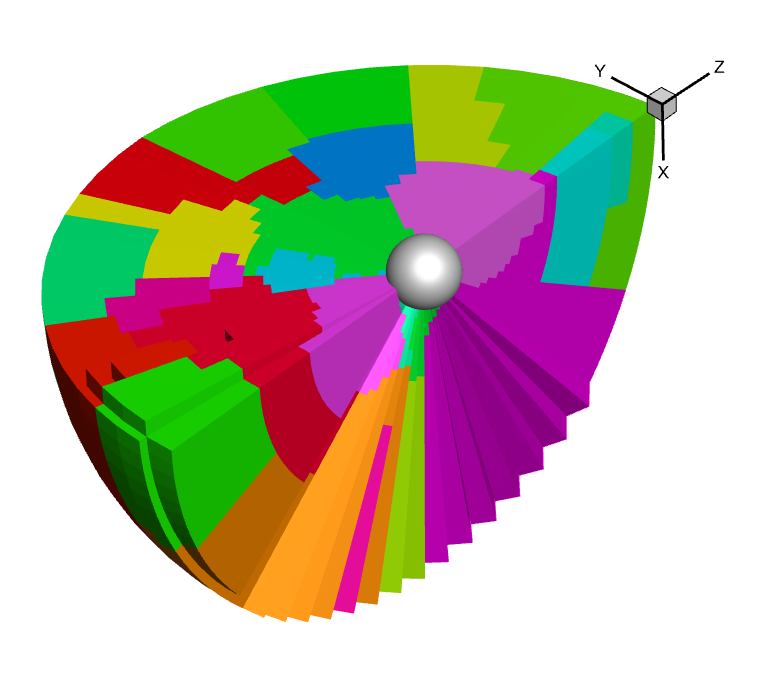}}\\
	\subfigure[$\eta_L^\text{crit}=2.5$]{\includegraphics[width=0.48\textwidth]{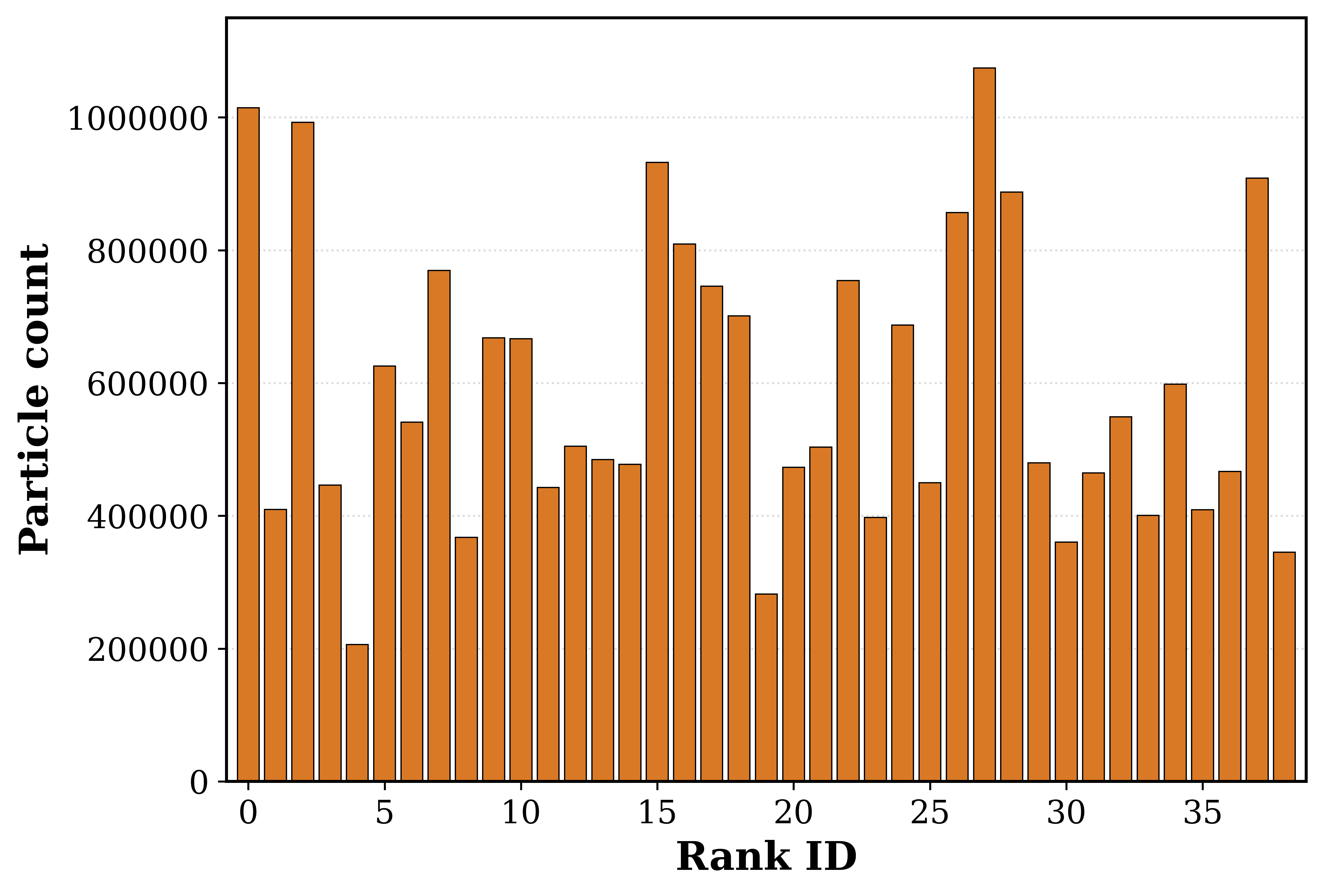}}
    \subfigure[$\eta_L^\text{crit}=1.05$]{
		\includegraphics[width=0.48\textwidth]{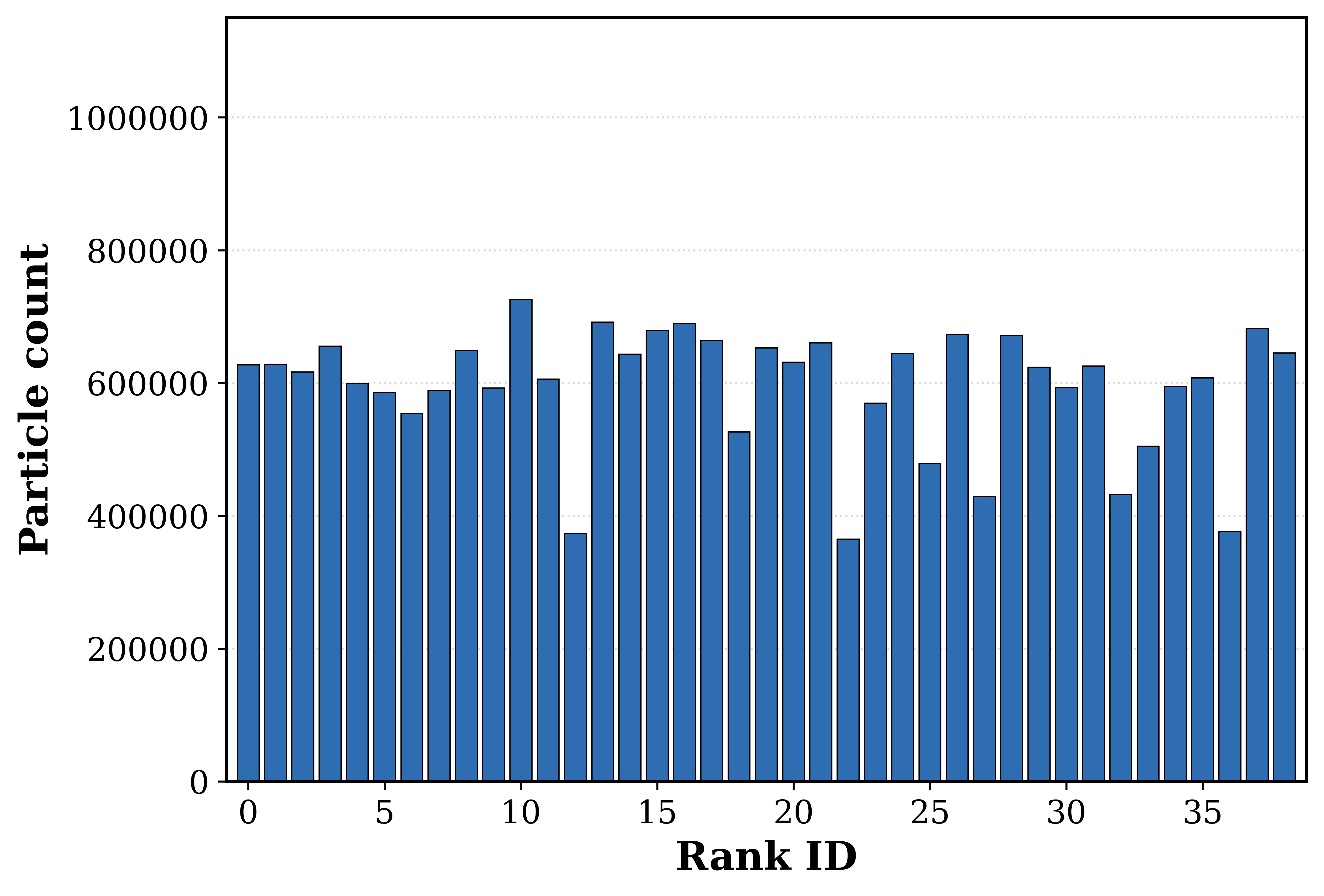}}
	\caption{Dynamic mesh re-partitioning for hypersonic flow over a sphere by 40 MPI processes. (a) and (b) present the initial and final state mesh partitioning results with the load imbalance ratio threshold $\eta_L^\text{crit}=1.05$. (c) and (d) compare the particle-count distributions among MPI processes obtained with different load imbalance thresholds, $\eta_L^\text{crit}=2.5$ and $1.05$, respectively.}
	\label{fig:ch5_sphere_dynamic_partition}
\end{figure}

Moreover, the evolution of the mesh partition during the simulation has been examined.
Starting from the initial mesh partition shown in Fig.~\ref{fig:ch5_sphere_dynamic_partition}(a), the mesh is dynamically re-partitioned every $N_\text{bal}=10$ DIG steps during the transient state.
As the solution approaches steady state, the workload distribution gradually stabilizes, and the mesh partition consequently converges to a nearly unchanged configuration, as shown in Fig.~\ref{fig:ch5_sphere_dynamic_partition}(b). 
To further examine the influence of the load imbalance ratio, Figs.~\ref{fig:ch5_sphere_dynamic_partition}(c) and (d) compare the particle-count distributions among MPI processes obtained with $\eta_L^\text{crit}=2.5$ and $1.05$, respectively. A smaller value of $\eta_L^\text{crit}$ triggers mesh re-partitioning more frequently, resulting in a more uniform particle distribution among MPI processes and hence a better balanced computational workload. 
As shown in Fig.~\ref{fig:ch5_sphere_dynamic_partition}(c) and (d), the particle counts are distributed much more uniformly when $\eta_L^\text{crit}=1.05$ than $\eta_L^\text{crit}=2.5$, indicating a more balanced workload across MPI processes.
However, $\eta_L^\text{crit}$ should not be chosen excessively small, since frequent mesh re-partitioning process increases the cost of inter-process particle migration and particle data redistribution, thereby reducing the overall computational efficiency.

\subsection{Apollo Reentry} 

The hypersonic flow around an Apollo reentry capsule is further tested. Compared with the sphere case, the capsule geometry and the finite angle of attack lead to stronger flow asymmetry, including a compressed windward shock layer, an expanded leeward region, and a low density wake behind the body. The freestream Mach number is $Ma_\infty=5$, and the angle of attack is $30^\circ$. The reference length is $L_0=3.912\,\mathrm{m}$, the freestream temperature is $T_\infty=142.2\,\mathrm{K}$, and the wall temperature is $T_w=300\,\mathrm{K}$.

\begin{figure}[t!]
	\centering
\subfigure[local Mach number]{\includegraphics[width=0.4\linewidth]{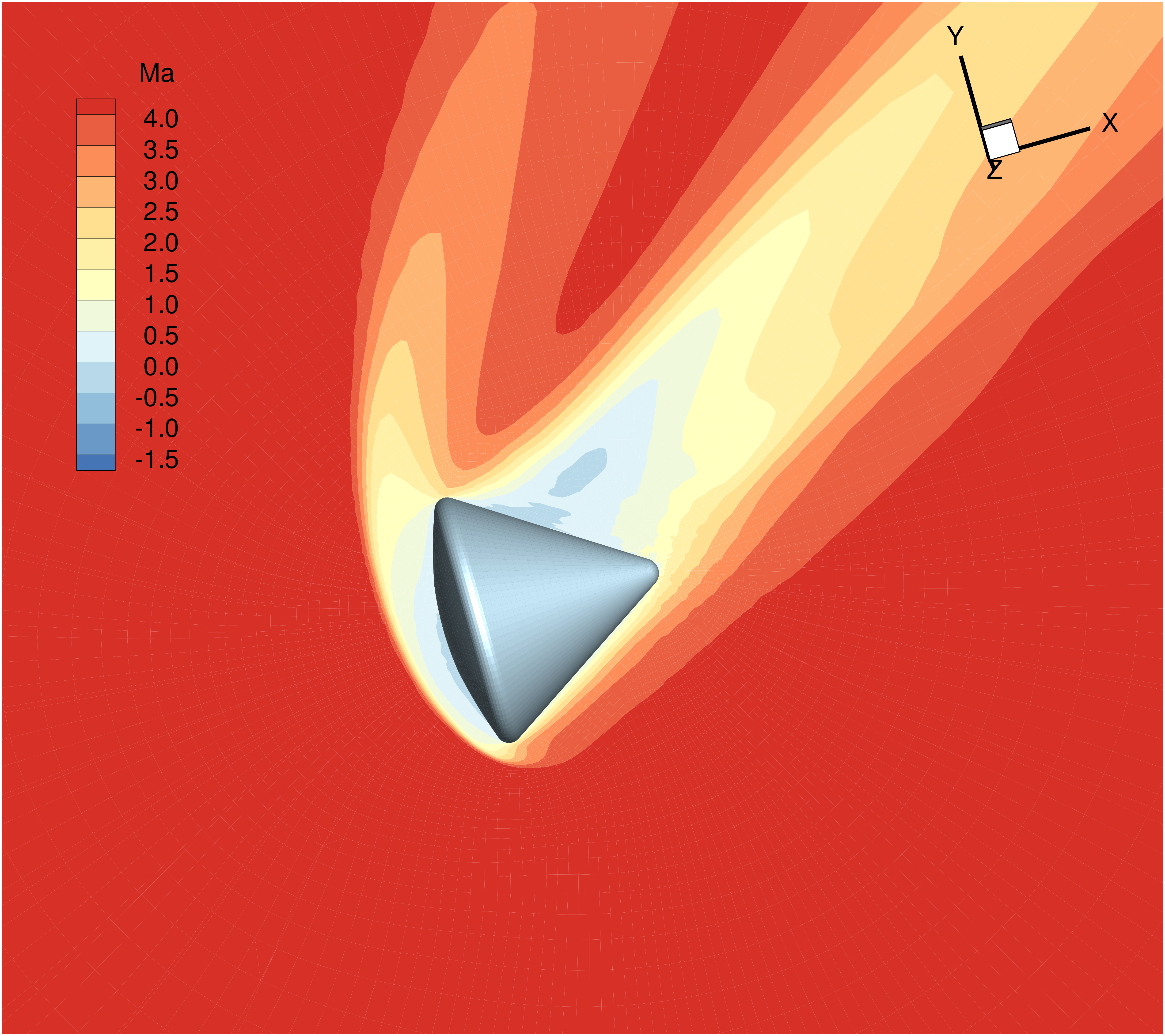}}
\subfigure[translational temperature]{\includegraphics[width=0.4\linewidth]{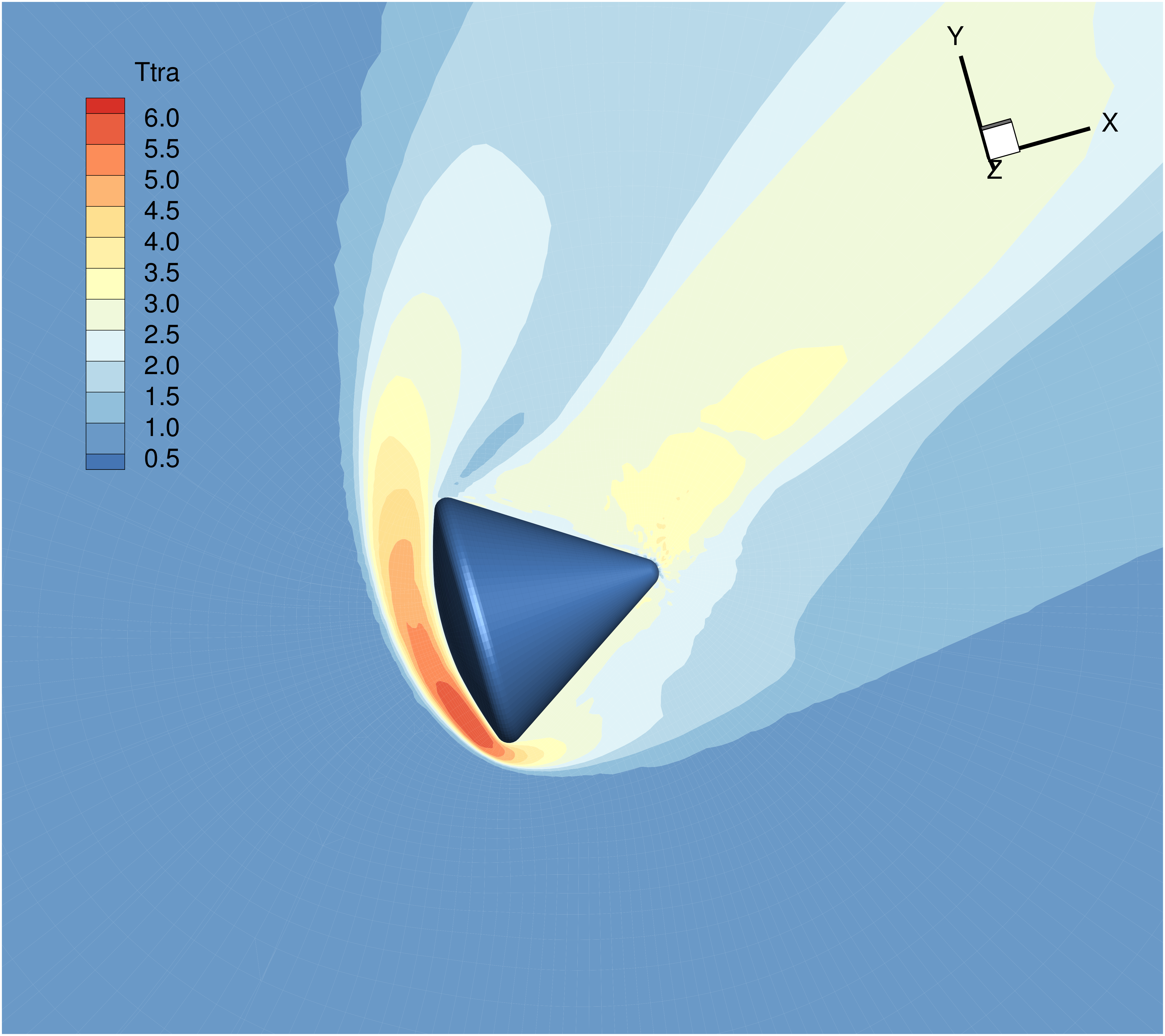}}\\
\subfigure[local cell Knudsen number in Eq.~\eqref{eq:cell_kn_stability}]{
{\includegraphics[width=0.4\linewidth]{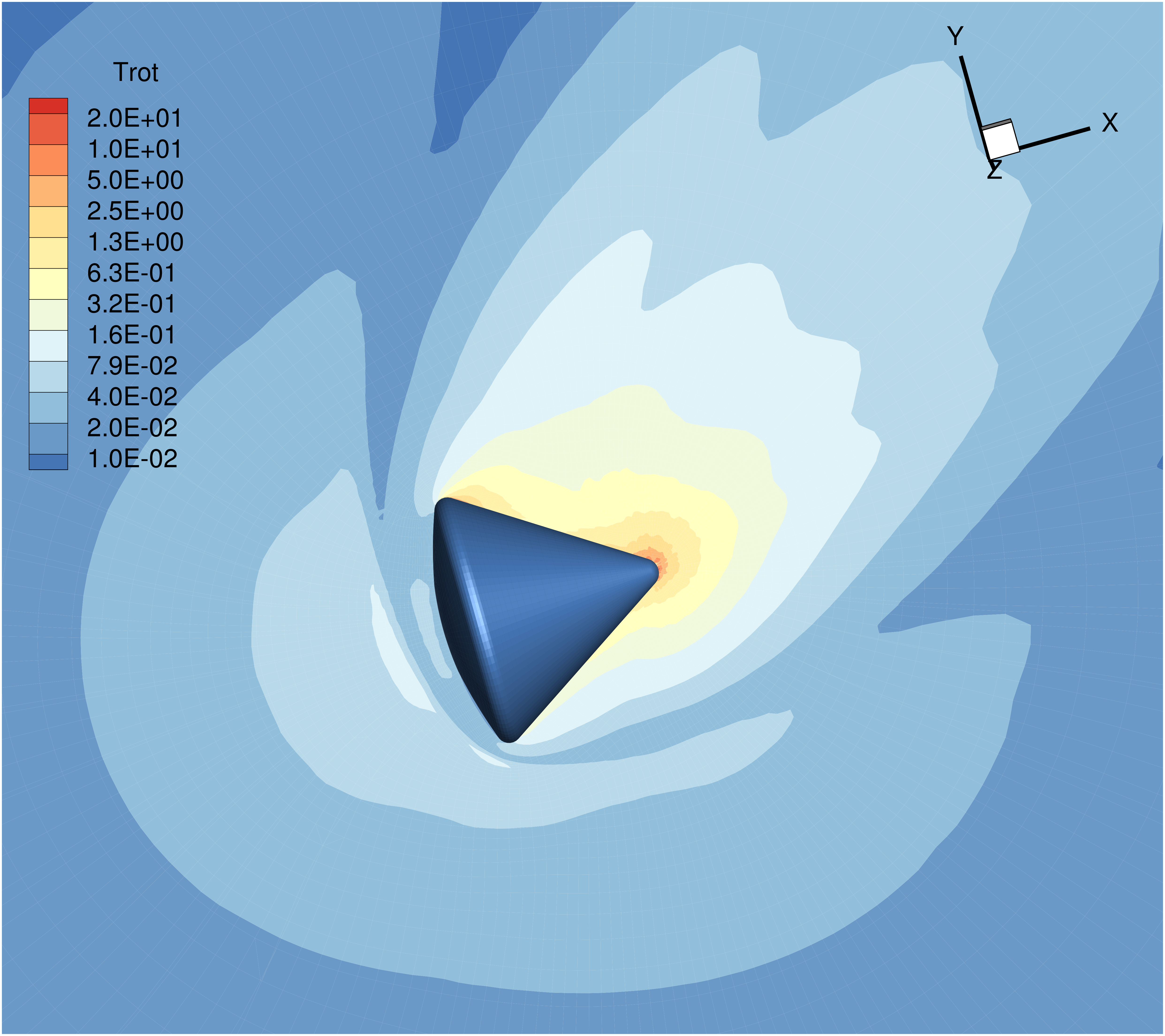}} }
	\caption{DIG results in the hypersonic flow around an Apollo reentry capsule for $Kn=0.01$. }
	\label{fig:ch5_apollo_flowfield_kn001}
\end{figure}

The DIG uses a 3D unstructured mesh with $372{,}500$ computational cells. 
Figure~\ref{fig:ch5_apollo_flowfield_kn001} shows the representative macroscopic fields when $Kn=0.01$. A thin bow shock is formed ahead of the windward side, and the variations of  Mach number and temperature are concentrated in a narrow region. 
Figure~\ref{fig:ch5_apollo_flowfield_kn001}(c) further shows the local cell Knudsen number \eqref{eq:cell_kn_stability}. Large values are mainly distributed in the shock layer, near the capsule shoulder, and in the wake, which is consistent with the regions where the sampled macroscopic input becomes more sensitive and the stability enhancement procedure in Section~\ref{subsec:stability_enhancement} is more frequently activated.

\begin{figure}[t!]
	\centering
	\begin{minipage}[t]{0.32\textwidth}
		\centering
		\includegraphics[width=\linewidth,trim={30 10 50 50}]{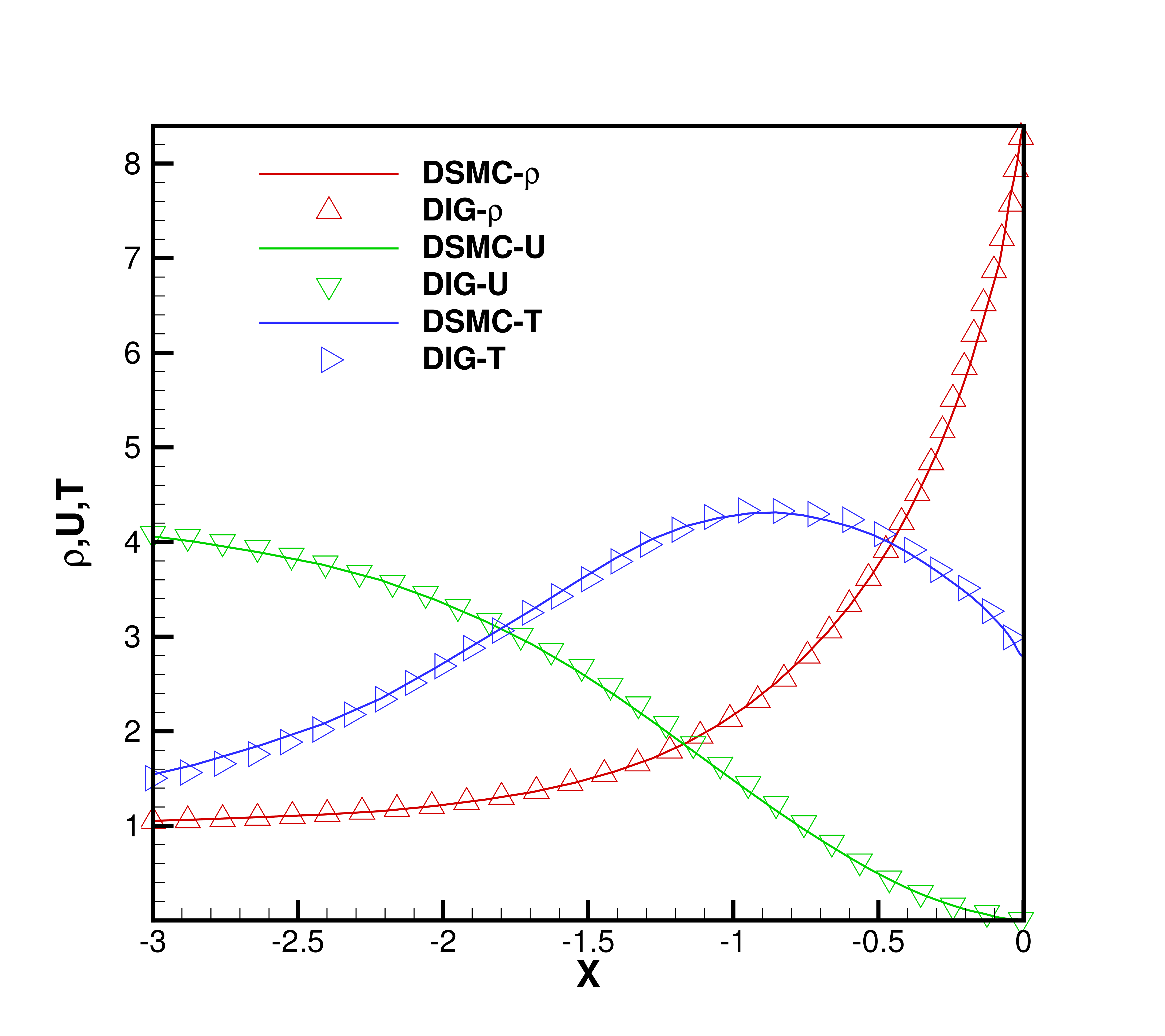}\\[-0.1cm]
		(a) $Kn=1$
	\end{minipage}
	\hfill
	\begin{minipage}[t]{0.32\textwidth}
		\centering
		\includegraphics[width=\linewidth,trim={30 10 50 50}]{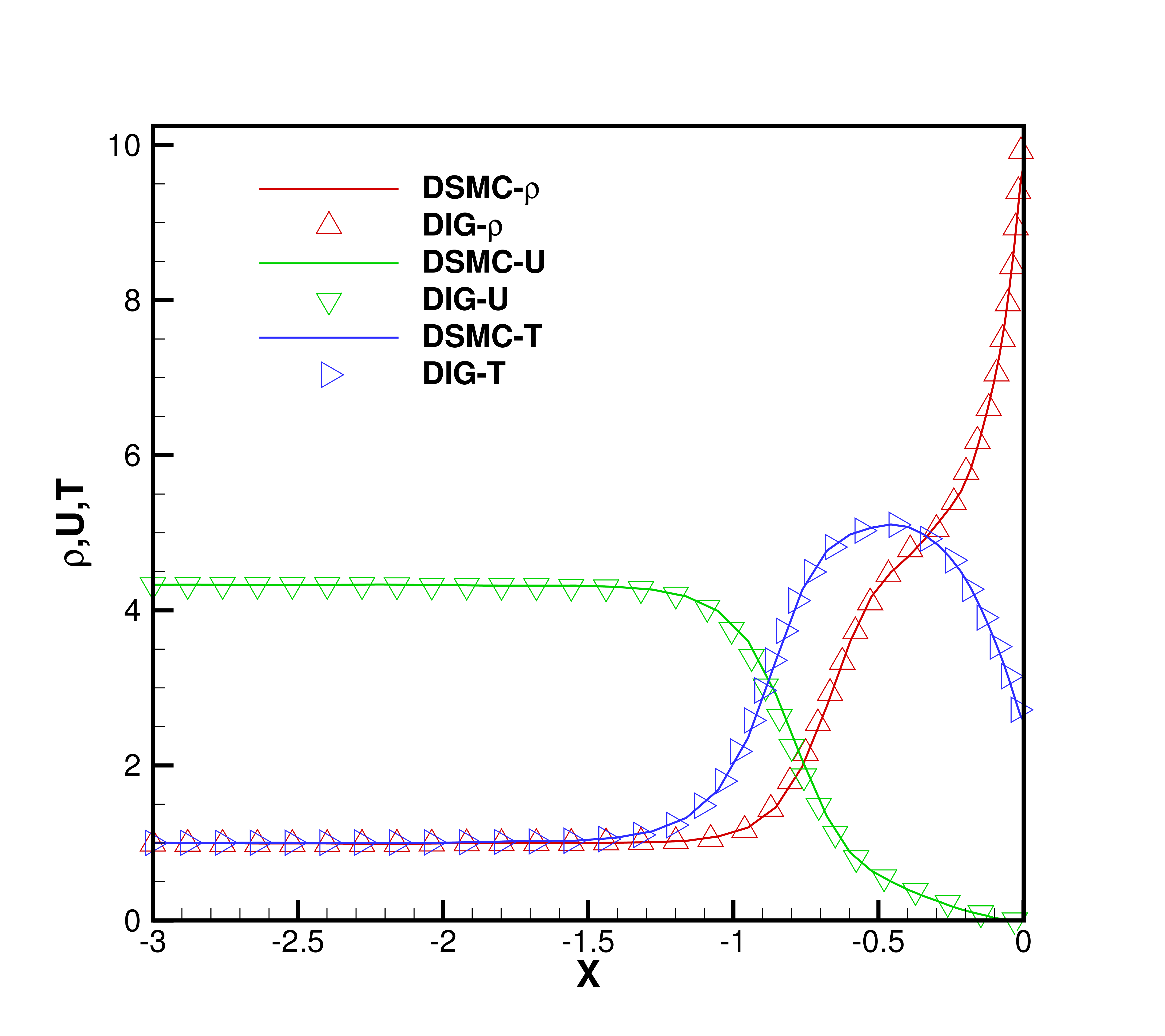}\\[-0.1cm]
		(b) $Kn=0.1$
	\end{minipage}
	\hfill
	\begin{minipage}[t]{0.32\textwidth}
		\centering
		\includegraphics[width=\linewidth,trim={30 10 50 50}]{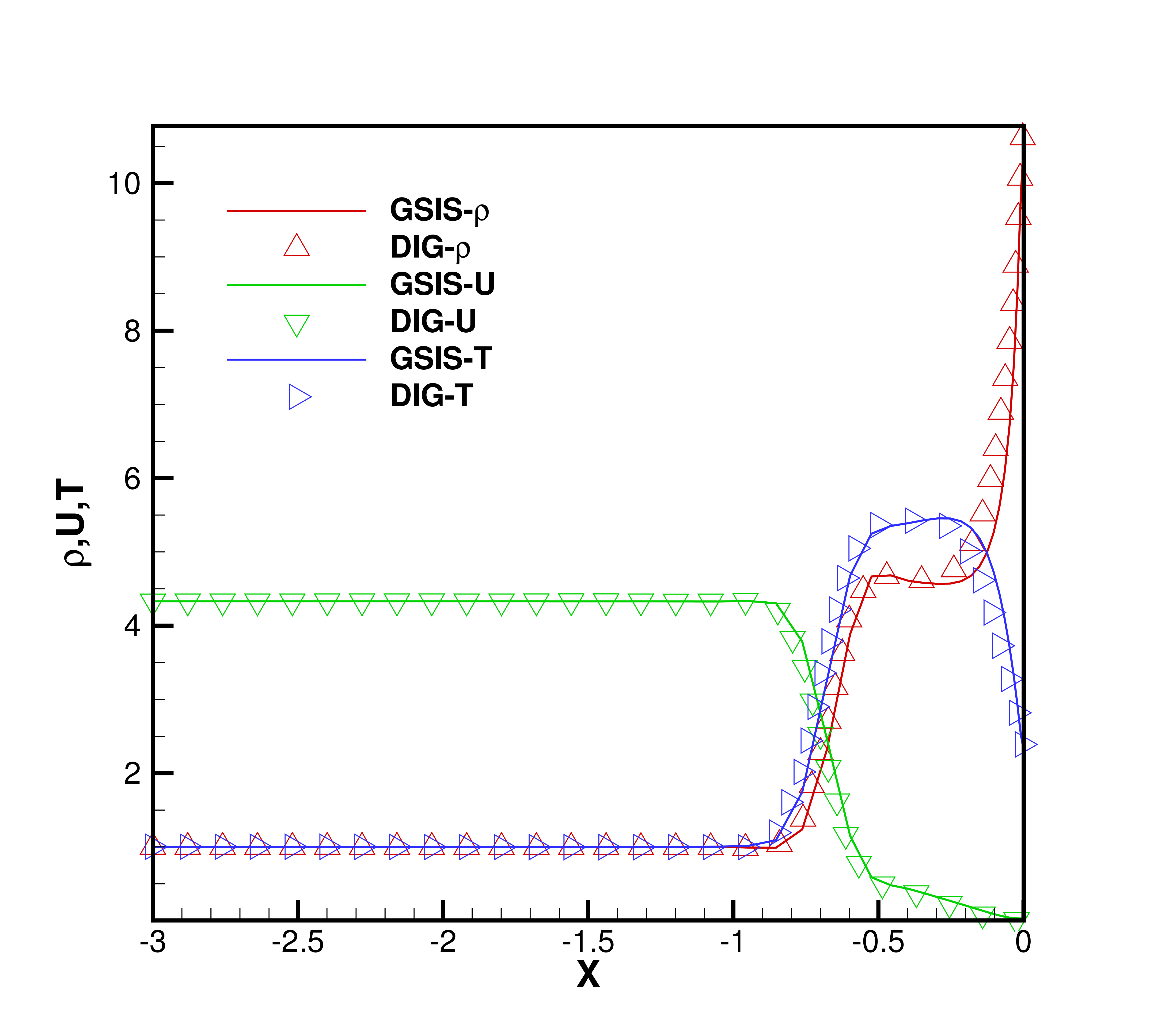}\\[-0.1cm]
		(c) $Kn=0.01$
	\end{minipage}
	\caption{Comparisons of density, velocity and temperature from $(x,y,z)=(-3,0,0)$ to $(0,0,0)$, between the DIG and DSMC in (a,b) and the DIG and GSIS in (c). The GSIS is a deterministic kinetic solver based on the simplified kinetic model of the Boltzmann equation~\cite{zhang2024parallelGSIS}. }
	\label{fig:ch5_apollo_profiles}
\end{figure}

The macroscopic profiles along a representative line on the symmetry plane of the Apollo capsule are compared in Fig.~\ref{fig:ch5_apollo_profiles}. The reference data for $Kn=1$ and $Kn=0.1$ are obtained from direct 3D SPARTA calculations, whereas the $Kn=0.01$ reference is taken from the parallel deterministic GSIS result in Ref.~\cite{zhang2024parallelGSIS}. The profiles are relatively diffuse at $Kn=1$, become more concentrated at $Kn=0.1$, and show the sharpest near-wall variation at $Kn=0.01$. In all cases the DIG results agree well with the reference solutions.

Table~\ref{tab:ch5_apollo_efficiency} summarizes the computational cost. The DIG uses the same $372{,}500$ physical cells for all three Knudsen numbers, whereas the SPARTA Cartesian grid is refined substantially as the flow approaches the near continuum regime. For $Kn=1$, the wall-clock times of SPARTA and DIG are comparable. For $Kn=0.1$, SPARTA uses $50{,}109{,}772$ cells and $600$ CPU cores, giving about $2670$ core hours, whereas the DIG uses $160$ CPU cores and $200$ core hours. The CPU cost is therefore reduced by about $13$ times, and the wall-clock time decreases from $4.45\,\mathrm{h}$ to $1.25\,\mathrm{h}$. At the same Knudsen number, the memory usage is reduced from $265\,\mathrm{GB}$ in SPARTA to about $62\,\mathrm{GB}$ in DIG, corresponding to a reduction by about $4.3$ times.

\begin{table}[t]
	\centering
	\caption{
		Computational cost in the hypersonic flow simulation around an Apollo reentry capsule. 
        The last row lists the GSIS results from a deterministic kinetic solver built on the simplified Boltzmann kinetic model~\cite{zhang2024parallelGSIS}, where $N_{\mathrm{step}}=47$ denotes the total kinetic equation solves, enabled by the noise-free deterministic formulation.
	}
	\label{tab:ch5_apollo_efficiency}
	\small
	\begin{tabular}{ccccccc}
		\hline
		$Kn$ & Method & $N_{\mathrm{cell}}$ & $N_{\mathrm{core}}$ & $N_{\mathrm{step}}$ & $t_w$ (h) & $N_p$ \\
		\hline
		\multirow{2}{*}{$1$}
		& SPARTA
		& $6{,}593{,}551$
		& $600$
		& $10{,}000+30{,}000$
		& $0.97$
		& $98{,}903{,}265$ \\
		{}
		& DIG
		& $372{,}500$
		& $160$
		& $3{,}000+27{,}000$
		& $1.03$
		& $37{,}622{,}500$ \\
		\hline
		\multirow{2}{*}{$0.1$}
		& SPARTA
		& $50{,}109{,}772$
		& $600$
		& $20{,}000+40{,}000$
		& $4.45$
		& $50{,}109{,}772$ \\
		{}
		& DIG
		& $372{,}500$
		& $160$
		& $3{,}000+27{,}000$
		& $1.25$
		& $37{,}622{,}500$ \\
		\hline
		\multirow{2}{*}{$0.01$}
		& SPARTA
		& --
		& --
		& --
		& --
		& -- \\
		{}
		& DIG
		& $372{,}500$
		& $160$
		& $5{,}000+5{,}000$
		& $0.75$
		& $37{,}250{,}000$ \\
        {}
        & GSIS~\cite{zhang2024parallelGSIS}
        & $372{,}500$
        & 128
        & 47
        & 0.36
        & -\\
		\hline
	\end{tabular}
\end{table}

When $Kn=0.01$, 10,000 evolution steps are simulated in DIG, where the first $2{,}000$ steps are used to establish stable exponentially averaged macroscopic moments, and the final $5{,}000$ steps are used for sampling. The latter value is fewer than the $27{,}000$ sampling steps used in the $Kn=1$ and $Kn=0.1$ calculations. This reduction stems from the enhanced macroscopic composite correction at low Knudsen numbers. Fourier stability analysis~\cite{hu2025fastconverging} demonstrates that the spectral radius of DIG iterations declines with decreasing Knudsen number, yielding accelerated convergence.
The cost of the direct 3D SPARTA simulation is not listed because it becomes prohibitively large. Instead, the parallel GSIS result~\cite{zhang2024parallelGSIS} provides a deterministic reference on the same $372{,}500$-cell Apollo mesh. As a deterministic solver, the GSIS method achieves convergence within 47 iterations. For DIG calculations, the GSIS-based macroscopic correction is implemented every $N_s=100$ DSMC time steps, yielding approximately 80 effective DIG correction cycles, which is on the same order of magnitude as the standalone GSIS. In terms of computational cost, GSIS is faster: it reports $0.36\,\mathrm{h}$ wall-clock time on $128$ CPU cores, while the DIG uses $0.75\,\mathrm{h}$ wall-clock time on $160$ CPU cores. However, GSIS stores the velocity-space distribution function and requires $487\,\mathrm{GB}$ of memory~\cite{zhang2024parallelGSIS}, while DIG requires $59.4\,\mathrm{GB}$. Thus, DIG reduces the memory demand by about 8 times while retaining a particle-based DSMC description. 

\begin{figure}[p]
	\centering
    \includegraphics[width=0.35\linewidth,trim={240 120 20 20}]{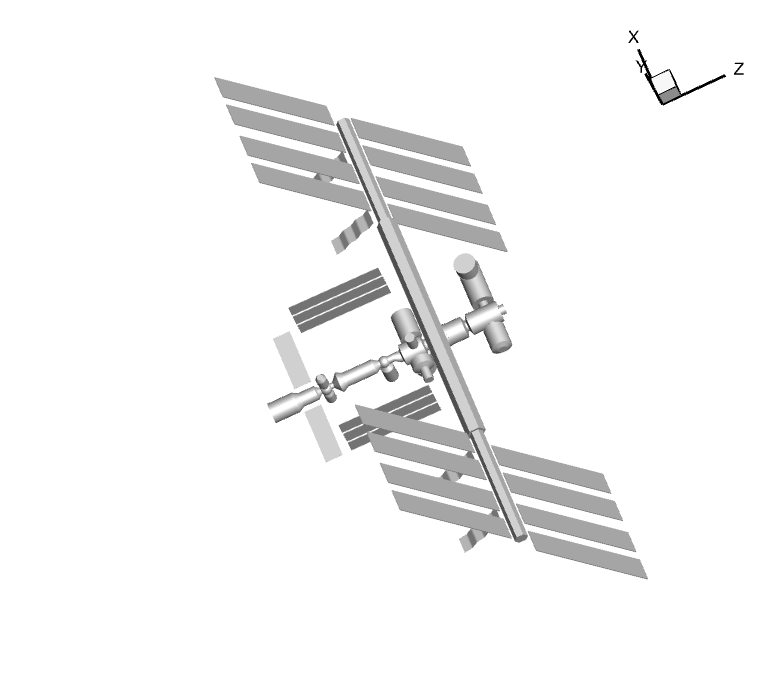}
    \hspace{0.5cm}
    \includegraphics[width=0.4\linewidth]{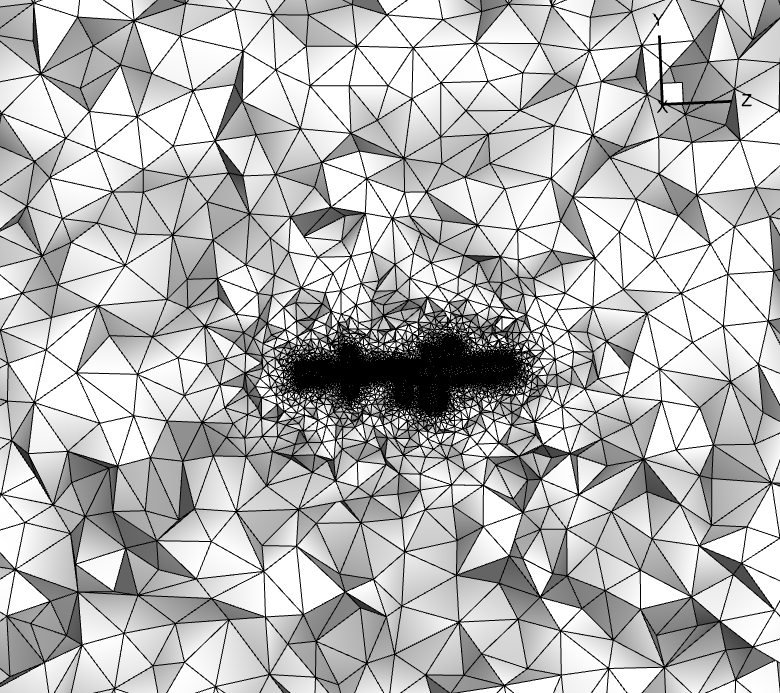}\\
    \vspace{0.5cm}
    \includegraphics[width=0.42\linewidth]{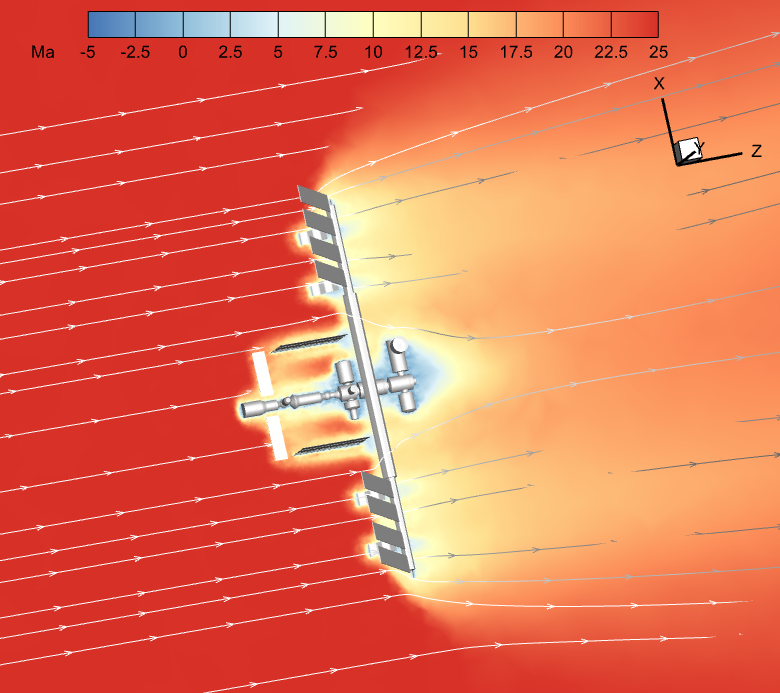}
	\includegraphics[width=0.42\linewidth]{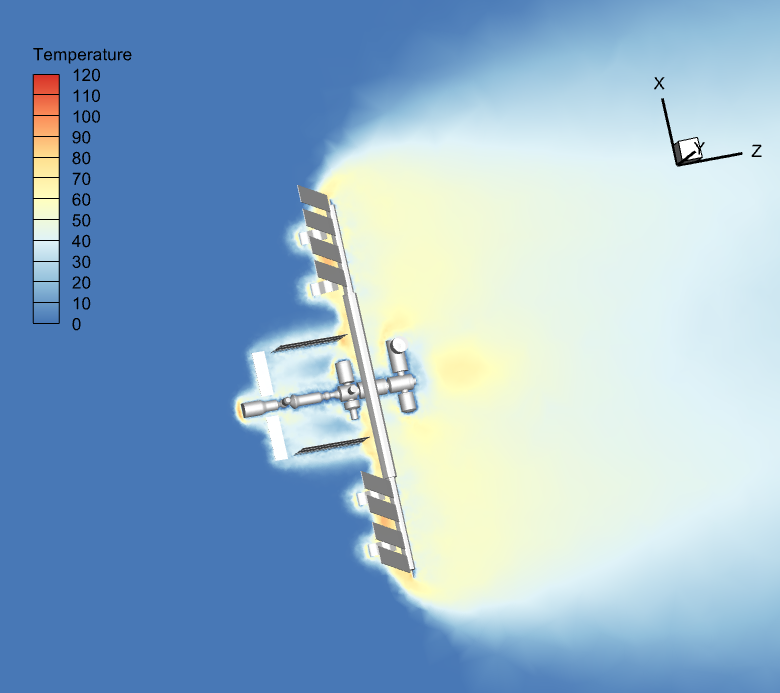}\\
    \vspace{0.5cm}
	\includegraphics[width=0.42\linewidth]{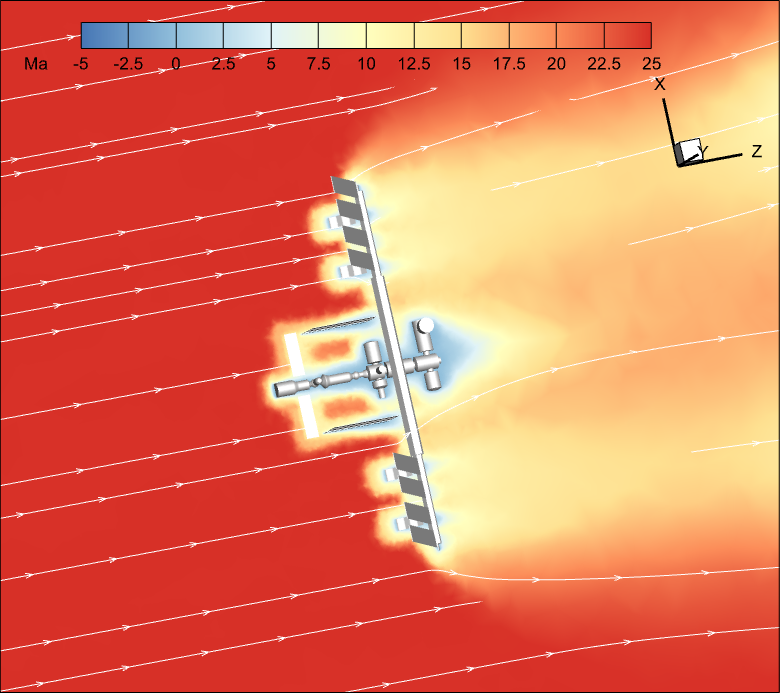}
	\includegraphics[width=0.42\linewidth]{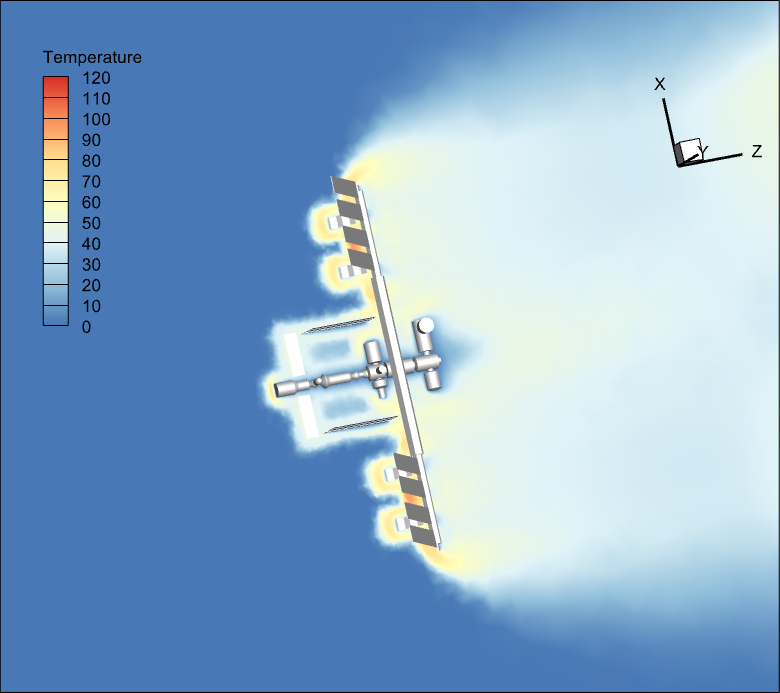}
	\caption{First row: Geometric model and surface mesh for the hypersonic flow of $Ma_{\infty}=25$  around the International Space Station~\cite{zhang2024parallelGSIS}.
    Second and third rows: Contours of the local Mach number and total temperature obtained by DIG for $Kn=0.01$ and  $Kn=1$, respectively.
    }
	\label{fig:ch5_space_station_geometry}
\end{figure}

\subsection{Flow passing the International Space Station} 
\label{subsec:ch5_space_station}

Finally, large-scale simulation of the hypersonic flow over the International Space Station is tested. The corresponding geometric model and surface mesh are illustrated in Fig.~\ref{fig:ch5_space_station_geometry}. The model contains multiple connected modules and thin solar panels. 
The computational domain consists of a total of 5,640,776 physical cells, including tetrahedral, pentahedral, triangular-prism, and hexahedral elements, and the local mesh is refined around the main body and structural details.  In DIG calculation, each cell is initialized with 100 simulation particles on average.
The freestream is imposed along the positive $z$ direction with the Mach number $Ma_{\infty}=25$ at the reference temperature $T_0=T_{\infty}=142.2\,\mathrm{K}$.
The wall temperature is $T_w=500\,\mathrm{K}$.

The local Mach number distribution in Fig.~\ref{fig:ch5_space_station_geometry} shows a strong compressed region on the windward side and a broad wake behind the space station. The total temperature field is highly nonuniform, with elevated values in the compressed and separated regions. The flow structures around the connected modules and appendages interact with each other, producing a strongly 3D wake.

The computational cost of the DIG is compared with the deterministic GSIS solver~\cite{zhang2024parallelGSIS} at $Kn=0.01$ (the reference length is $L_0=0.01\,\mathrm{m}$). The DIG calculation is performed on 200 CPU cores, and the timing is measured over $10{,}000$ DSMC time steps. The wall clock time is $4.88\,\mathrm{h}$ and the memory usage is $339\,\mathrm{GB}$. For the corresponding GSIS calculation, it costs a wall clock time of $0.87\,\mathrm{h}$ on 9216 CPU cores and a memory usage of $21.5\,\mathrm{TB}$. Thus, GSIS gives a shorter wall clock time for this case, whereas DIG reduces the CPU cost by about $8.2$ times and the memory usage by about $63$ times. The high cost of GSIS is related to the large velocity space discretization required by high-speed flow. This result indicates that the parallel DIG has the capability for large-scale 3D computations over complex configurations.

\subsubsection{Parallel efficiency}

The parallel scalability of the DIG method is further examined on a large scale unstructured mesh containing 5,640,776 cells. A strong scaling test is then performed for the statistically steady $Kn=1$ and $Ma_{\infty}=25$ case using the same processor organization as in Fig.~\ref{fig:parallel-organization-overall}. Starting from this steady particle state, $10{,}000$ DSMC time steps are advanced and the corresponding wall clock time is recorded. This test is used to assess the parallel performance of the DSMC stage, with the particle load set to $10\%$ of that used in the production calculation.

\begin{table}[t]
	\centering
	\caption{Parallel efficiency of the DIG for hypersonic flow over the International Space Station for $Kn=1$ and $Ma_{\infty}=25$.
		The wall clock time is measured for $10{,}000$ DSMC time steps after the flow reaches a statistically steady state.
	}
	\label{tab:ch5_space_station_parallel_efficiency}
	\begin{tabular}{lcccccc}
		\hline
		Cores & $120$ & $200$ & $400$ & $600$ & $800$ & $1000$ \\
		\hline
		$t_w$ (s) & 16,622 & 9,885 & 6,052 & 4,977 & 3,815 & 3,743 \\
		$S_p$ & 1.0 & 1.7 & 2.7 & 3.3 & 4.4 & 4.4 \\
		ideal speedup & 1.0 & 1.7 & 3.3 & 5.0 & 6.7 & 8.3 \\
		$\eta_p$ (\%) & 100 & 100.9 & 82.4 & 66.8 & 65.4 & 53.3 \\
		\hline
	\end{tabular}
\end{table}

Table~\ref{tab:ch5_space_station_parallel_efficiency} lists the wall clock time for $10{,}000$ DSMC time steps together with the corresponding actual speedup and parallel efficiency, using the 120 core result as the reference. The wall clock time decreases from $16{,}622$ s on 120 cores to $3{,}743$ s on 1000 cores, corresponding to an actual speedup of $4.4$ and a parallel efficiency of $53.3\%$. The efficiency remains above $80\%$ up to 400 cores and stays around $65\%$--$67\%$ at 600 and 800 cores, indicating that the DIG method remains effective on this large scale unstructured mesh. When the number of CPU cores is further increased, the additional gain becomes weaker because the work assigned to each process becomes smaller, while the relative cost of MPI communication, particle transfer, synchronization, and residual load imbalance becomes more important.

The particle distribution on the space station mesh remains highly nonuniform because of the strong compression regions, the broad wake, and the numerous geometric details. Compared with the sphere case, the higher Mach number and the more complex configuration make residual load imbalance more difficult to remove completely at larger core counts and further reduce the benefit of additional repartitioning. The efficiency slightly above $100\%$ at 200 cores is attributed to the stochastic fluctuation of the particle method.

\section{Conclusions}\label{sec:conclusion}

In summary, this work develops a parallel DIG-augmented DSMC solver tailored for 3D rarefied gas flow simulations on unstructured meshes, which integrates a complete set of novel algorithms including intermittent micro-macro coupling, virtual-cell-based batch particle migration, and graph-partitioning dynamic load balancing. The core contributions and key findings are summarized as follows:
\begin{enumerate}

\item A hybrid MPI parallel architecture is implemented for unstructured grids, separating one dedicated storage process and multiple compute processes. Auxiliary virtual cells are introduced to enable complete local particle tracing within each time step, deferring inter-rank communication and enabling batch particle migration to mitigate frequent small MPI messages. A weighted graph partitioning strategy powered by METIS is integrated for dynamic load balancing, where cell weights are defined by either particle population or measured runtime timing, and edge weights reflect inter-cell particle exchange intensity. The Kuhn–Munkres assignment further minimizes redundant data redistribution during re-partitioning, effectively alleviating workload imbalance induced by highly non-uniform particle distributions across shock layers, wall boundary layers and wake regions. Parallel scaling tests demonstrate satisfactory parallel scalability.

\item  A multi-layer stability enhancement strategy is designed for irregular unstructured meshes with tiny cells prone to severe statistical noise. It combines cell admissibility screening with a cell Knudsen number threshold check, local moment accumulation for under-sampled cells, and logarithmic-averaged field reconstruction from reliable neighboring cells to guarantee robust micro-macro coupling. Exponentially weighted moving averaging is applied to smooth noisy sampled moments, balancing statistical noise suppression and response speed to transient flow variations.

\item The asymptotic-preserving characteristic of DIG fundamentally eliminates the stringent kinetic-scale constraints of conventional DSMC. The solver allows vastly coarser spatial discretization and drastically fewer statistical sampling steps, leading to dramatic reductions in mesh count, memory footprint and total computational overhead, especially prominent in the near-continuum regime. Four representative 3D test cases are adopted for comprehensive numerical validation across a wide Knudsen number spectrum. Macroscopic quantities predicted by the DIG solver show excellent agreement with benchmark SPARTA DSMC results. For the 3D lid-driven cavity benchmark at \(Kn=0.01\), the DIG solver cuts core-hour consumption by a factor of approximately 362 and reduces memory by 50 times compared with the reference SPARTA DSMC code. For complex reentry capsule geometry at \(Kn=0.1\), DIG reduces core-hour cost by around 13.4 times and memory consumption by 4.3 times relative to SPARTA. When 
$Kn=0.01$, running a full 3D SPARTA DSMC simulation requires impractically massive computational resources. As a substitute reference, the deterministic GSIS method is adopted for comparison: the DIG solver achieves similar convergence speeds while reducing memory overhead by roughly a factor of 8.2.

\end{enumerate}

Future work will target full support for chemically reacting gas systems; although preliminary DIG extensions addressing gas mixtures and reactive flows have been documented, the present parallel unstructured-mesh solver will be upgraded to comprehensively model multi-component transport alongside dissociation, ionization reactions and even radiation representative of hypersonic reentry environments, while preserving the high-fidelity collision and energy exchange physics native to standard DSMC.

\section*{Declaration of competing interest}
The authors declare that they have no known competing financial interests or personal relationships that could have appeared to influence the work reported in this paper.

\section*{Acknowledgments}
This work was supported by the National Natural Science Foundation of China (Grant No.~12450002). Special thanks are given to the Center for Computational Science and Engineering at the Southern University of Science and Technology.

\bibliographystyle{elsarticle-num}
\bibliography{ref}

\end{CJK}
\end{document}